\documentclass[a4paper,11pt]{article}
\pdfoutput=1 

\usepackage{jheppub} 

\usepackage{amsmath}
\usepackage{slashed}
\usepackage[normalem]{ulem}
\usepackage{graphicx}
\usepackage[utf8]{inputenc}
\usepackage{snapshot}

\newcommand{\beq}{\begin{equation}}
\newcommand{\eeq}{\end{equation}}
\newcommand{\beqn}{\begin{eqnarray}}
\newcommand{\eeqn}{\end{eqnarray}}

\preprint{\begin{flushright}Nikhef 2019-045\\UWTHPH-2019-30\\ MCnet-19-23\end{flushright}}

\title{On the phenomenology of sphaleron-induced processes at the LHC and beyond}

\author[a,b]{Andreas Papaefstathiou,}
\author[c]{Simon Pl\"atzer,}
\author[d]{Kazuki Sakurai,}

\affiliation[a]{Institute for Theoretical Physics Amsterdam and Delta Institute for Theoretical Physics, University of Amsterdam, Science Park 904, 1098 XH Amsterdam, The Netherlands.}
\affiliation[b]{Nikhef, Theory Group, Science Park 105, 1098 XG,
  Amsterdam, The Netherlands.}
\affiliation[c]{Particle Physics, Faculty of Physics, University of Vienna, Vienna, Austria.}
\affiliation[d]{Institute of Theoretical Physics, Faculty of Physics, University of Warsaw, ul. Pasteura 5, PL–02–093 Warsaw, Poland.}

\emailAdd{apapaefs@cern.ch}
\emailAdd{simon.plaetzer@univie.ac.at}
\emailAdd{kazuki.sakurai@fuw.edu.pl}

\abstract{We investigate the phenomenological aspects of
  non-perturbative baryon- and lepton-number-violating processes at
  hadron colliders.  Such processes, induced by instanton/sphaleron
  configurations of the electroweak gauge fields, are believed
  to play a crucial role in the generation of baryon asymmetry in the
  early Universe at finite temperature.  On the other hand, at
  colliders (that represent the zero-temperature high-energy regime) the rate and
  observability of such processes are still under debate.  Motivated
  by current theoretical considerations, we construct a modern event
  generator within the general-purpose \texttt{Herwig} Monte Carlo
  framework, that aims to capture the most relevant features of the
  dominant processes. We perform a detailed phenomenological analysis
  focussing on the Large Hadron Collider, at 13~TeV proton-proton
  centre-of-mass energy, a potential high-energy upgrade at 27~TeV and
  the proposed Future Circular Collider (FCC-hh) at 100~TeV. We derive
  constraints on the expected rates for various parametrisations of
  our model. We find that all three colliders are capable of providing
  meaningful information on the nature of instanton/sphaleron-induced
  processes at various energy scales.}

\begin{document} 
\maketitle
\flushbottom

\section{Introduction}
\label{sec:intro}

Discerning the details of dynamics of baryon-number violation would be a crucial step towards an ab initio understanding of the observed baryon asymmetry of the Universe. 
In particular, a class of baryon-number-changing processes associated with electroweak theory 
has been long studied~\cite{tHooft:1976rip}. 
The computation of amplitudes for such transitions employs approximate classical solutions of the electroweak theory, known as instantons. At zero temperature, with zero energy, the amplitudes for such tunnelling processes can be estimated to be of order $\exp{[-2\pi/\alpha_w]} \sim \mathcal{O}(10^{-82})$, where $\alpha_w \sim 1/30$ is the SU(2) coupling constant. Evidently this would have rendered the processes in question phenomenologically irrelevant and this article particularly short.

\begin{figure}[!t]
\centering 
\includegraphics[width=.55\textwidth,origin=c,angle=0]{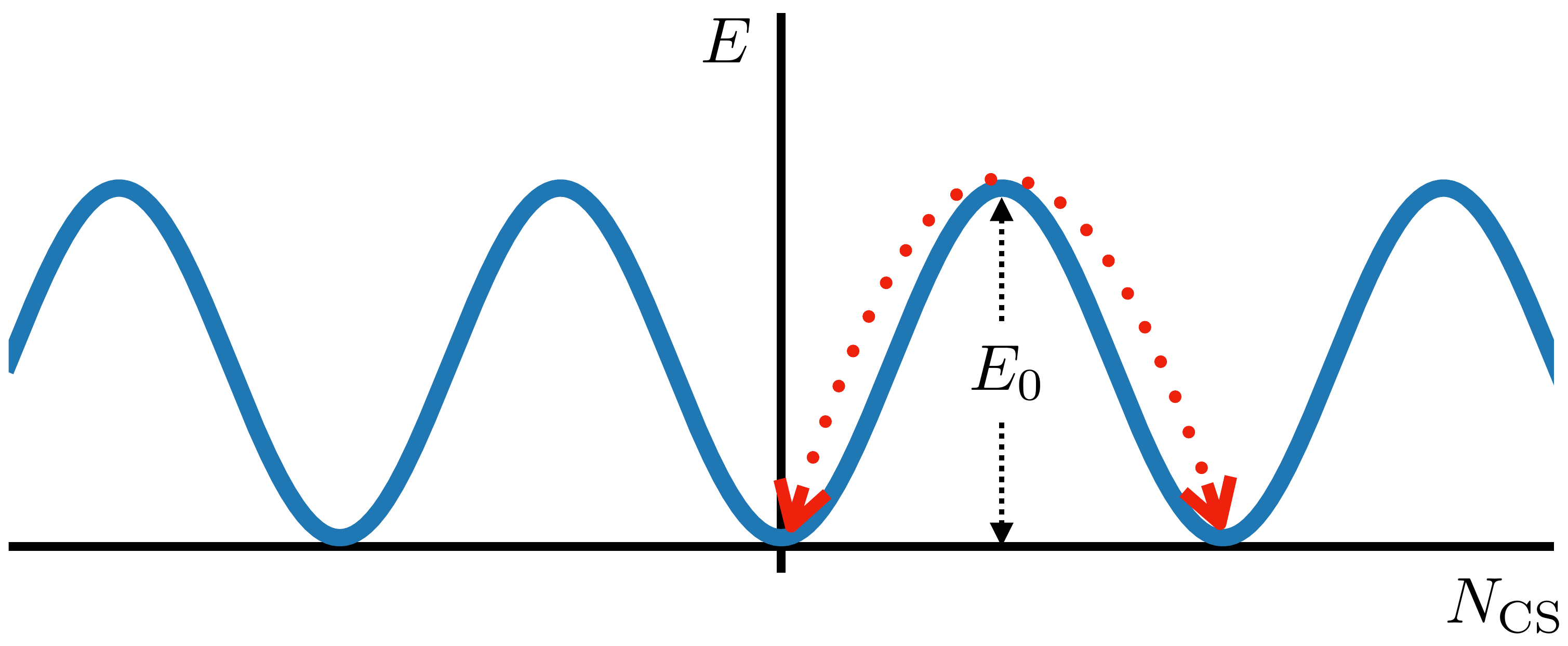}
\caption{\label{fig:barrier} A schematic diagram of the degenerate Chern-Simons vacua, separated by a barrier of energy $E_0$. The result of a sphaleron transition between two of these vacua is illustrated.}
\end{figure}

Nevertheless per aspera ad astra,\footnote{Latin, ``Through hardships to the stars"~\cite{wiki:001}.} and a number of subsequent calculations (see e.g.~\cite{Ringwald:1989ee, Espinosa:1989qn, Arnold:1987zg, McLerran:1989ab, McLerran:1990ed,Mattis:1991bj}) have shown that the rate of instanton-induced processes exponentially grows 
with energy $E$, albeit in the limit $E \ll E_0$, where $E_0$ is the energy scale at which the instanton approximation itself breaks down. More specifically, $E_0$ is the height of the barrier that separates sectors of the electroweak vacuum, characterised by different values of the so-called Chern-Simons (or winding) number, $N_{\rm CS}$, as demonstrated schematically in Fig.~\ref{fig:barrier}. Furthermore, there exist static solutions of the classical equations of motion that are unstable and sit on top of the barrier. These are the so-called ``sphalerons''. 
The existence of these solutions allows a transition from one EW vacuum to another dynamically, going over the barrier
with energies larger than $E_0$,
as schematically illustrated in Fig.~\ref{fig:barrier}.
Unlike instantons, this type of Chern-Simons number changing processes is not quantum tunneling,
and thus not necessarily exponentially suppressed.  

There have been several attempts to estimate the rate of 
the instanton/sphaleron-induced EW vacuum transition processes 
in the high-energy regime ($E \gtrsim E_0$). 
For example, studies exploiting a semiclassical approximation concluded 
that the exponential suppression persists even at energy higher than 250 TeV \cite{Bezrukov:2003er}.
As was pointed out in the seminal paper by Klinkhamer and Manton~\cite{Klinkhamer:1984di},
this may be due to a ``few-to-many'' suppression, which stems from a necessity, and difficulty, 
of assembling (highly-coherent and extended) instanton/sphaleron configurations
from ordinary two particle states in the collision.\footnote{See also \cite{Funakubo:2016xgd}.} 
On the other hand, other estimations based on the optical theorem suggest that the EW vacuum transition rate 
may become unsuppressed at energies around or above 20 TeV \cite{Ringwald:2002sw,Ringwald:2003ns}.
As pointed out in \cite{Tye:2017hfv}, the aforementioned few-to-many suppression may not be present
because emitting one virtual gauge boson contributes a factor $g^{-1}$ to the amplitude in the instanton background, 
rather than $g$ as in the perturbative vacuum,
and many gauge bosons can relatively easily be produced and assembled into a coherent state.
A more recent study pointed out that 
it may be important to take the periodicity of the EW potential (see Fig.~\ref{fig:barrier}) into account,
since the vacuum transition rate can be enhanced due to the resonant tunneling effect. 
They have estimated the EW vacuum transition rate by analysing the band structure of the spectrum
and concluded that the instanton/sphaleron processes may become observably large 
at energies around or above 9 TeV \cite{Tye:2015tva,Tye:2017hfv}.
Motivated by these encouraging estimates,
several phenomenological studies on the zero-temperature instanton/sphaleron-induced processes 
have been carried out recently  
\cite{Ellis:2016ast,Ellis:2016dgb,Brooijmans:2016lfv,Spannowsky:2016ile,Jho:2018dvt,Cerdeno:2018dqk,Ringwald:2018gpv,Anchordoqui:2018ssd}.  

\begin{figure}[!htp]
\centering 
\includegraphics[width=.45\textwidth,origin=c,angle=0]{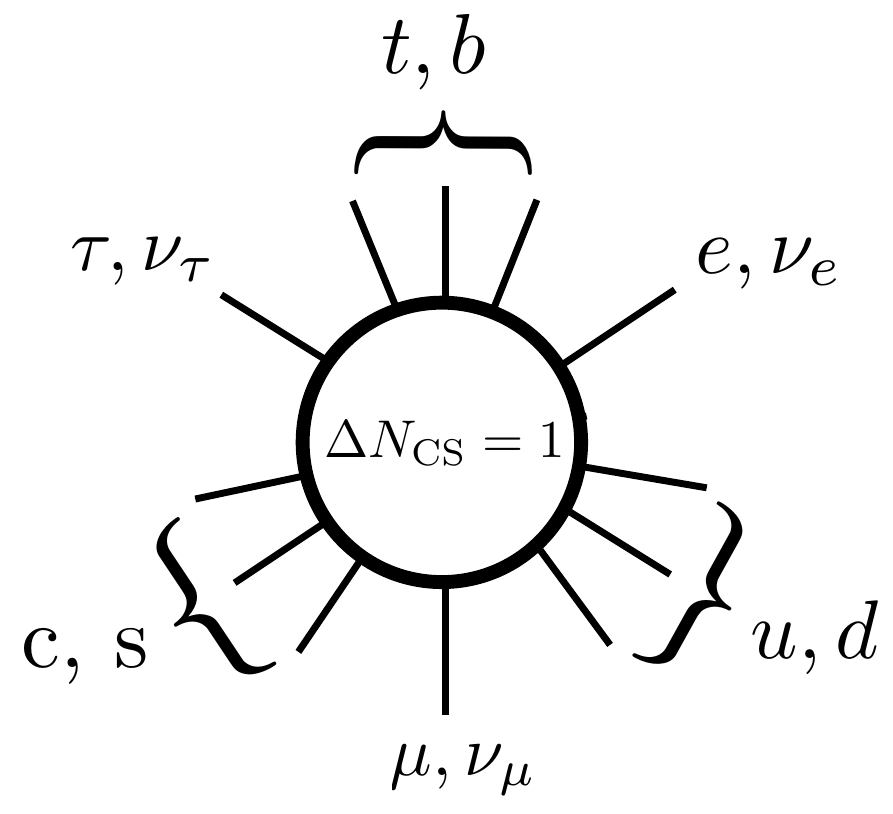}
\caption{\label{fig:vertex} A schematic diagram of flavour structure of the process related to a unit change in Chern-Simons number, $\Delta N_{\rm CS}  = 1$.}
\end{figure}

Although the potential for observing instanton/sphaleron-induced processes at colliders is not theoretically clear, 
one can instead turn to experiments to address the issue.
Compared to the large uncertainty on the event rate, the signatures of such processes are relatively well understood.
Due to the coupling of the fermions to the SU(2) gauge fields and the presence of the anomalous divergence of the axial-vector current, a change in $N_{\rm CS}$ implies a change in baryon ($B$) and lepton ($L$) numbers $\Delta L_e = \Delta L_\mu = \Delta L_\tau = \frac{1}{3} \Delta B =  \Delta N_{\rm CS}$ as in Fig.~\ref{fig:vertex}. The enhancement is expected to occur when the process involves a large number ($\sim 1/\alpha_w$) of gauge and/or Higgs bosons.\footnote{The relation between the change in the fermion and Chern-Simons numbers, as well as the enhancement associated with the large number of bosons, are illustrated rather neatly in the context of the two-dimensional Abelian Higgs model, see e.g.~\cite{Arnold:1987zg, McLerran:1989ab}.}
The basic process would involve 12 left-handed fermions: 3 quarks for each generation and one lepton for each generation (Fig.~\ref{fig:vertex}). Hence, processes that could be observable at hadron colliders, schematically, would be of the form:
\begin{equation}\label{eq:sphaleronproc}
q + q \rightarrow 7 \bar{q} + 3 \bar{\ell} + n_B W/Z/\gamma/H\;,
\end{equation}
where $q$, $\bar{q}$, $\bar{\ell}$ denote a quarks, anti-quarks and anti-leptons, respectively, and $n_B$ is the total number of gauge and Higgs bosons. The charge structure of this process is explained in detail in the next section. From here on, we will refer to such processes as being ``sphaleron'' induced, emphasising the phenomenological nature of our analysis. We consider various parametrisations of the distributions of gauge bosons. 
We note, however, that we neglect the helicity of the produced fermions in our Monte Carlo (MC) simulation. 
We expect this to have a negligible effect on angular distributions that would be washed out by hadronization and other effects.\footnote{Indeed, it is already challenging to determine the helicity of top quarks, that do not hadronize, even in processes that are less populated, see e.g.~\cite{Papaefstathiou:2011kd}.}
 
The paper is organised as follows: we describe the MC simulation of sphaleron-induced processes in section~\ref{sec:simulation}. There, we also present a discussion of the uncertainties present in our parametrisation. In section~\ref{sec:pheno} we present a study of the phenomenology of the processes at hadron colliders such as the CERN LHC at centre-of-mass energy of 13 TeV, a potential future upgrade to 27 TeV and a potential Future Circular Collider colliding protons at 100~TeV. 

\section{Monte Carlo simulations of sphaleron-induced processes}
\label{sec:simulation}

We begin by describing the details of the MC event generator for sphaleron-induced processes that we have built, including the flavour structure, colour factors and the generation of the phase space. We also discuss the cross section in the context of unitarity and the parametrisation of the distribution of the number of gauge bosons.

\subsection{Process generation}

\begin{figure}[tbp]
\centering 
\includegraphics[width=.35\textwidth]{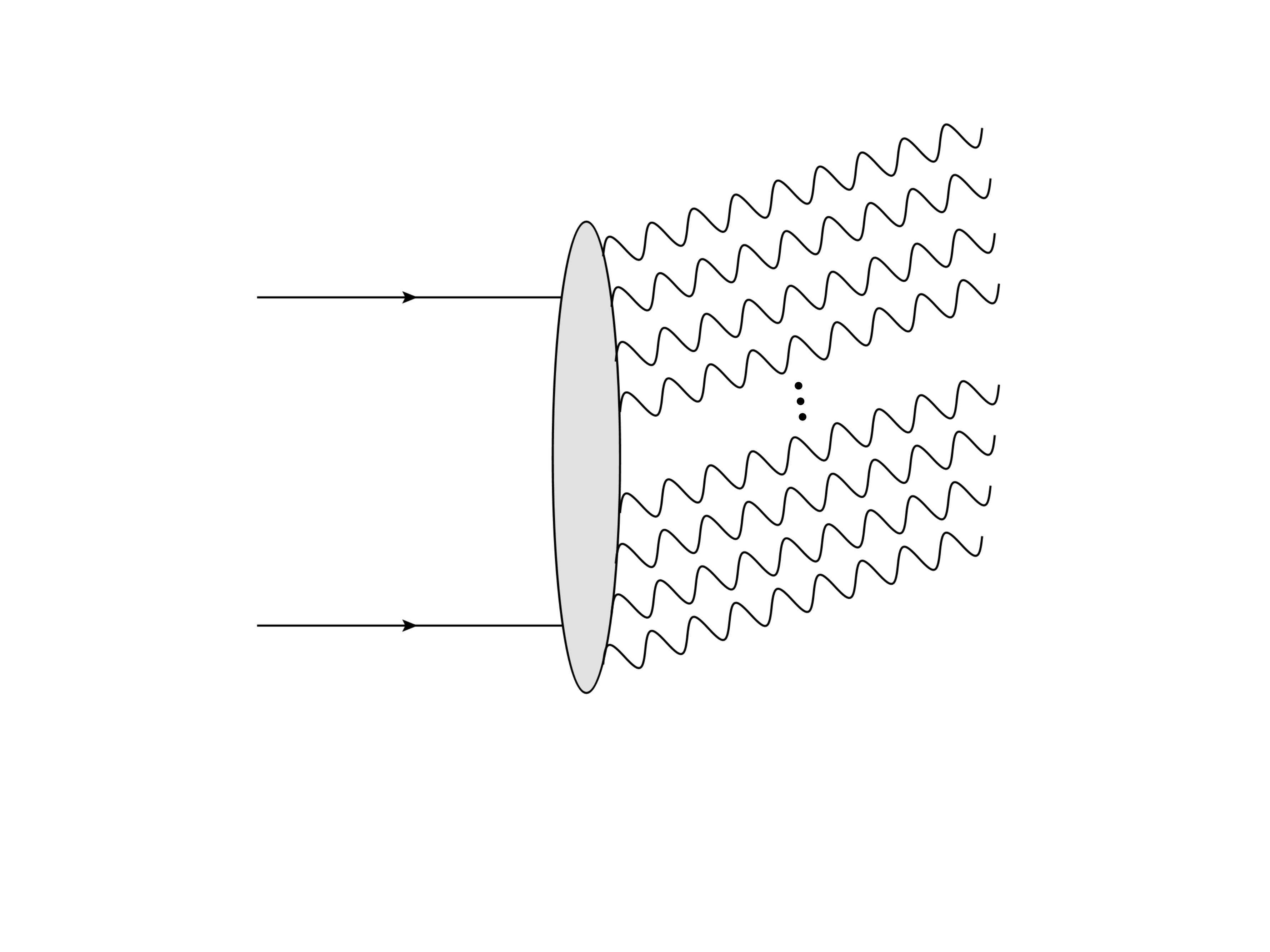}
\hfill
\caption{\label{fig:blob} A schematic diagram of a $2\rightarrow n$ processes that can be simulated via the ``blob'' matrix elements within the \texttt{Herwig} general-purpose event generator.}
\end{figure}

To facilitate the generation of $2\rightarrow n$ processes in all generality, we have constructed customised infrastructure within the general-purpose \texttt{Herwig} Monte Carlo event generator~\cite{Bahr:2008pv, Gieseke:2011na, Arnold:2012fq, Bellm:2013hwb, Bellm:2015jjp, Bellm:2017bvx}. This allows for the generation of ``blob'' matrix elements (MEs) through the definition of any $2\rightarrow n$ process that can be viewed as a contact interaction.\footnote{This functionality should be particularly useful in the simulation of final states such as QCD instantons (e.g.~\cite{Gibbs:1995bz, Moch:1996bs, Ringwald:1999jb}), microscopic black holes (e.g.~\cite{Harris:2003db, Harris:2004xt, Frost:2009cf, Dai:2007ki}) and other non-perturbative multi-particle processes~\cite{Khoze:2017tjt}.} An example of such a process is shown in Fig.~\ref{fig:blob}. We note that \texttt{Herwig} possesses the necessary infrastructure to handle the SU(3) colour source and sinks that appear in the $B$-violating processes~\cite{Richardson:2000nt}. We defer the description of the technical details to a future release note of \texttt{Herwig}. 

For the simulation of sphaleron-induced processes, we only include the dominant $qq$ initial state given by Eq.~\eqref{eq:sphaleronproc}, where two valence quarks collide. 
The conjugate process is suppressed in the $pp$ collider due to a small ratio of the luminosity functions 
between valence-quark ($qq$) and sea-quark ($\bar q \bar q$) initial states at $\sqrt{\hat s} \gtrsim {\cal O}(10)$ TeV.
We take the initial-state partons to consist only of the light quarks $u$ and $d$ and we allow all quark generations in the produced final states, including the top and bottom quarks.  
The quark and lepton content of the process can be thought of as originating in a class of operators
of the form
\beq
\mathcal{O}_\mathrm{\slashed{B}\slashed{L}} \sim (Q_1 Q_2 Q_3) (Q'_1 Q'_2 Q'_3) (Q''_1 Q''_2 Q''_3) L^1 L^2 L^3
, 
\label{eq:op}
\eeq
where $Q$ and $L$ are the left-handed quark and lepton doublets, respectively. The anomaly argument suggests that all the degrees of freedom that are orthogonal to the SU(2)$_L$ gauge group, i.e.\ colour and flavour, must appear at least and only once.  With this condition, the colour indices of the quark fields must be constructed properly, making three ``baryonic'' (anti-symmetric) colour singlet combinations $(QQQ)$. Furthermore, the SU(2)$_L$ indices of all quarks and leptons must be constructed in such a way that the operator is a SU(2)$_L$ singlet. In terms of the left-handed doublet components, i.e.\ $(u, d)_L$ and $(\ell, \nu)_L$, this condition is equivalent to requiring the net electro-magnetic charge of the above operators is zero.\footnote{See the original article by t'Hooft~\cite{tHooft:1976rip} or e.g.~\cite{McLerran:1989ab} for further details.}
In the hard process we generate colour flows compatible with this pattern on an equally likely, random basis.\footnote{Future work, however, should investigate this issue in more detail, specifically in presence of instanton-induced processes with a large number of additional gluons.}


To be concrete, e.g.\ the combination $(uct)(dsb)(uct)e \mu \tau$ would form an allowed operator, but the combinations $(udt)(csb)(uct) e\mu\nu_\tau$, $(uut)(dsb)(uct) e\mu\nu_\tau$ or $(uct)(dsb)(uct)ee\nu_\tau$ would \textit{not} be allowed, each violating one of the aforementioned conditions. 
In the event generation, we take into account all the possible allowed permutations of the quark flavours within the operator, that would lead to identical flavour content in the final state, by multiplying by an appropriate combinatorial ``colour'' factor.
In other words, e.g.\ the final state $ud \rightarrow \bar{c} \bar{t} \bar{d} \bar{s} \bar{b} \bar{u} \bar{c}\bar{t} e^+ \mu^+ \bar{\nu}_\tau$ appears only once in the \texttt{Herwig} process generation, with its weight multiplied by the appropriate combinatorial colour factor. Explicitly, e.g.\ in the case of the $ud \rightarrow \bar{c} \bar{t} \bar{d} \bar{s} \bar{b} \bar{u} \bar{c}\bar{t} e^+ \mu^+ \bar{\nu}_\tau$ process, this factor is given by $ [1/(3!/2!)] \times (3!/2!)^3 = 9$, where the first factor takes into account that two of the first generation quarks have to be chosen to form the initial state and there is a factor of $(3!/2!)$ for each of the generations. At the stage of MC event generation, the colour configuration which determines which combinations of quarks form the colour singlets is chosen uniformly at random from the allowed ones.

\subsection{Phase space generation and integration}

Due to the lack of a detailed calculation describing the
non-perturbative sphaleron-induced processes differentially, we
distribute the particles uniformly in the available phase space. Two
methods are particularly suited to the sampling of ``flat'' phase
space at high multiplicities: the \texttt{RAMBO}~\cite{Kleiss:1985gy}
algorithm as described in~\cite{Platzer:2013esa}, and the
\texttt{MAMBO} algorithm according to~\cite{Kleiss:1991rn}. The former
was found to be inefficient for large numbers of massive
particles,\footnote{The low efficiency was in fact the motivation for
  the \texttt{MAMBO} algorithm of~\cite{Kleiss:1991rn}.} and therefore
for all the phenomenological studies of the present article we have
employed the \texttt{MAMBO} algorithm. Given the structure of the
cross section, a detailed adaptive Monte Carlo integration is not
required and we choose to predetermine a constant cross section value
to select processes of equal final-state multiplicity. This further
speeds up the phase space generation and integration. A sequence of
different phase space generators might be considered in the future in order to
attach definitive decays, or initial state splittings attached to the
``blob'' matrix element's legs.

\subsection{Leading-order matrix element, unitarity and boson distribution}

Our estimation for the boson multiplicity is based on the cross
section formula from the leading-order matrix element (LOME) approach
\cite{Ringwald:1989ee,Espinosa:1989qn,Khoze:1990bm},
\beqn \hat
\sigma_{\rm LOME}(\bar n_B, \bar n_H) ~=~ C {\cal G}^2 2^{n_B} v^{-2n}
\left[ \frac{\Gamma(n + 103/12)}{\Gamma(103/12)} \right]^2
\frac{1}{\bar n_B! \bar n_H! }  \\ \nonumber \int \prod_{i=1}^{10}
\frac{d^3 p_i E_i}{(2 \pi)^3 2 E_i} \prod_{j=1}^{\bar n_B} \frac{d^3
  p_j }{(2 \pi)^3 2 E_j} \frac{2 (4 E_j^2 - m_W^2) }{m_W^2}
\prod_{k=1}^{n_H} \frac{d^3 p_k}{(2 \pi)^3 2 E_k} 
\\ \nonumber
(2 \pi)^4 
\delta^4( P_{\rm in} - \sum_{i=1}^{10} p_i - \sum_{j=1}^{\bar n_B} p_j - \sum_{k=1}^{\bar n_H} p_k ) \,,
\label{eq:LOME}
\eeqn
where $\bar n_B$ is the number of electroweak (EW) gauge bosons ($W$, $Z$, $\gamma$), $\bar n_H$ is the number of Higgs bosons in the final state.
Also, $n_B \equiv \bar n_B + \bar n_H$, $n=n_B+10$ is the total number of final state particles,
$v$ represents the VEV of the Higgs field, ${\cal G} \equiv 1.6 \cdot 10^{-101}\,{\rm GeV}^{-14}$ is an effective coupling constant, and $C$ is some unknown constant.

While the above expression gives the cross section as a function of $\sqrt{\hat s}$ and
the multiplicity of bosons, there are several issues.
Firstly, this formula is only valid in a low energy regime, $\sqrt{\hat s} \ll M_0$, where
$M_0 \equiv \sqrt{6}\pi m_W \alpha_W^{-1} \sim 18$ TeV.
In fact, the cross section grows exponentially in this regime and eventually exceeds the 
$s$-wave
unitarity bound\footnote{Note that both the instanton and sphaleron field configurations are approximately spherically symmetric, hence
the $s$-wave unitarity bound is expected to apply.}
\beq
\hat \sigma^s_{\rm unitary}(\sqrt{\hat s}) \,=\, \frac{16 \pi}{\hat s} \;,
\eeq
for $\sqrt{\hat s} \gtrsim M_0$.
This motivates the phenomenological parametrisation of the sphaleron cross section:
\beq
\hat \sigma(\sqrt{\hat s}) \,=\, \min( \hat \sigma_0, \, \hat \sigma^{s}_{\rm unitary} ) \;,
\label{eq:sighat}
\eeq
with 
\beq
\hat \sigma_0(\sqrt{\hat s}) \,=\, \frac{p_{\rm sph}}{m^2_W} \Theta( \sqrt{\hat s} - E_{\rm thr}) \;,
\label{eq:sig0}
\eeq
where $\Theta(x)$ is the Heaviside step function.
\begin{figure}[tbp]
\centering 
\includegraphics[width=.5\textwidth,clip]{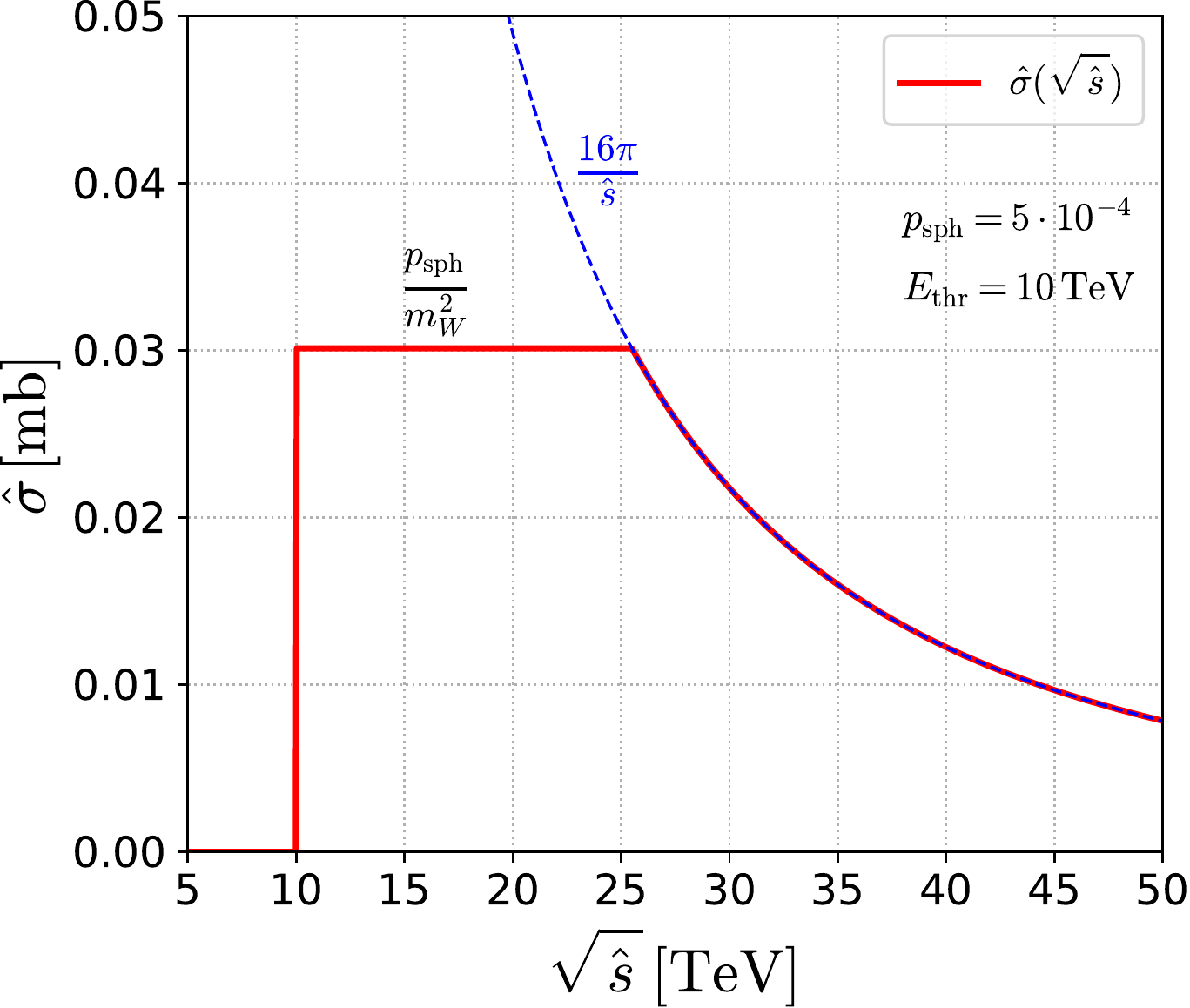}
\caption{\label{fig:unitarity} The partonic $(B+L)$-violating cross section, $\hat \sigma$, as a function of the centre of mass energy, $\sqrt{\hat s}$.
The blue dashed line represents the $s$-wave unitarity bound.}
\end{figure}
The $(B+L)$-violating partonic cross section is parametrised by two unknown parameters, $p_{\rm sph}$ and $E_{\rm thr}$.
The former is a dimensionless coefficient, which controls the overall size of the cross section.
The latter represents the threshold energy where the partonic cross section is turned on,
i.e.\ this is the scale at which the exponential suppression in the instanton formula is overcome 
by the exponential growth of cross section with the energy.
Our cross section formula Eq.~\eqref{eq:sighat} always respects the $s$-wave unitarity bound,
as illustrated in Fig.~\ref{fig:unitarity}.
This is in contrast to the previous phenomenological studies,
e.g.~\cite{Ellis:2016dgb,Ellis:2016ast,Sirunyan:2018xwt,Ringwald:2018gpv}, 
where only an equation of the form of Eq.~\eqref{eq:sig0} was used on its own to describe the partonic cross section.

\subsection{Parametrisation of boson distributions}

Our estimation of the boson multiplicity distributions is based on the LOME formula given by Eq.~\eqref{eq:LOME}.
This formula predicts the total number of bosons to be $\langle n_B \rangle \gtrsim \frac{3}{2} \frac{\pi}{\alpha_W} \left( \frac{\sqrt{\hat s}}{M_0} \right)^{4/3}$ on average \cite{Gibbs:1995bt}.
As was done in \cite{Farrar:1990vb, Gibbs:1995bt}, we fit the boson distributions generated from Eq.~\eqref{eq:LOME}
with the Gaussian:%
\footnote{We note that the sign of the exponent in Eq.~(17) of~\cite{Gibbs:1994cw} should be $(n-a)$ instead of $(n+a)$.}
\begin{equation}
\mathcal{P} (n_B, \hat{s}/\mu^2) = \exp \left( - \frac{(n_B-a(\hat{s}/\mu^2)^2}{b(\hat{s}/\mu^2)} \right)\;,
\end{equation}
where $n_B$ is the total boson multiplicity (i.e.\ $W$, $Z$, $\gamma$ and Higgs bosons)
and $a$ and $b$ are functions of $\hat{s}$ in terms of some reference scale $\mu^2$. We note that for the the fitted parameter values that were given in Ref.~\cite{Gibbs:1995bt}, $b$ becomes negative at $\sqrt{\hat{s}} \simeq 57$~TeV.
Therefore, if those values are used, an absolute technical cut-off should be imposed at $\sqrt{\hat{s}} \simeq 50$~TeV. However, we have performed an independent fit of the three curves at $\sqrt{\hat{s}} = 16, 17, 18$~TeV, appearing in Fig.~2 of Ref.~\cite{Farrar:1990vb}, obtaining the values:
\begin{eqnarray}
a &=& -19 + 6\times 10^{-3} \sqrt{\hat{s}/{\rm GeV}^2}\;, \nonumber \\
b &=& 7.4 + 3 \times 10^{-3}  \sqrt{\hat{s}/{\rm GeV}^2} \ .
\end{eqnarray}
We use the above values in the phenomenological studies that involve a variable number of bosons. We note that sphaleron-induced final states are dominated by EW gauge boson production, and in all the studies of the present article we have set the number of Higgs bosons produced to zero~\cite{Gibbs:1994cw}.

As mentioned above, the LOME formula we rely on is valid only for $\sqrt{\hat s} \ll M_0 \sim 18$ TeV.
Around and beyond this energy, no reliable theoretical estimation for the boson multiplicity is known. To obtain a phenomenological treatment of this uncertainty, we introduce an energy scale $E_\mathrm{freeze}$ that acts as a transition point. 
That is, beyond $E_\mathrm{freeze} > M_0$, the $a$ and $b$ are fixed to their values at $\sqrt{\hat{s}} = E_\mathrm{freeze}$. 
Since the boson multiplicity monotonically grows with $\sqrt{\hat s}$ in the LOME formula, 
this effectively introduces an upper bound to the total number of bosons that can be produced.

\begin{figure}[tbp]
\centering 
\includegraphics[width=.45\textwidth,clip]{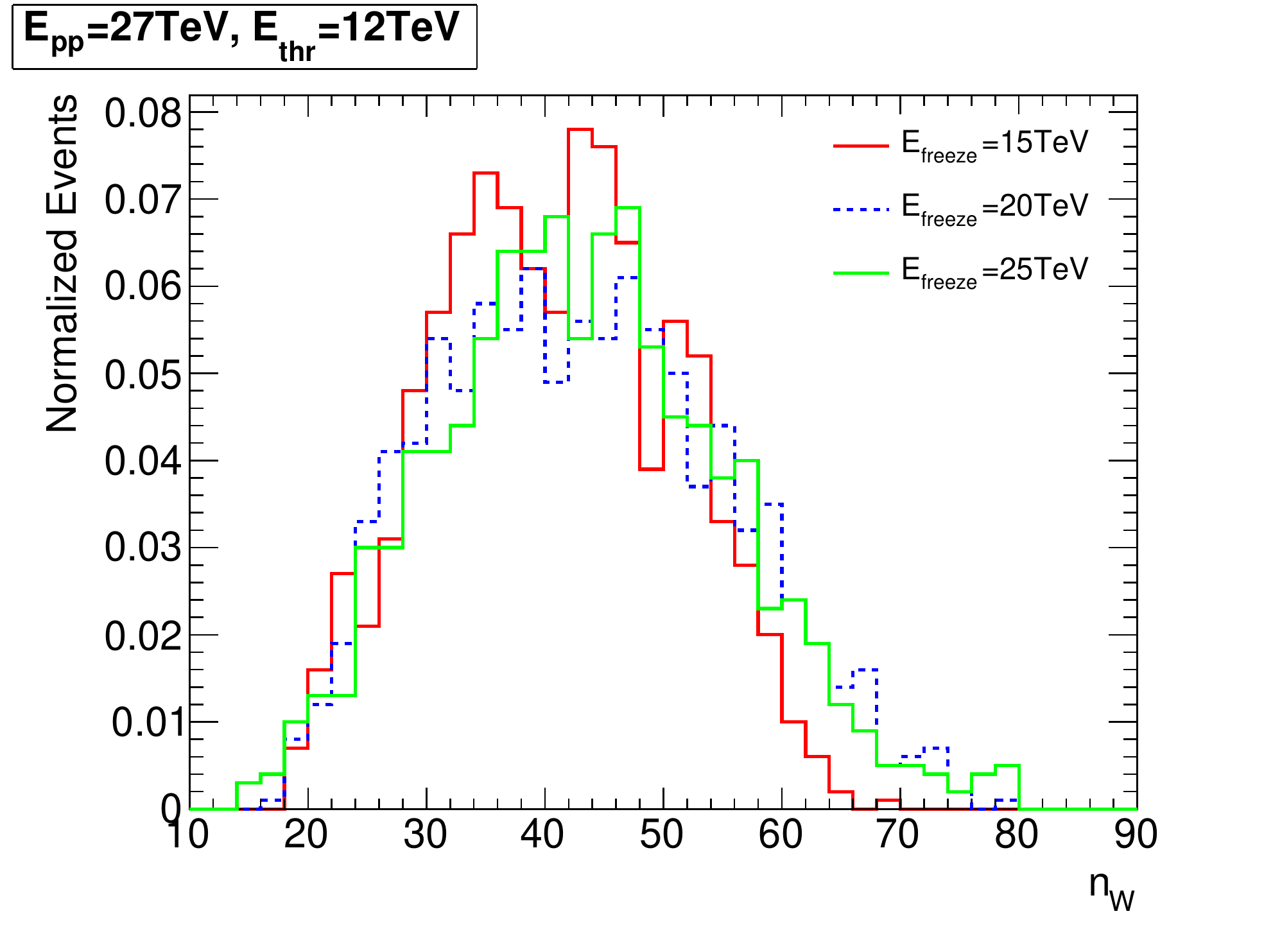}
\includegraphics[width=.45\textwidth,clip]{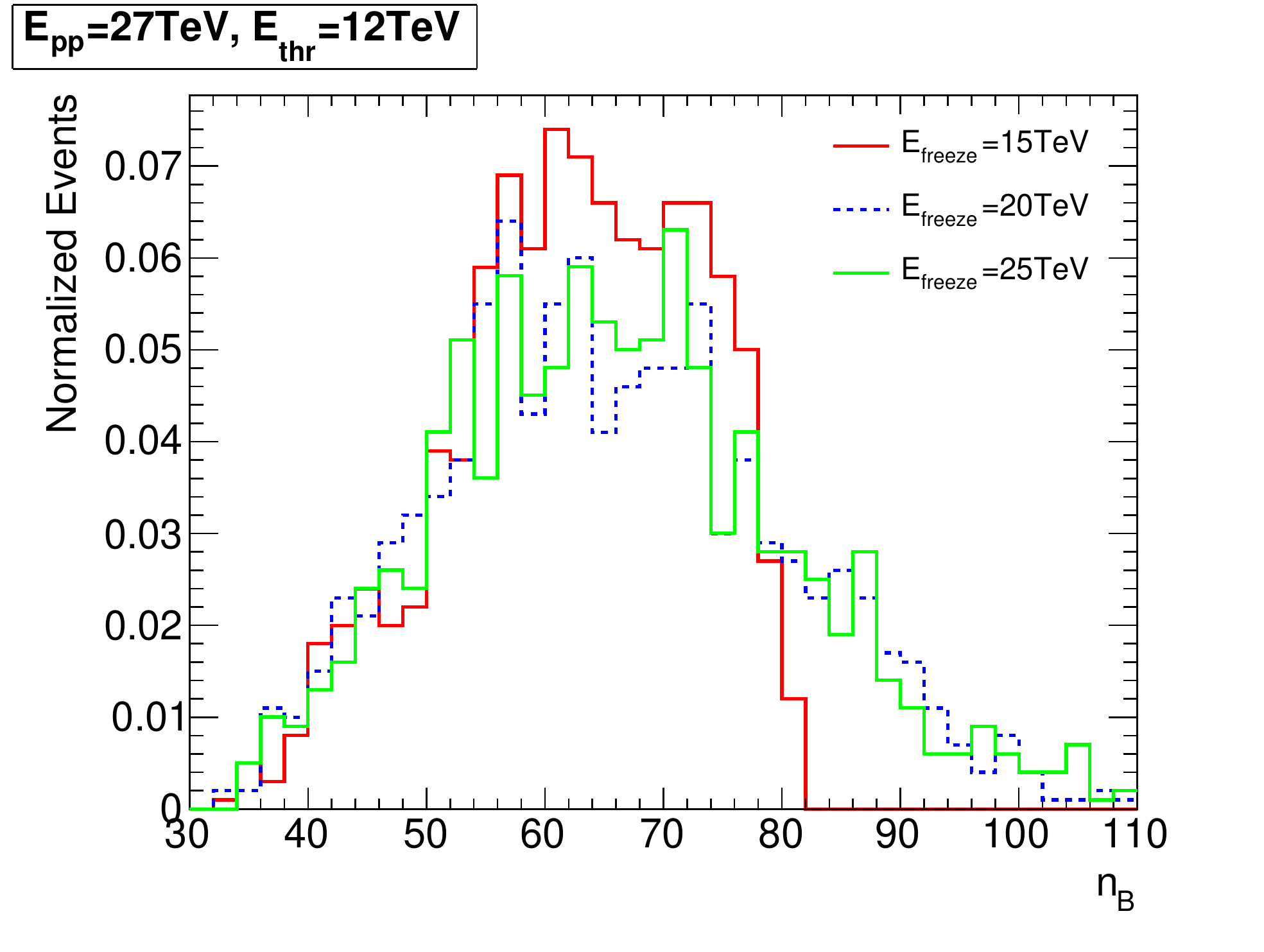}
\caption{\label{fig:freeze} Normalised distributions of the number of $W$ bosons (left) and all bosons (right) for
$\sqrt{s}=27$\;TeV, $E_{\rm thr} = 12$\;TeV.
The red-solid, blue-dashed and green-solid histograms correspond to $E_{\rm freeze} = 15$, 20\;TeV and 25 TeV, respectively.
}
\end{figure}

To see the effect of varying $E_{\rm freeze}$, we show in Fig.~\ref{fig:freeze}
normalised distributions of the number of $W$ bosons (left) and all bosons (right) for
$\sqrt{s}=27$\;TeV, $E_{\rm thr} = 12$\;TeV with three different values of $E_{\rm freeze}$:
$E_{\rm freeze} = 15$\;TeV  (red-solid), 20\;TeV (blue-dashed) and 25\;TeV (green-solid).
One can see that the tail of the distributions become more suppressed as $E_{\rm freeze}$ becomes smaller. 
This is because the LOME formula predicts larger multiplicities for larger $\sqrt{\hat s}$
but the boson multiplicity distribution is frozen for $\sqrt{\hat s} > E_{\rm freeze}$ at $E_{\rm freeze}$.  
This effect is more visible for $E_{\rm freeze} \sim E_{\rm thr}$ (red vs green histograms).
For $E_{\rm freeze} \gg E_{\rm thr}$, the effect of $E_{\rm freeze}$ is very small 
because the partonic collisions are dominated by $\sqrt{\hat s} \sim E_{\rm thr}$ due to suppression of the parton density functions (PDF) (green vs blue histograms).\footnote{For the event generation we use the {\tt PDF4LHC15\_nlo\_mc} PDF sets~\cite{Butterworth:2015oua}.}

\subsection{Multi-boson $B$- and $L$-conserving processes}
To investigate the potential for observability of baryon- and lepton-number-violating processes at colliders, we must consider the relevant backgrounds that do not exhibit this property. Since sphaleron-induced processes are expected to contain a large number of bosons $n_B \gtrsim 1/\alpha_w \sim 30$, they will subsequently decay to large numbers of jets -- see the quantitative analysis of the next section. This implies that any Standard Model cross sections of perturbative background processes will be extremely small. However, it is expected that if the ``signal'' sphaleron-induced processes exist, there would also be accompanied with $B$- and $L$-conserving non-perturbative multi-boson processes with a similarly-large number of bosons~\cite{Gibbs:1994cw}. The behaviour of such ``multi-boson'' processes is as uncertain as those of the sphaleron signal, but may occur at a similar rate. 

We have also constructed a MC simulation of the multi-boson processes which behaves in an identical manner to that of sphaleron-induced processes in all aspects, apart from the quark ``content'' of the participating sub-processes. In this case, these are simply $2\rightarrow 2$ scatterings, where we only allow $u$ and $d$ quarks in the initial state as in the sphaleron case. Since multi-boson processes are $B$- and $L$-conserving, this implies that only the same $u$ and $d$ quarks appear in the final state. The number of colour configurations is accounted for as in the sphaleron case, i.e.\ by multiplying the event weight by an appropriate factor and then sampling uniformly at random. We do not investigate these processes as backgrounds in the phenomenological analysis that follows, but future studies should re-asses the results of Ref.~\cite{Gibbs:1994cw} which has considered discrimination between the $B$- and $L$-violating and conserving processes. These processes are available in our MC event generator~\cite{sphaleronrepo}. 

\section{Phenomenology at hadron colliders}
\label{sec:pheno}

In this section we examine the experimental signatures of sphaleron-induced processes.
In order to take into account the detector effects the generated hadronic event samples are passed to {\tt Delphes}~\cite{Ovyn:2009tx, Selvaggi:2014mya}. 
Jets are reconstructed with the anti-$k_T$ algorithm~\cite{Cacciari:2008gp} with distance parameter $R=0.4$ via the \texttt{FastJet} package~\cite{Cacciari:2011ma}.
Leptons ($e^\pm$ and $\mu^\pm$) and photons are required to be sufficiently isolated from other energetic particles around them.
We calculate the scalar sum, ${\cal P}$, of the transverse momenta, $p_T$, of the neighbouring particles within $\Delta R = \sqrt{\Delta \eta^2 + \Delta \phi^2} < r_{\rm max}$,
and demand ${\cal P}/p_T^{i} < {\cal I}_{\rm max}$ for isolation, where $p_T^{i}$ is the magnitude of transverse momentum of the lepton or photon to be isolated.
We take 
$r_{\rm max} = 0.3$ for electrons and photons and 0.4 for muons.  
${\cal I}_{\rm max}$ is taken to be 0.10, 0.12 and 0.15 for electrons, photons and muons, respectively. 
Finally, we require all objects (jets, leptons and photons) to have $p_T > 30$\;GeV and pseudo-rapidity $|\eta|< 5$ (jets), $< 2.4$ (muons) and $< 2.5$ (electrons and photons).


\subsection{Reconstructed distributions}
\label{sec:dist} 

\begin{figure}[tbp]
\centering 
\includegraphics[width=.45\textwidth,clip]{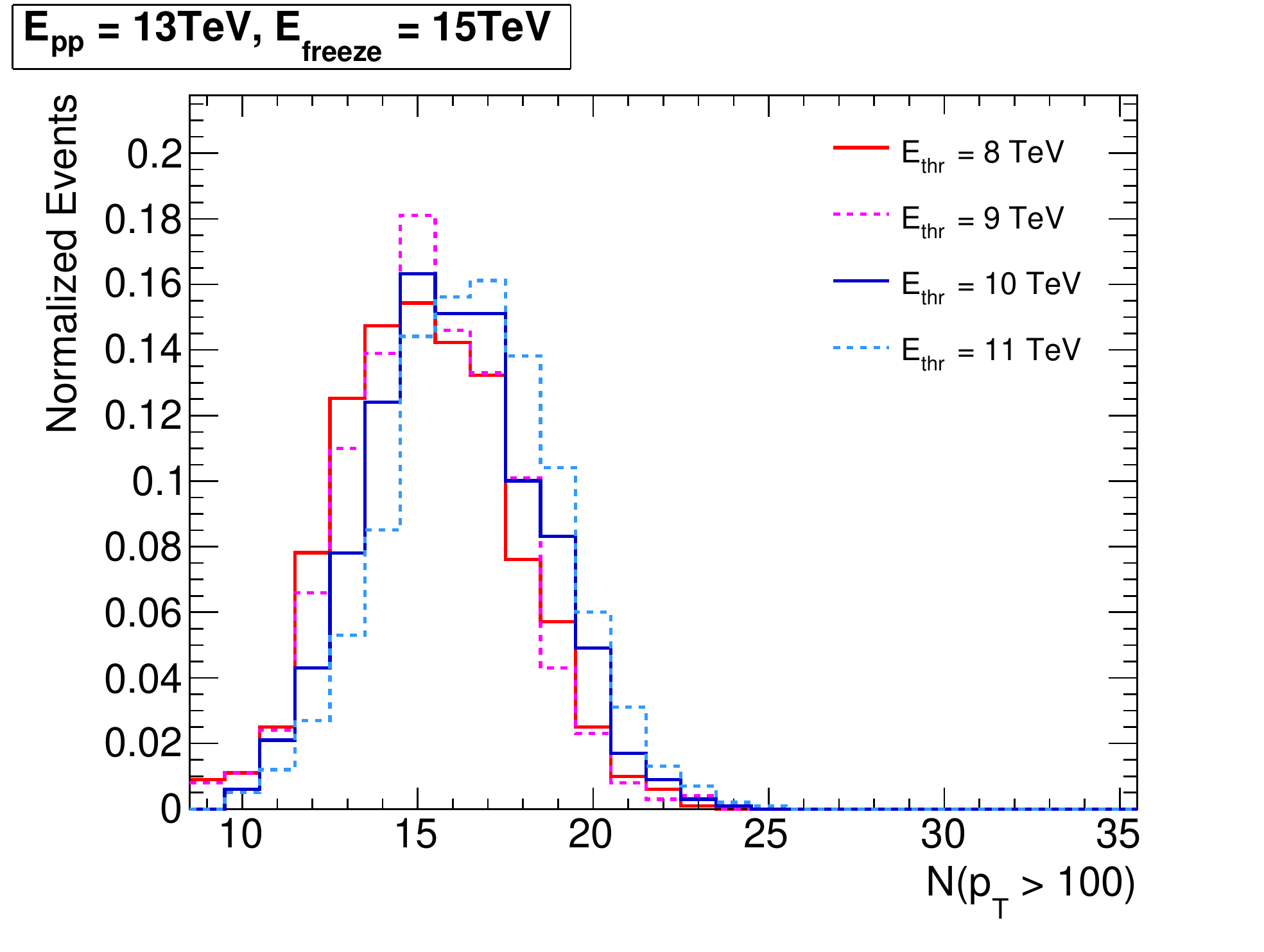}
\includegraphics[width=.45\textwidth,clip]{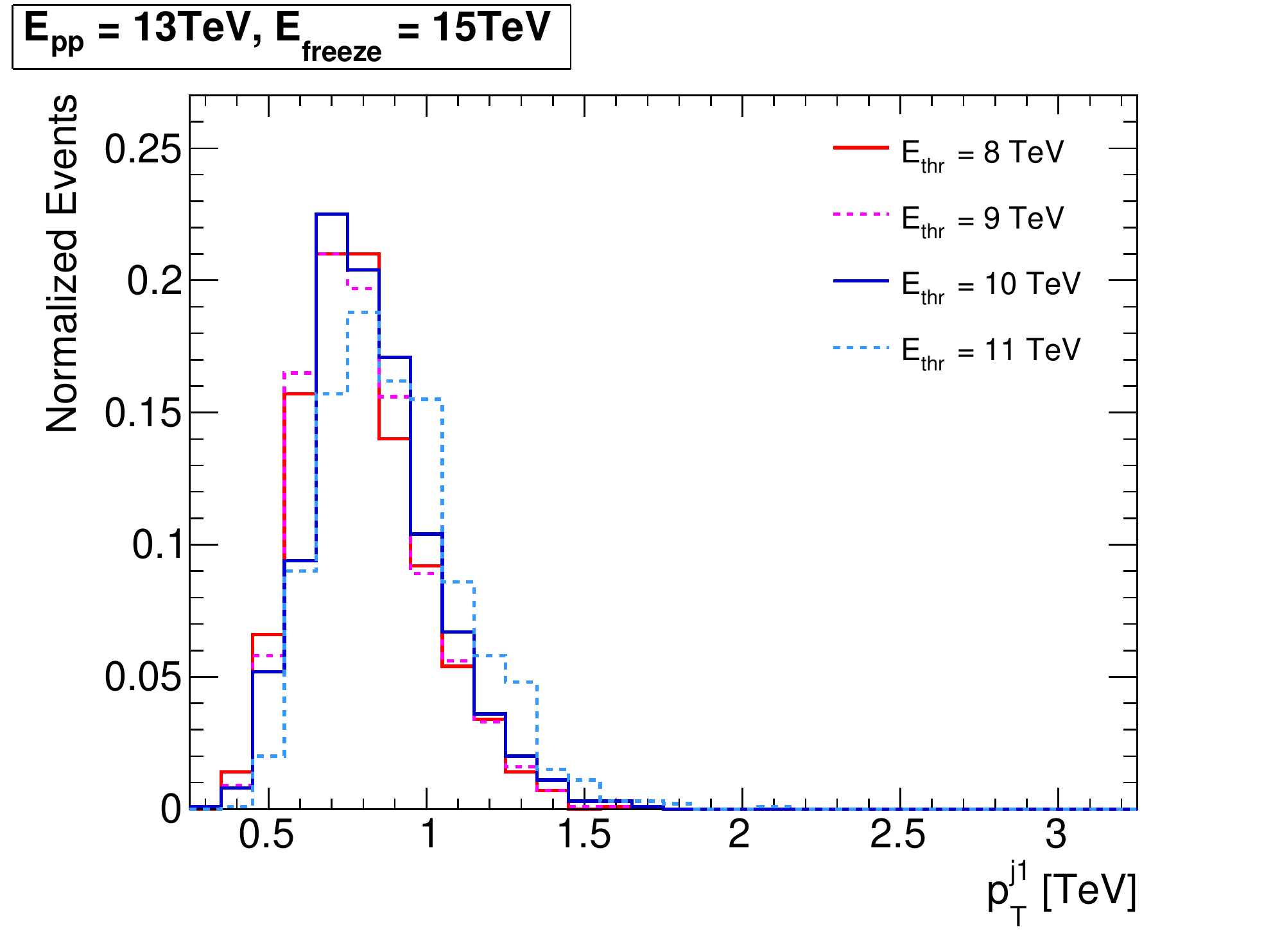}
\includegraphics[width=.45\textwidth,clip]{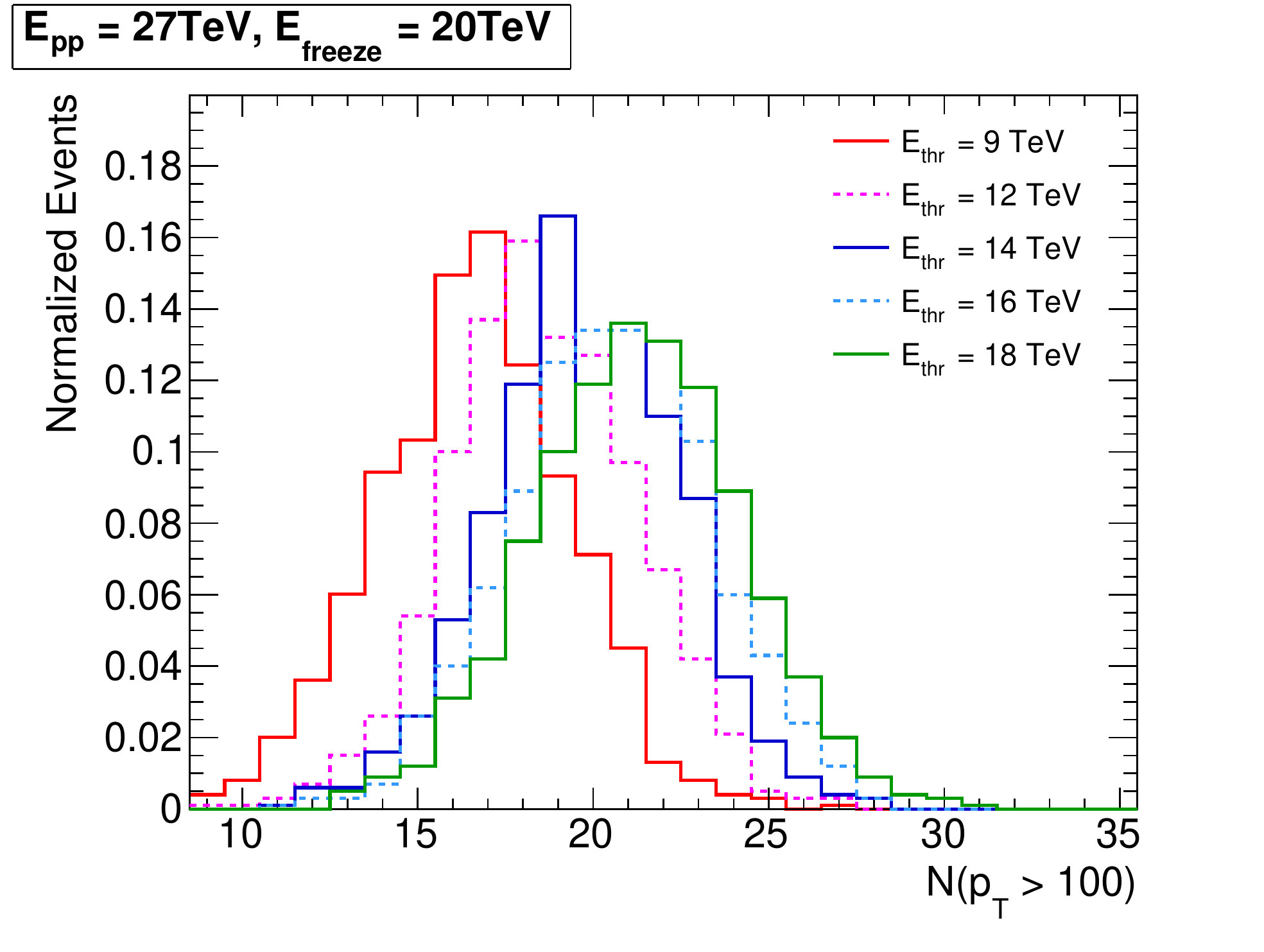}
\includegraphics[width=.45\textwidth,clip]{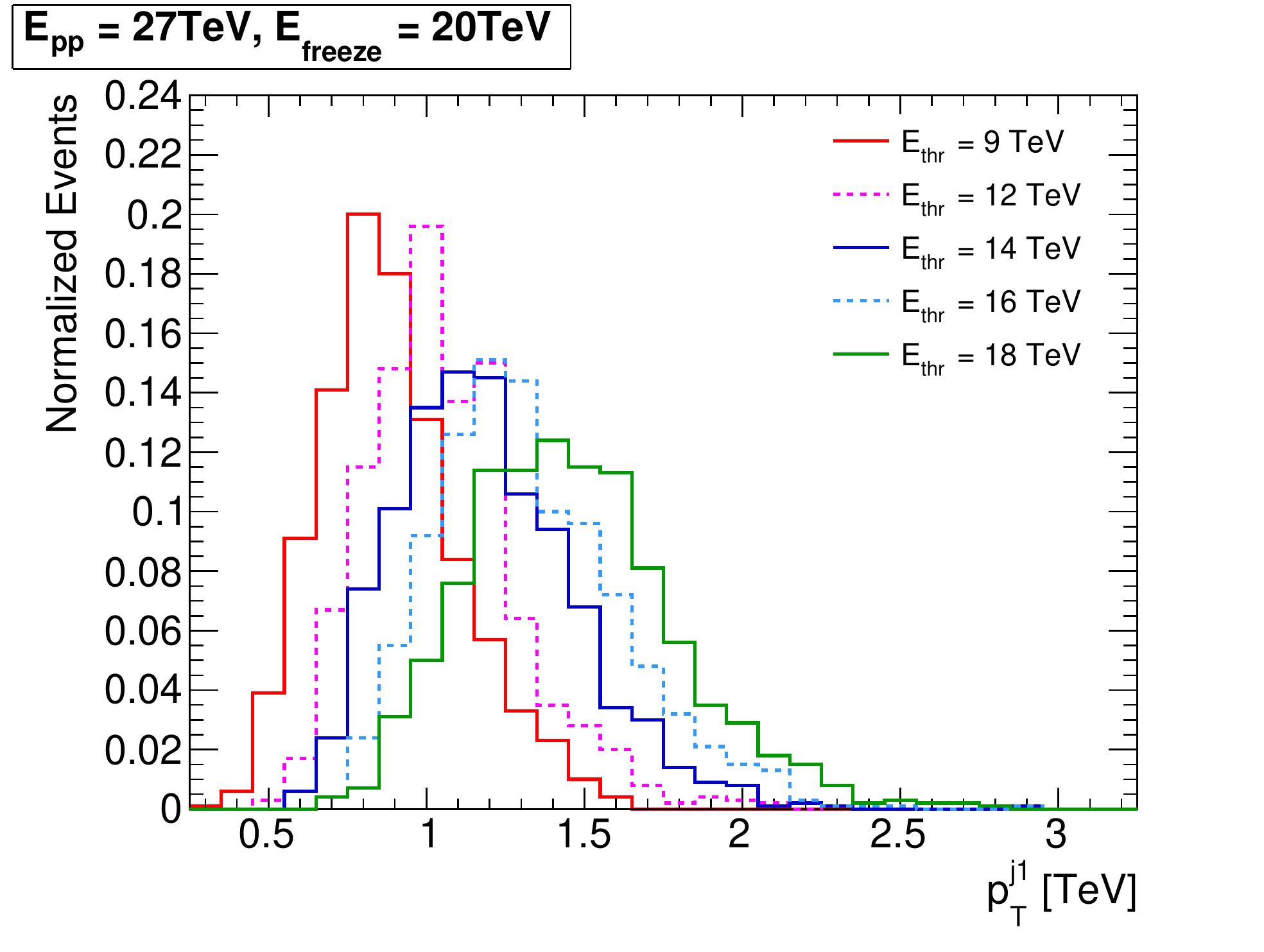}
\includegraphics[width=.45\textwidth,clip]{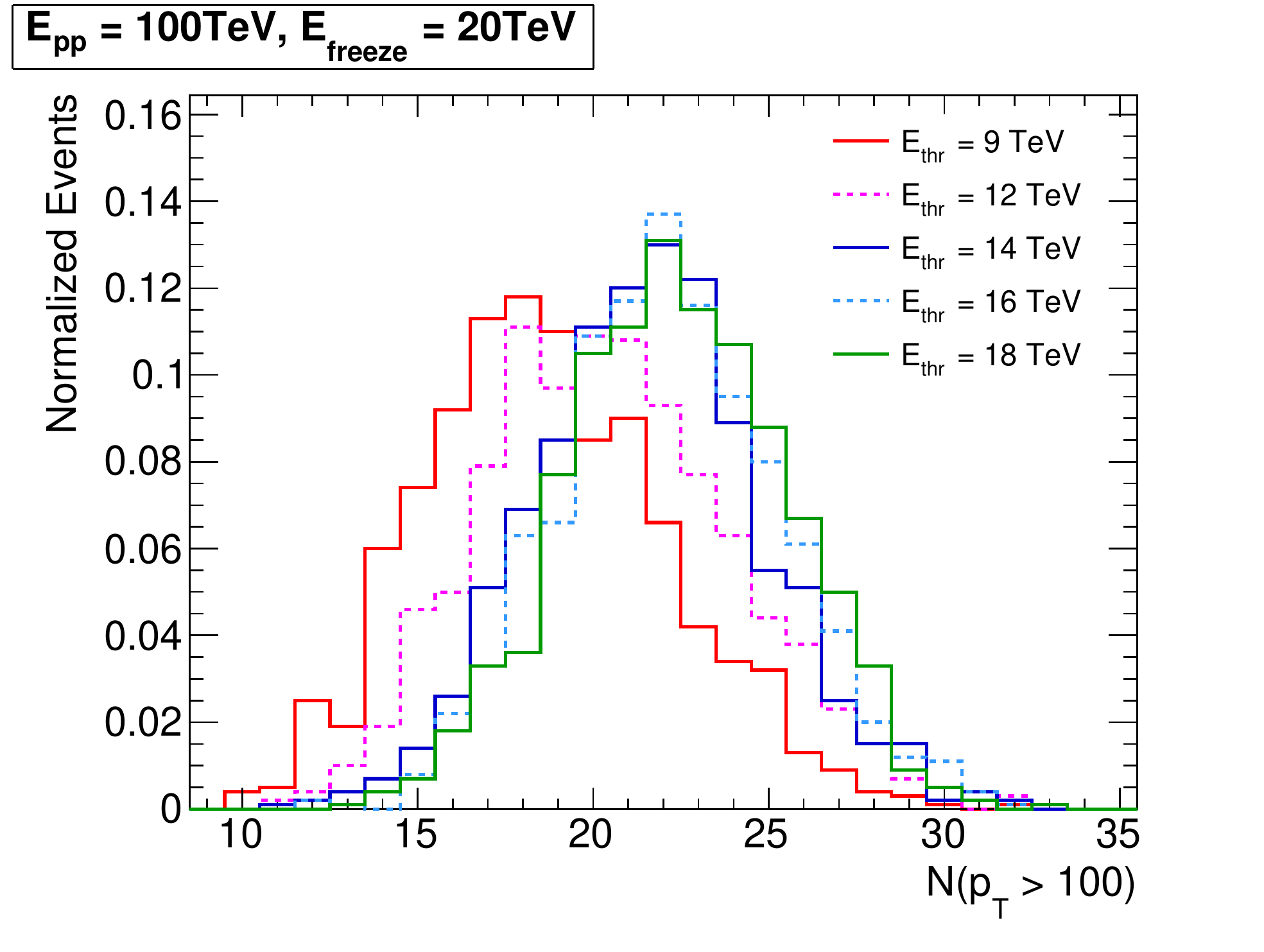}
\includegraphics[width=.45\textwidth,clip]{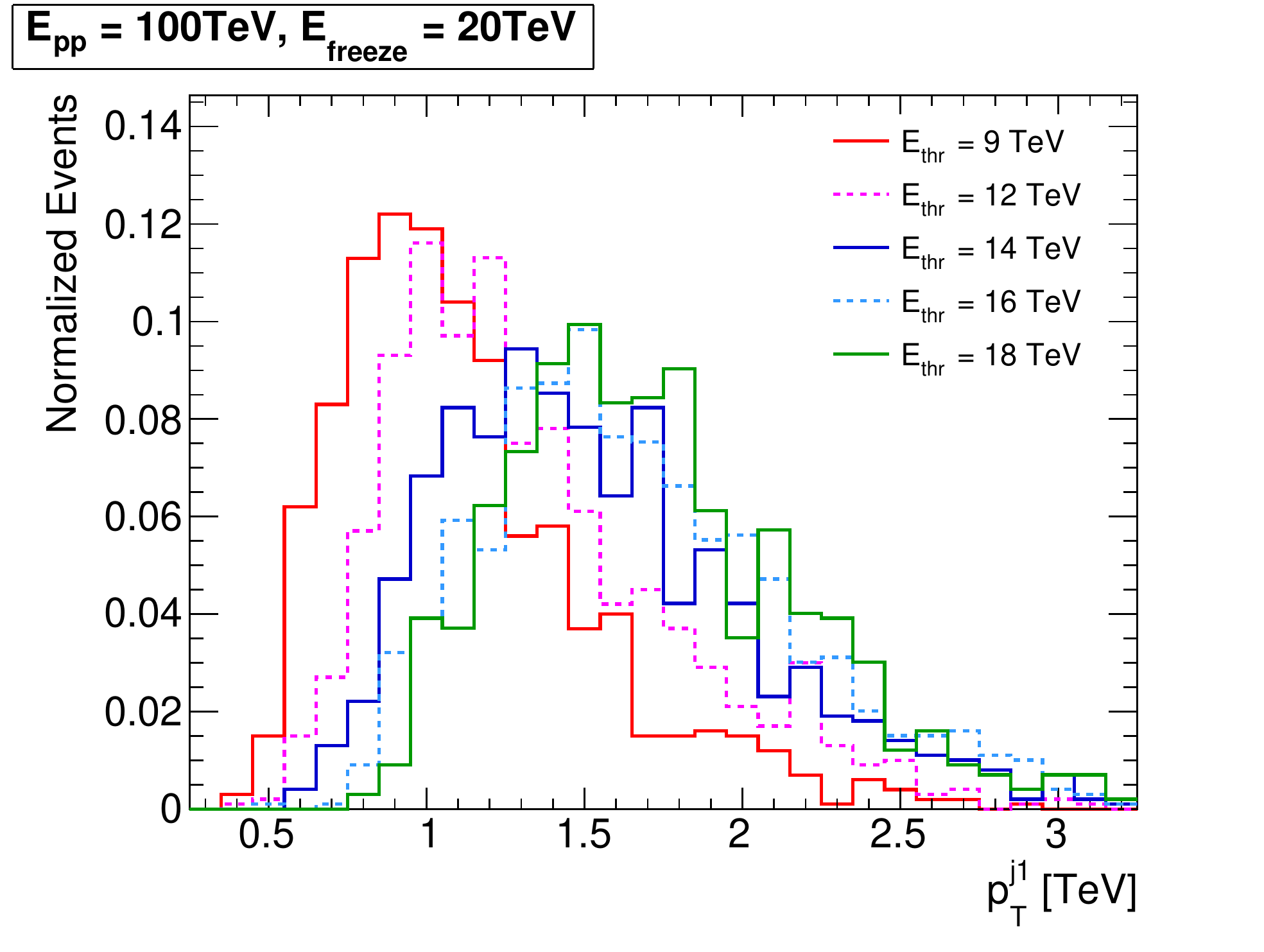}
\caption{\label{fig:n100_ptj1} Distributions of the inclusive reconstructed object multiplicity $N(p_T > 100)$ (left) and the $p_T$ of the hardest jet in the events (right) for $\sqrt{s} = 13$\;TeV (top), 27\;TeV (middle) and 100\;TeV (bottom).
We set $E_{\rm freeze} = 15$\;TeV for $\sqrt{s} = 13$\;TeV, whereas for $\sqrt{s} = 27$ and 100 TeV
we adopt $E_{\rm freeze} = 20$ TeV.
}
\end{figure}

In this subsection we study the impact of some parameters that are present in our phenomenological model, such as $\sqrt{s}$ and $E_{\rm thr}$,
on several multiplicity and kinematic distributions.
We start by showing in Fig.~\ref{fig:n100_ptj1}, the normalised distributions of the inclusive multiplicity (left) and the transverse momentum of the hardest jet, $p_T^{j1}$, in the events (right) for collider energies $\sqrt{s} = 13$\;TeV (top), 27\;TeV (middle) and 100\;TeV (bottom).
We take $E_{\rm freeze} = 15$\;TeV for $\sqrt{s} = 13$\;TeV and  $E_{\rm freeze} = 20$\;TeV for $27, 100$\;TeV, but the dependence on $E_{\rm freeze}$ of these distributions was found to be rather mild.
For $\sqrt{s} = 13$ TeV we show four histograms corresponding to $E_{\rm thr} = 8$\;TeV (red-solid), 9\;TeV (pink-dashed), 10\;TeV (blue-solid), 11\;TeV (cyan-dashed),
while for the 27\;TeV and 100 TeV colliders, we examined five values of $E_{\rm thr}$:
9\;TeV (red), 12\;TeV (pink-dashed), 14\;TeV (blue), 16\;TeV (cyan-dashed) and 18\;TeV (green-solid).

The inclusive multiplicity, denoted as $N(p_T > 100)$, is defined as the number of reconstructed objects 
(jets, leptons and photons satisfying the isolation criteria) with $p_T > 100$ GeV.   
These high-$p_T$ objects may originate either from direct anti-quark plus anti-lepton production in the hard interaction through the operator in Eq.~\eqref{eq:op},
or from secondary decays of the produced massive EW gauge bosons. 
For $\sqrt{s} = 13$ TeV (top-left) we see that the $N(p_T > 100)$ distribution peaks at around
15 -- 17 objects and the multiplicity is slightly larger for larger $E_{\rm thr}$ within the variation 
taken here: $E_{\rm thr} = (9 - 11)$ TeV.
For $\sqrt{s} = 27$\;TeV and 100\;TeV (middle-left and bottom-left, respectively),
the multiplicity distributions are shifted to larger values compared to those at the 13 TeV LHC.
For both the 27\;TeV and 100 TeV colliders, it peaks at $N(p_T > 100) \sim 17$ for $E_{\rm thr} = 9$ TeV
and at$N(p_T > 100) \sim 22$ for $E_{\rm thr} = 18$ TeV.
We also observe that the distributions are broader for $\sqrt{s} = 100$ TeV than for 27 TeV.

The right-hand-side plots of Fig.~\ref{fig:n100_ptj1} show the transverse momentum of the hardest jet, $p_T^{j1}$.
We expect that the hardest jet is likely to originate from one of the anti-quarks produced in the primary interaction,
through the operator in Eq.~\eqref{eq:op}.   
At the 13 TeV LHC (top-right), the distribution peaks around $p_T^{j1} \sim 700$\;GeV and the dependence of
$E_{\rm thr}$ is rather modest in the range we examined, $E_{\rm thr} \in (8 - 11)$ TeV.
Additionally, these distributions are somewhat narrow and confined below $p_T^{j1} \sim 1.5$\;TeV.
Contrary to them, the distributions for the 27\;TeV (middle-right) and 100\;TeV (bottom-right) colliders
are more dependent on $E_{\rm thr}$ in the range we considered, $E_{\rm thr} \in (9 - 18)$ TeV.
The peak position varies from $p_T^{j1} \sim 800\;$GeV to $p_T^{j1} \sim 1.5$\;TeV when $E_{\rm thr}$ is changed from 9 to 18 TeV.
We also note that the $p_T^{j1}$ distributions become broader for larger $E_{\rm thr}$, as well as for larger collider energies. This is due to the fact that the phase-space volume is bigger for larger $E_{\rm thr}$ and therefore it is easier for the hardest jet to attain a higher transverse momentum.
At higher collider energies, the distributions become broader because
the PDF suppression becomes milder at higher partonic collision energies.

\begin{figure}[tbp]
\centering 
\includegraphics[width=.45\textwidth,clip]{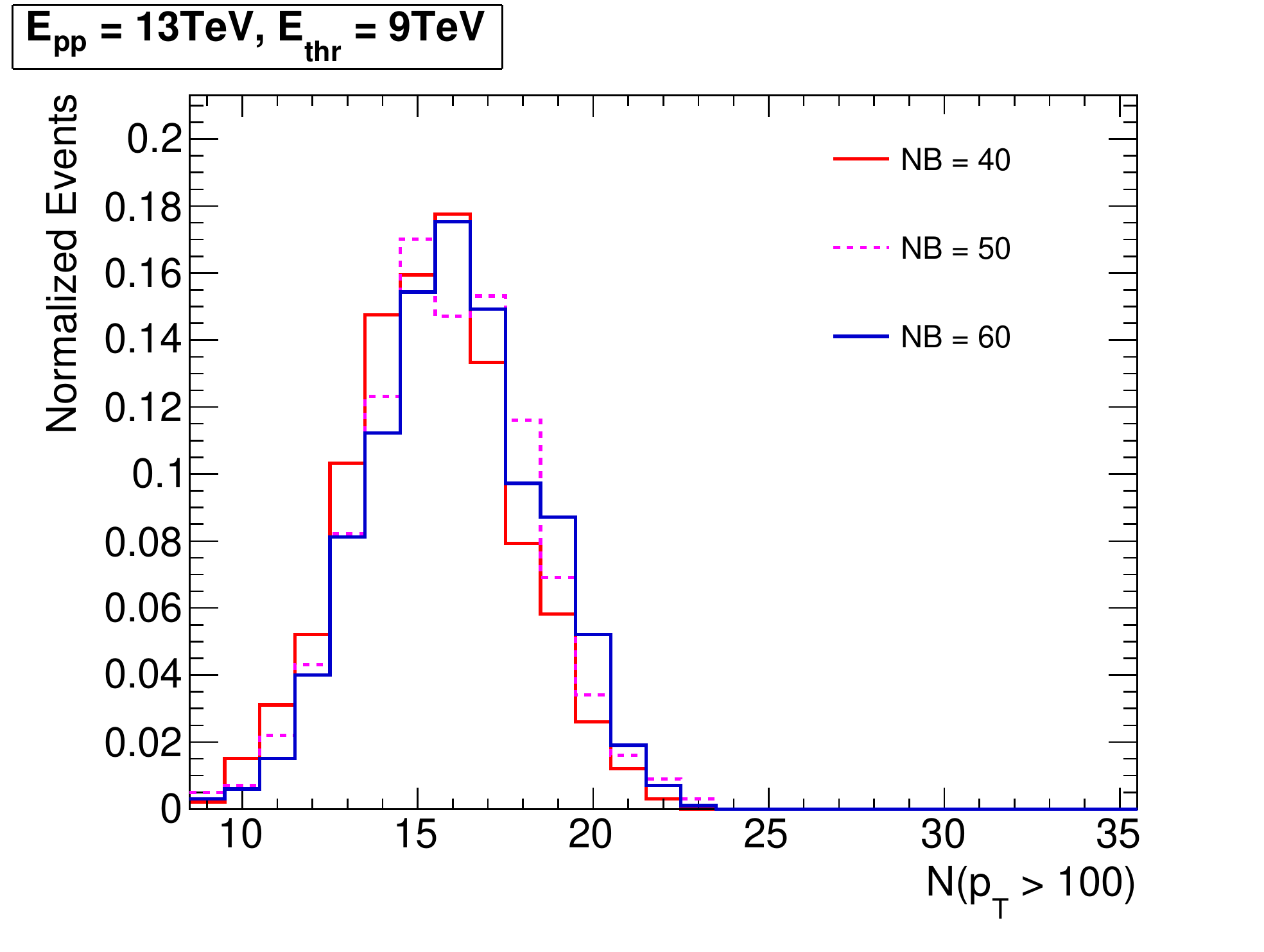}
\includegraphics[width=.45\textwidth,clip]{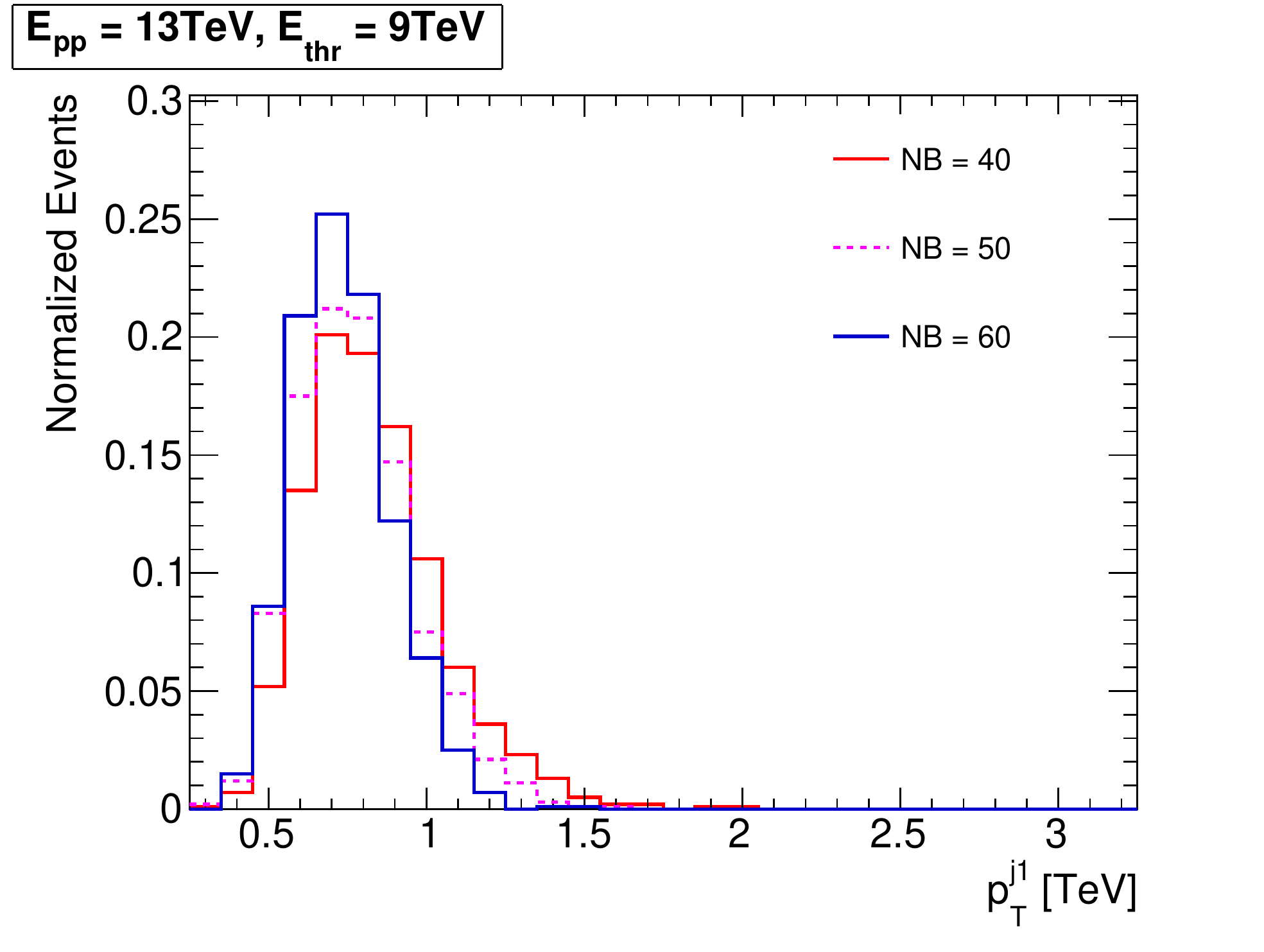}
\includegraphics[width=.45\textwidth,clip]{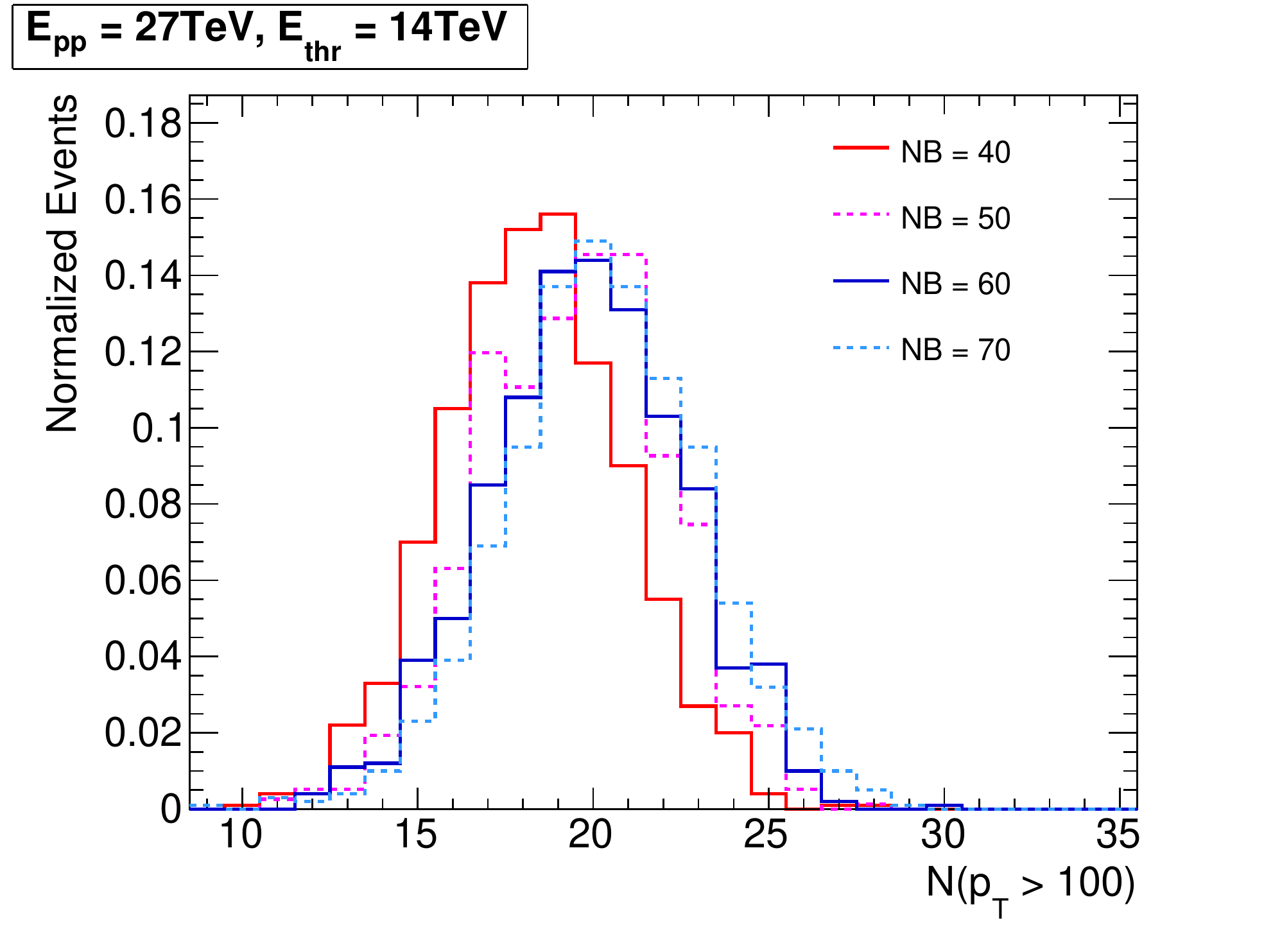}
\includegraphics[width=.45\textwidth,clip]{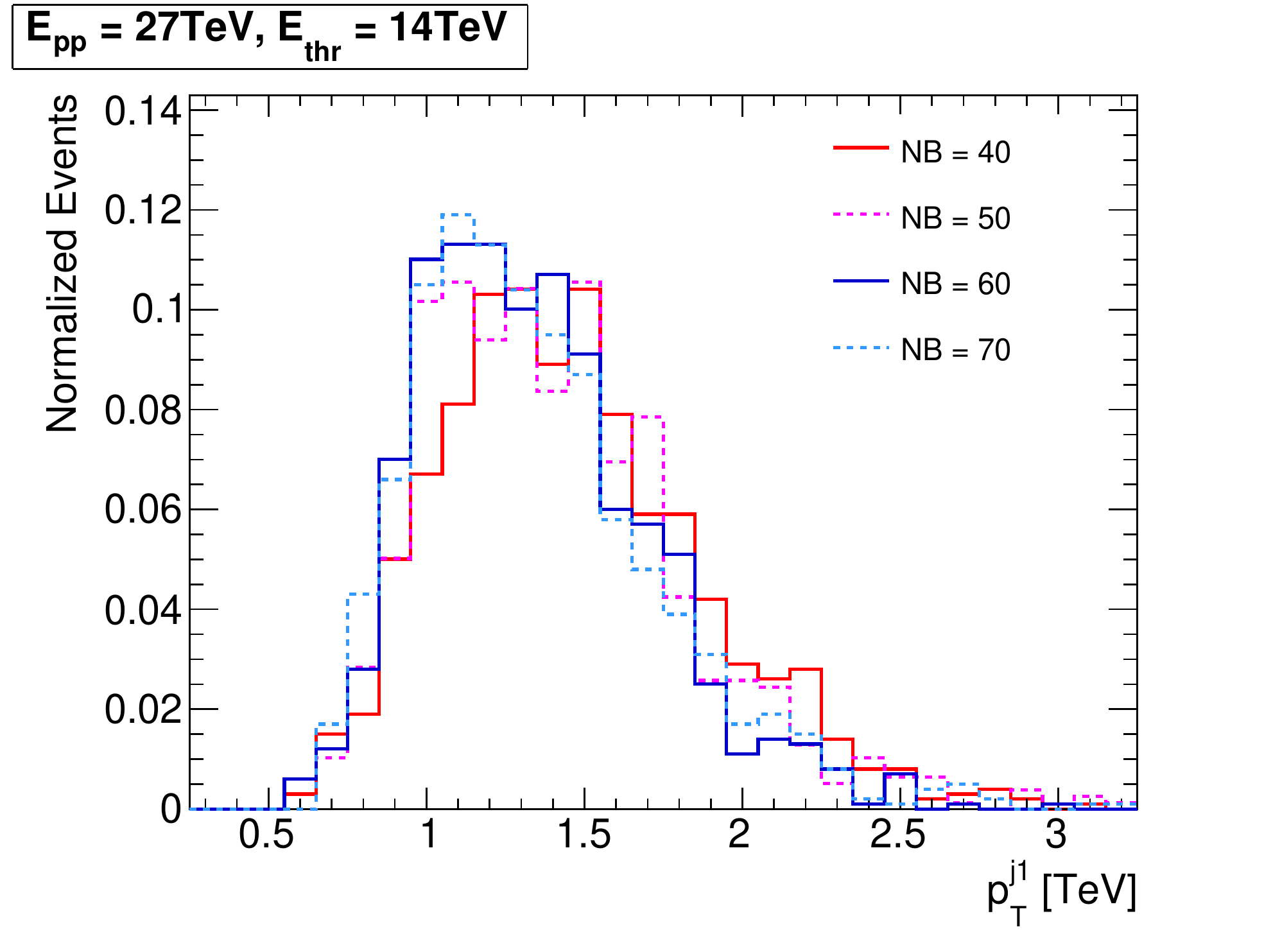}
\includegraphics[width=.45\textwidth,clip]{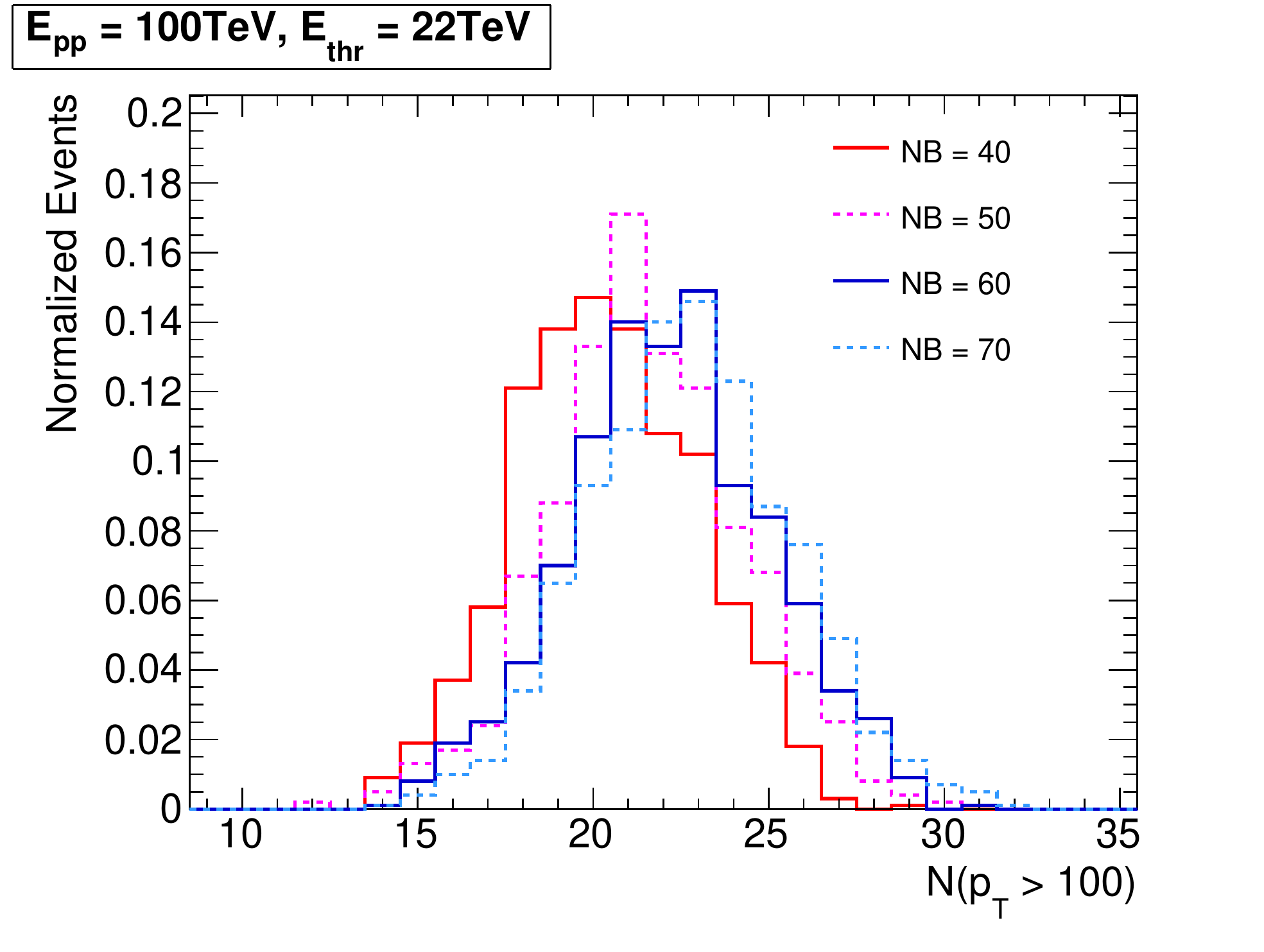}
\includegraphics[width=.45\textwidth,clip]{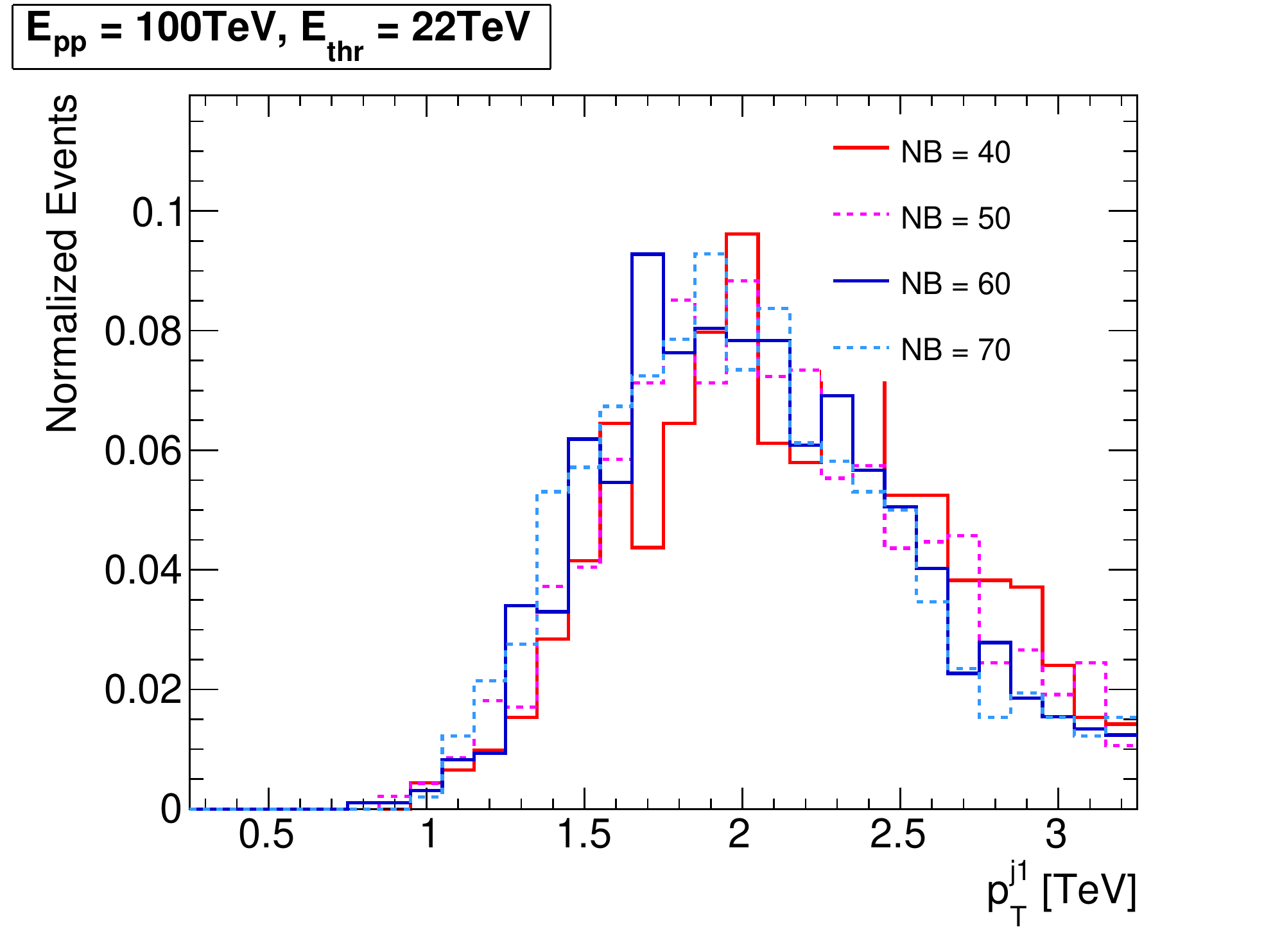}
\caption{\label{fig:n100_ptj1_nb} 
The same as Fig.~\ref{fig:n100_ptj1} but with the number of total gauge bosons fixed to 
$n_B = 40$ (red-solid), 50 (pink-dashed) and 60 (blue-solid) for the 13 TeV LHC.
In addition to these, the case $n_B = 70$ (cyan-dashed) is shown for $\sqrt{s} =$ 27 and 100 TeV. 
The threshold energy is taken as $E_{\rm thr} = 9$, 14 and 22 TeV for
$\sqrt{s} = 13$, 27 and 100 TeV.}
\end{figure}

So far we have generated the boson multiplicity according to the LOME formula.
To see the effect of the boson multiplicity on the jet transverse momentum and the number of high-$p_T$ objects,
we look at these distributions with a fixed boson multiplicity, $n_B$.
In Fig.~\ref{fig:n100_ptj1_nb}, we show the distributions of $N(p_T > 100)$ (left) and of the hardest jet $p_T$
for a fixed number of total EW gauge bosons.  
For the 13 TeV LHC (top panels), we show histograms for $n_B = 40$ (red-solid), 50 (pink-dashed) and 60 (blue-solid).
For 27\;TeV (middle) and 100\;TeV (bottom) colliders, in addition to those, the case $n_B = 70$ (cyan-dashed) is shown.
We take $E_{\rm thr} = 9$, 14\;TeV and 22\;TeV for $\sqrt{s} = 13$, 27 and 100 TeV, respectively.

In the right-hand plots of Fig.~\ref{fig:n100_ptj1_nb} we see that the number of high-$p_T$ reconstructed objects with $p_T > 100$\;GeV
has a mild dependence on $n_B$, though one can observe 
that $N(p_T > 100)$ increases with $n_B$.
This is expected because the EW gauge bosons decay into jets and leptons and as long as they are isolated 
and have $p_T > 100$\;GeV, they contribute to $N(p_T > 100)$.
On the other hand, $N(p_T > 100)$ rather strongly depends on the collider energy and $E_{\rm thr}$.
The peak of the $N(p_T > 100)$ distribution sits around 16, 20 and 22 objects for 
($\sqrt{s}$, $E_{\rm thr}$) = (13, 9), (27, 14), (100, 22) TeV, respectively.
We also note that the distributions are broader for larger $\sqrt{s}$ and $E_{\rm thr}$, 
as we have seen in Fig.~\ref{fig:n100_ptj1}.

The right-hand-side plots Fig.~\ref{fig:n100_ptj1_nb} show the distribution of the hardest jet transverse momentum, $p_T^{j1}$.
Just as for $N(p_T > 100)$, its dependence on $n_B$ is mild.
There is a tendency for the $p_T^{j1}$ to become softer for larger $n_B$.
This is because for a fixed partonic energy, larger $n_B$ implies a smaller amount of energy available to be shared among the primary particles (i.e.\ anti-fermions, EW gauge bosons).
As was the case for $N(p_T > 100)$, the distributions are broader and harder at larger collider energies and $E_{\rm thr}$.

\begin{figure}[tbp]
\centering 
\includegraphics[width=.45\textwidth,clip]{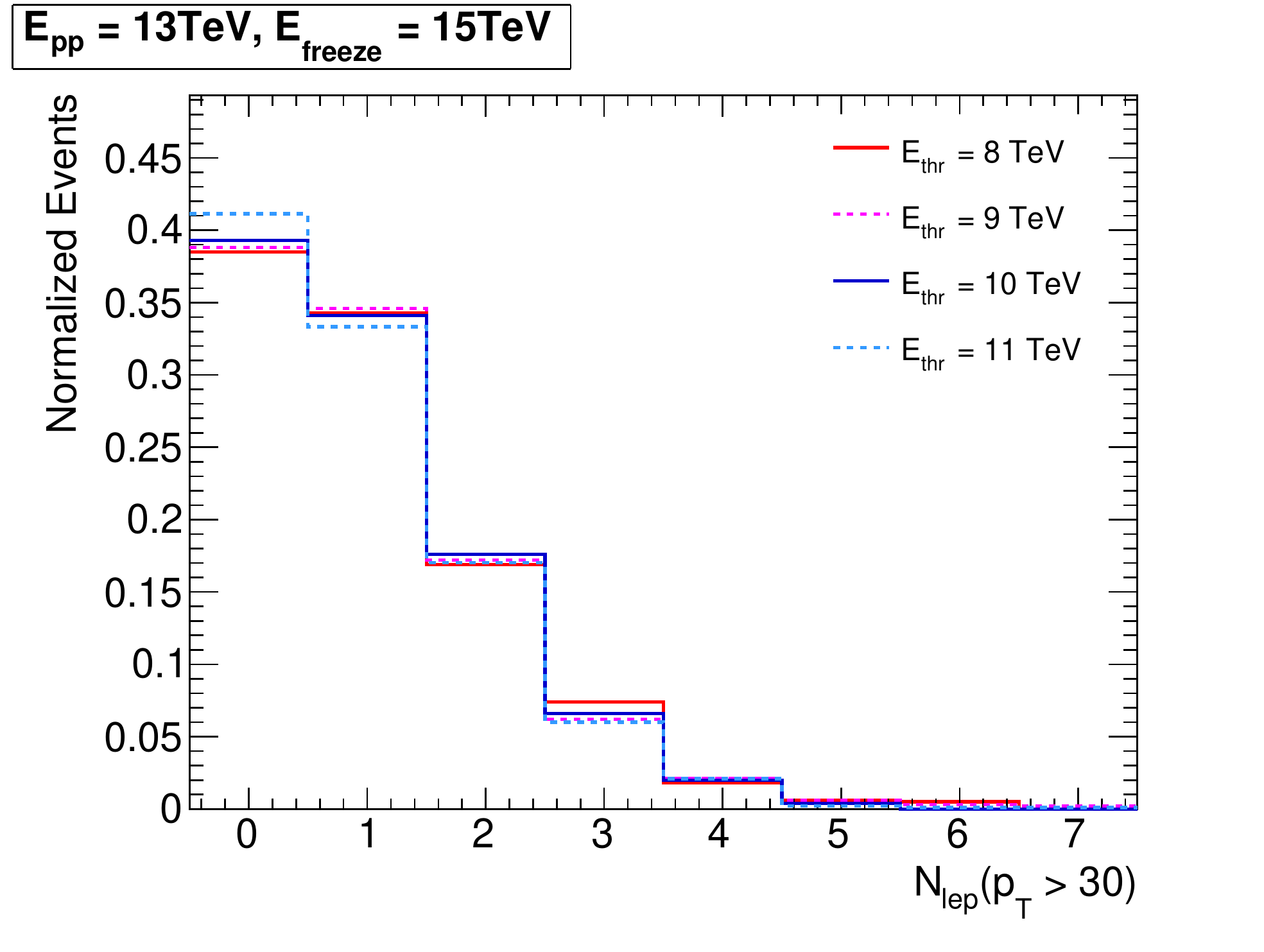}
\includegraphics[width=.45\textwidth,clip]{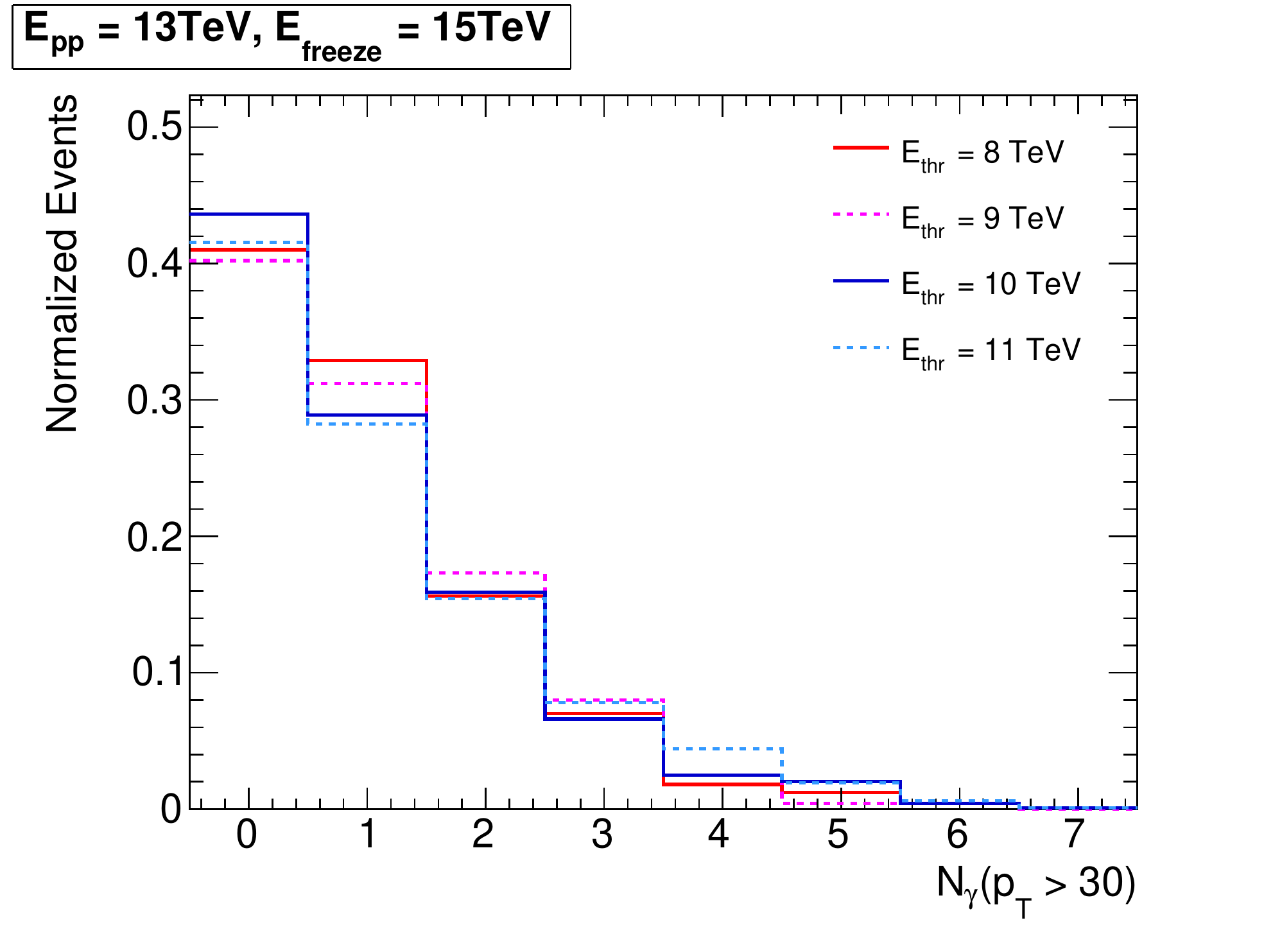}
\includegraphics[width=.45\textwidth,clip]{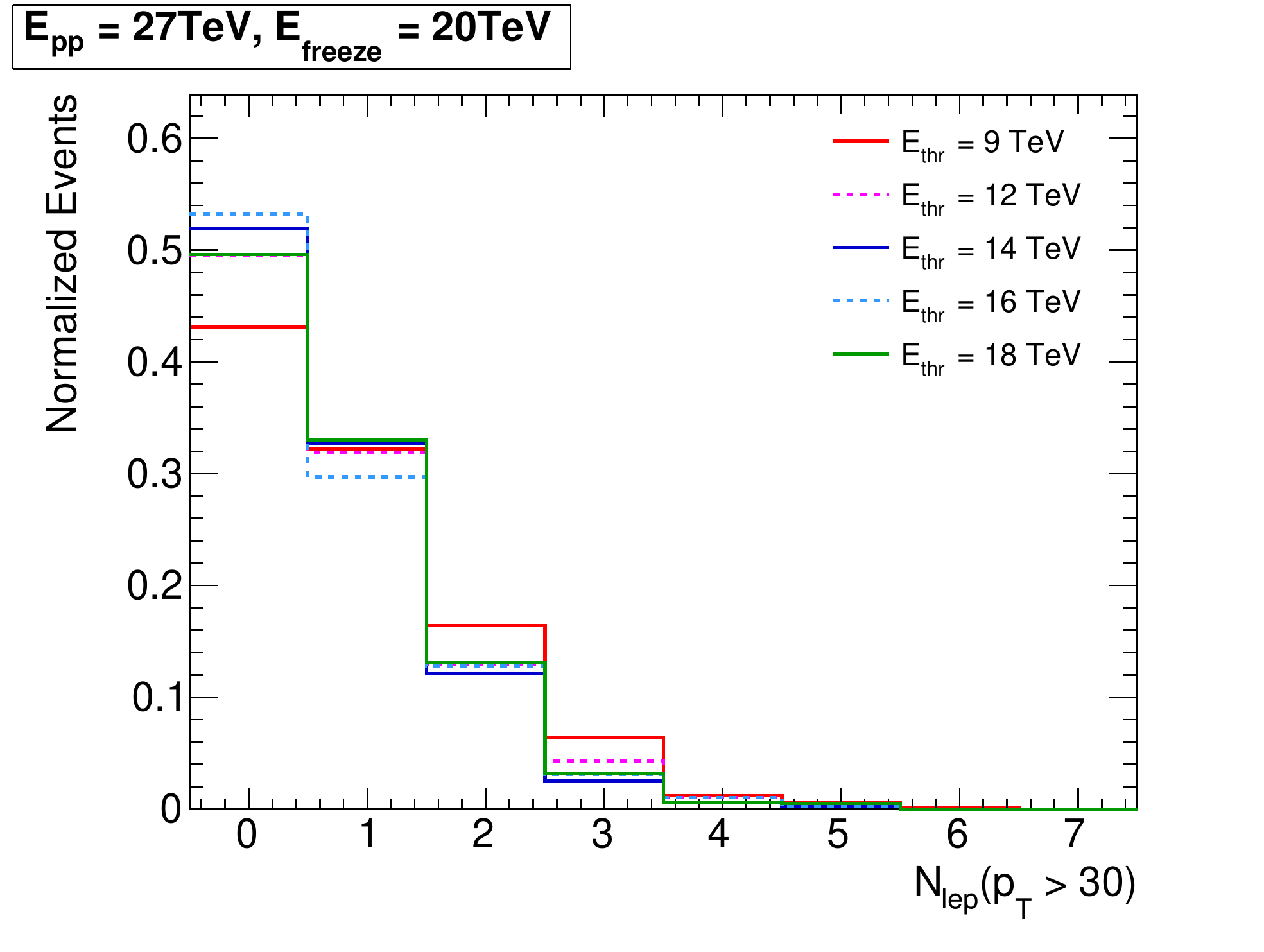}
\includegraphics[width=.45\textwidth,clip]{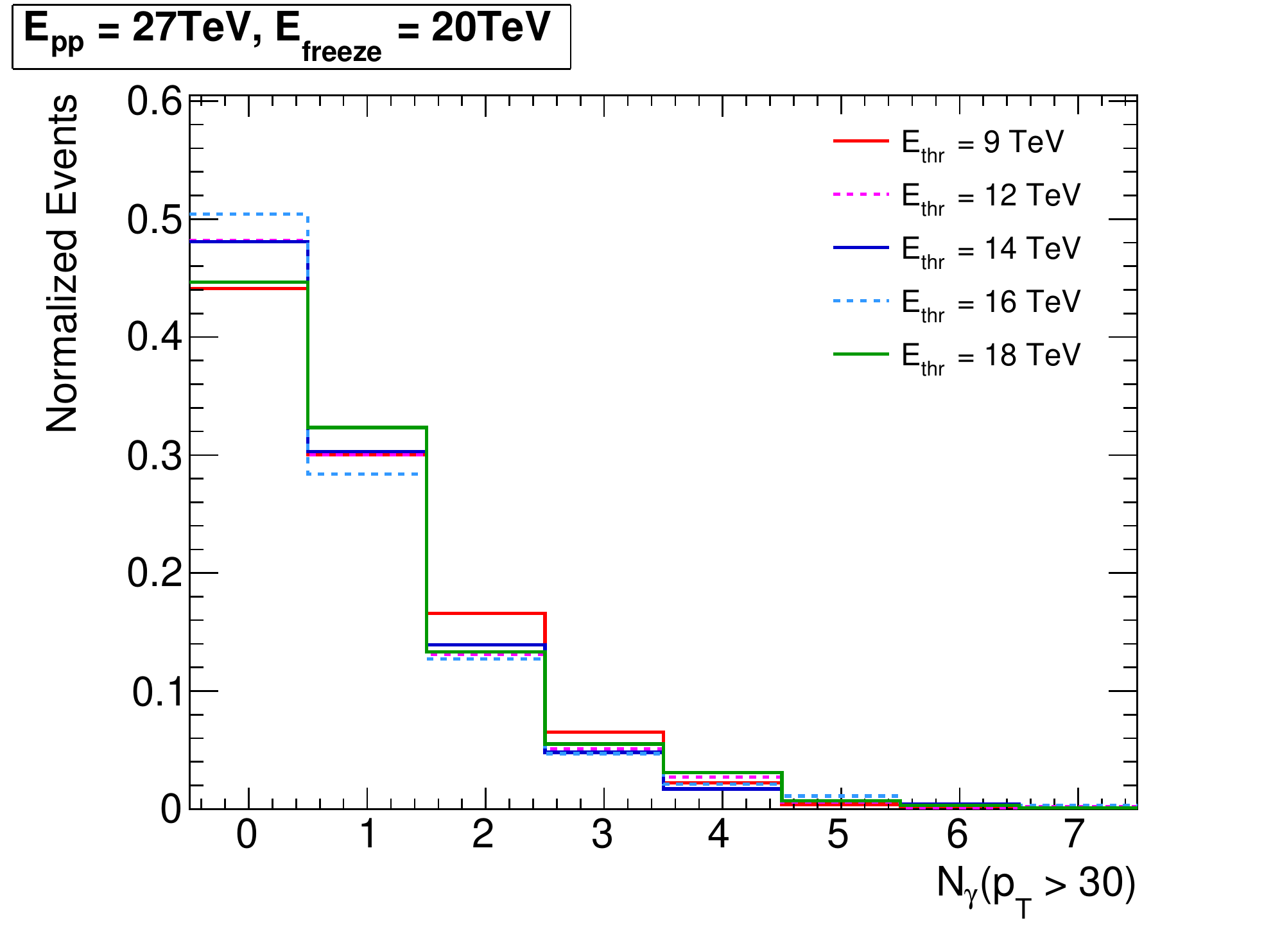}
\includegraphics[width=.45\textwidth,clip]{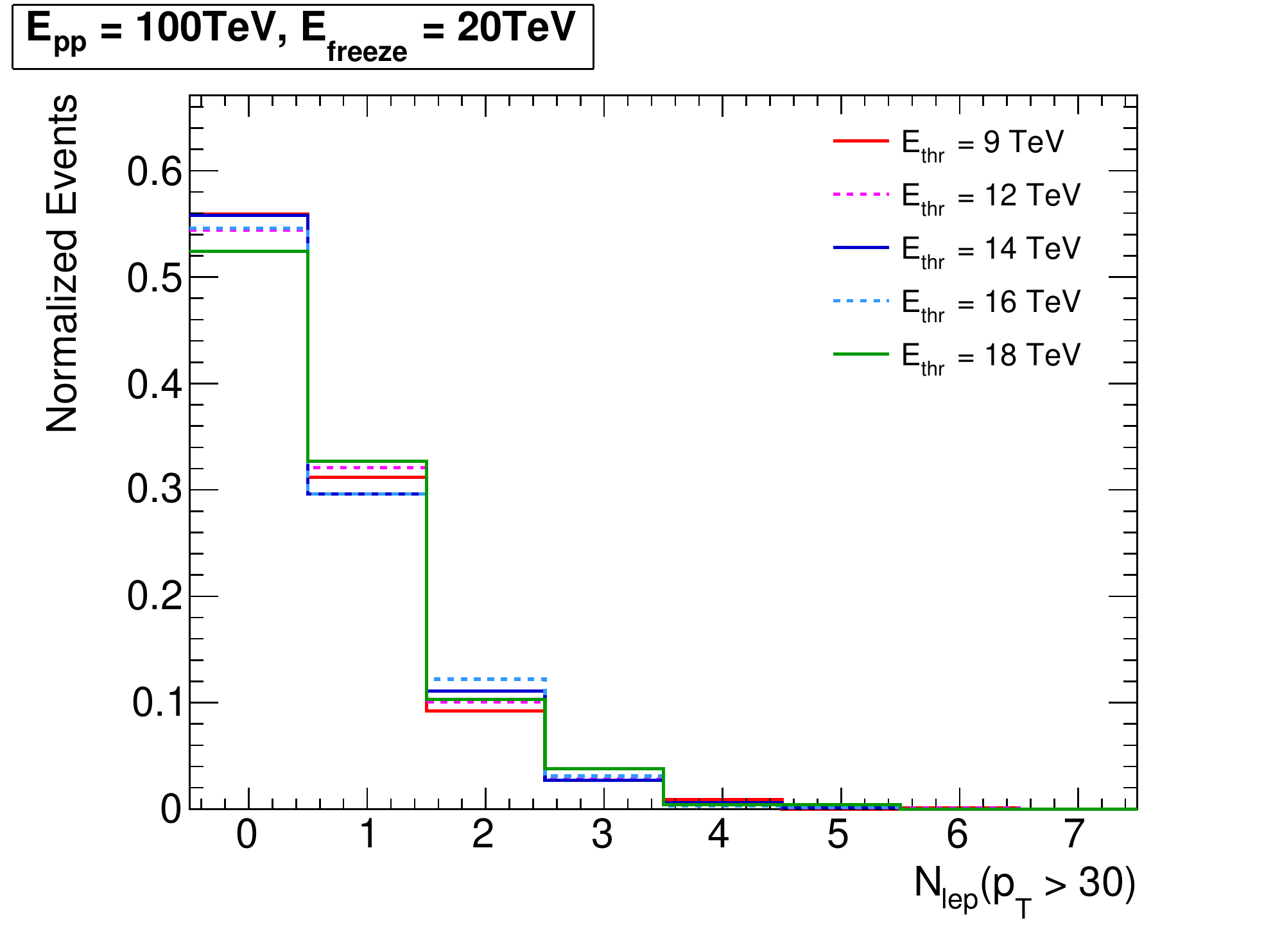}
\includegraphics[width=.45\textwidth,clip]{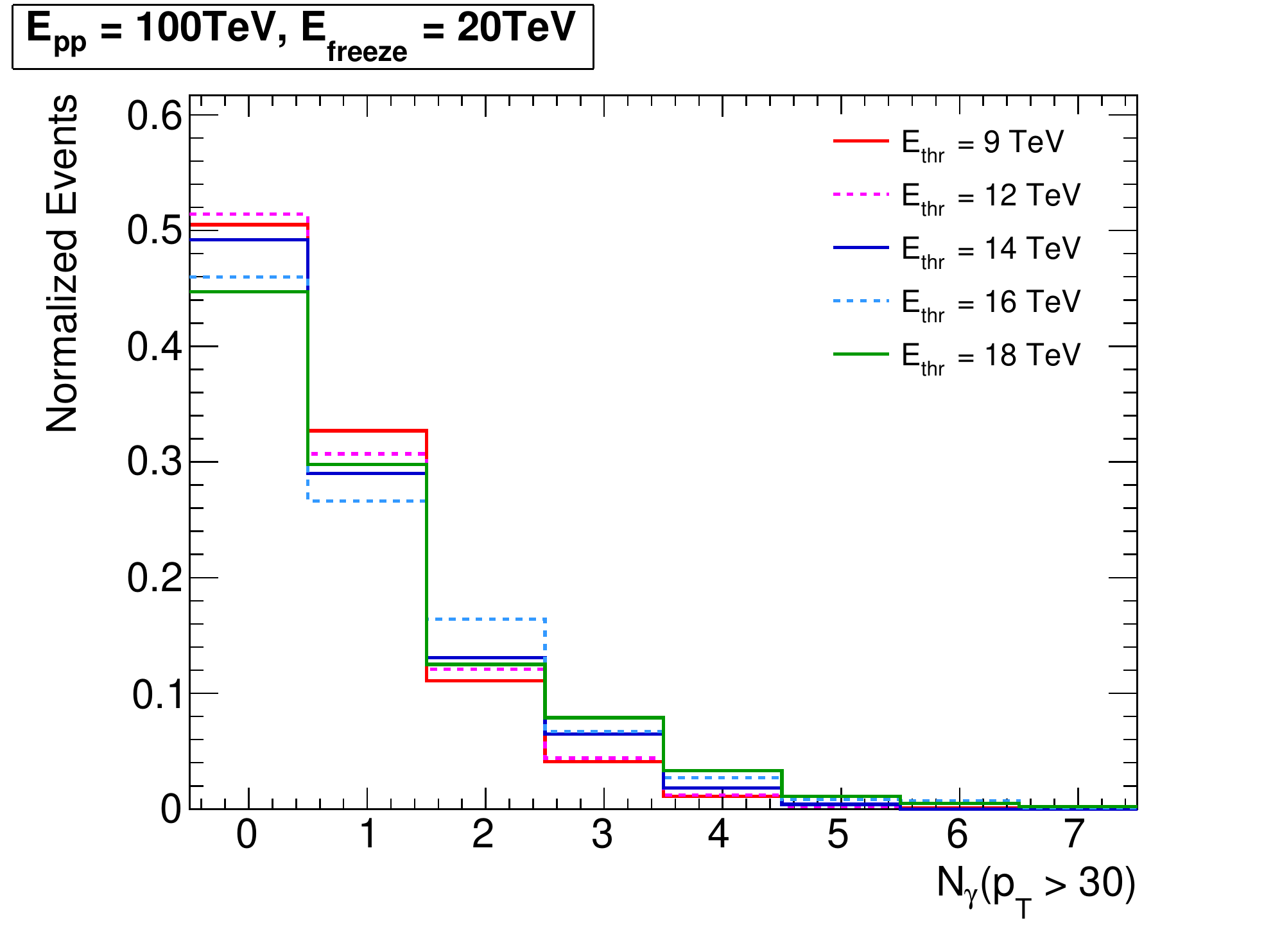}
\caption{\label{fig:nlep_nphot} Multiplicity distributions of isolated leptons (left) and photons (right) for $\sqrt{s} = 13$\;TeV (top), 27\;TeV (middle) and 100\;TeV (bottom).
The choice of values for $E_{\rm freeze}$ and $E_{\rm thr}$, as well as the line-styles and colour schemes are as in Fig.~\ref{fig:n100_ptj1}. 
}
\end{figure}

In Fig.~\ref{fig:nlep_nphot}, we show the distributions of the number of isolated leptons, $N_{\rm lep}$, 
(left) and photons, $N_\gamma$, (right) with $p_T>30$ GeV 
for $\sqrt{s} = 13$\;TeV (top), 27\;TeV (middle) and 100\;TeV (bottom).
The choice of $E_{\rm freeze}$ and $E_{\rm thr}$ and the line-style discriminating different $E_{\rm thr}$ are the same as in Fig.~\ref{fig:n100_ptj1}. 
In sphaleron events, high-$p_T$ leptons are produced either primarily from the hard interaction via the operator in Eq.~\eqref{eq:op}
or secondarily from the decay of massive EW bosons.\footnote{They can also be produced in the decays of tau-leptons 
and heavy hadrons.}
In the top-right plot, we see that the sphaleron events contain at least one isolated lepton with $p_T>30$ GeV more than half of the time.
Unlike the inclusive multiplicity $N(p_T>100)$, $N_{\rm lep}$ depends very weakly on the threshold energy $E_{\rm thr}$.
Comparing these distributions to those with higher collider energies (middle-left and bottom-left plots), 
we can observe that the number of isolated leptons {\it decreases}
as the $\sqrt{s}$ increases.  
This is, at first glance, counter-intuitive because, as we have seen above, events at a higher energy collider contain 
more massive EW gauge bosons, and one may expect there will be more leptons coming from decays of those.  
However, events with a very large number of massive EW bosons typically also end up containing a large number of jets coming from their decays. Consequently, in such a busy environment the lepton isolation criteria are more likely to fail, suppressing the total number of isolated leptons.
In the middle-left plot, with $\sqrt{s}=27$\;TeV, we also observe that $N_{\rm lep}$ is larger for $E_{\rm thr}=9$\;TeV than for $E_{\rm thr} = 14$\;TeV.  
This can also be understood by the same reasoning: the multiplicity of EW gauge bosons is larger in general for higher $E_{\rm thr} = 14$ and 16 TeV.
One can also notice that the lepton multiplicity is slightly larger for $E_{\rm thr} = 18$ TeV than
for $E_{\rm thr} = 14$ and 16 TeV in the middle-left and bottom-left plots.
This is because the number of EW bosons does not increase much from $E_{\rm thr} = 16$ to 18 TeV
since the multiplicity freezes out at $\sqrt{\hat s} = E_{\rm freeze } = 20$ TeV.
This implies that the leptons are more likely to be accepted for $E_{\rm thr} = 18$ TeV because 
they are more energetic on average and are more likely to satisfy the $p_T$ cut.

We now turn our attention to the right-hand-side plots of Fig.~\ref{fig:nlep_nphot}, in order to discuss the multiplicity of the isolated photons.
In sphaleron events, photons can be produced from the primary interaction.
Roughly speaking, a neutral $SU(2)$ gauge boson is converted to a photon or a $Z$-boson with
probabilities $\sin^2 \theta_W$ and $\cos^2 \theta_W$, respectively. 
Photons may also originate from initial and final state electromagnetic radiation and hadronic decays.
As can be seen, the multiplicity distribution of isolated photons is, by coincidence, similar to that of isolated leptons.
In general an event contains at least one photon about a half of the time. 
We see the same tendency as in the case of the leptons: the multiplicity becomes smaller at larger collider energies.
This stems from the same fact as for the case of the leptons; i.e.\ that the isolation criteria are likely to fail due to a busy environment.

\begin{figure}[tbp]
\centering 
\includegraphics[width=.45\textwidth,clip]{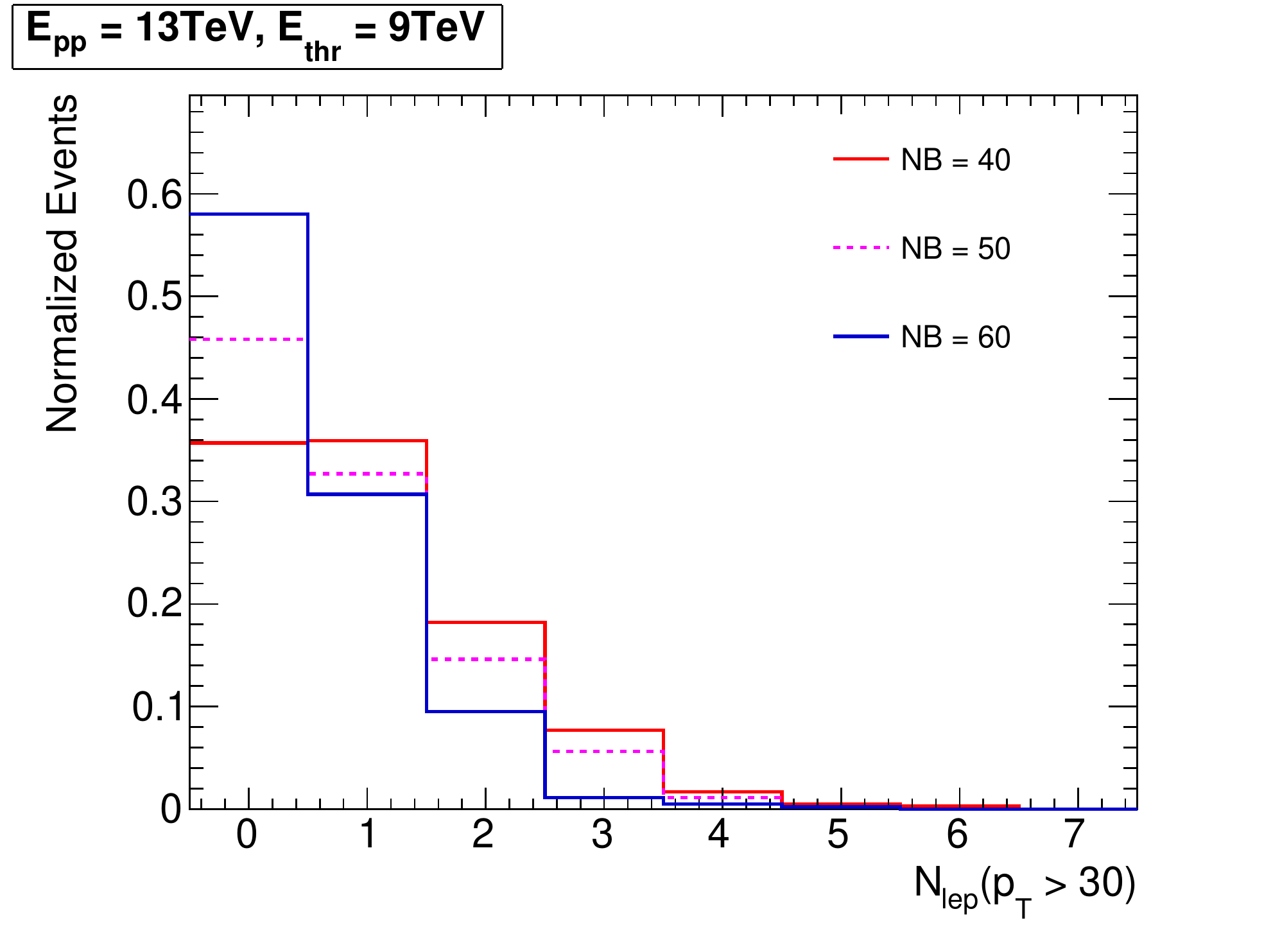}
\includegraphics[width=.45\textwidth,clip]{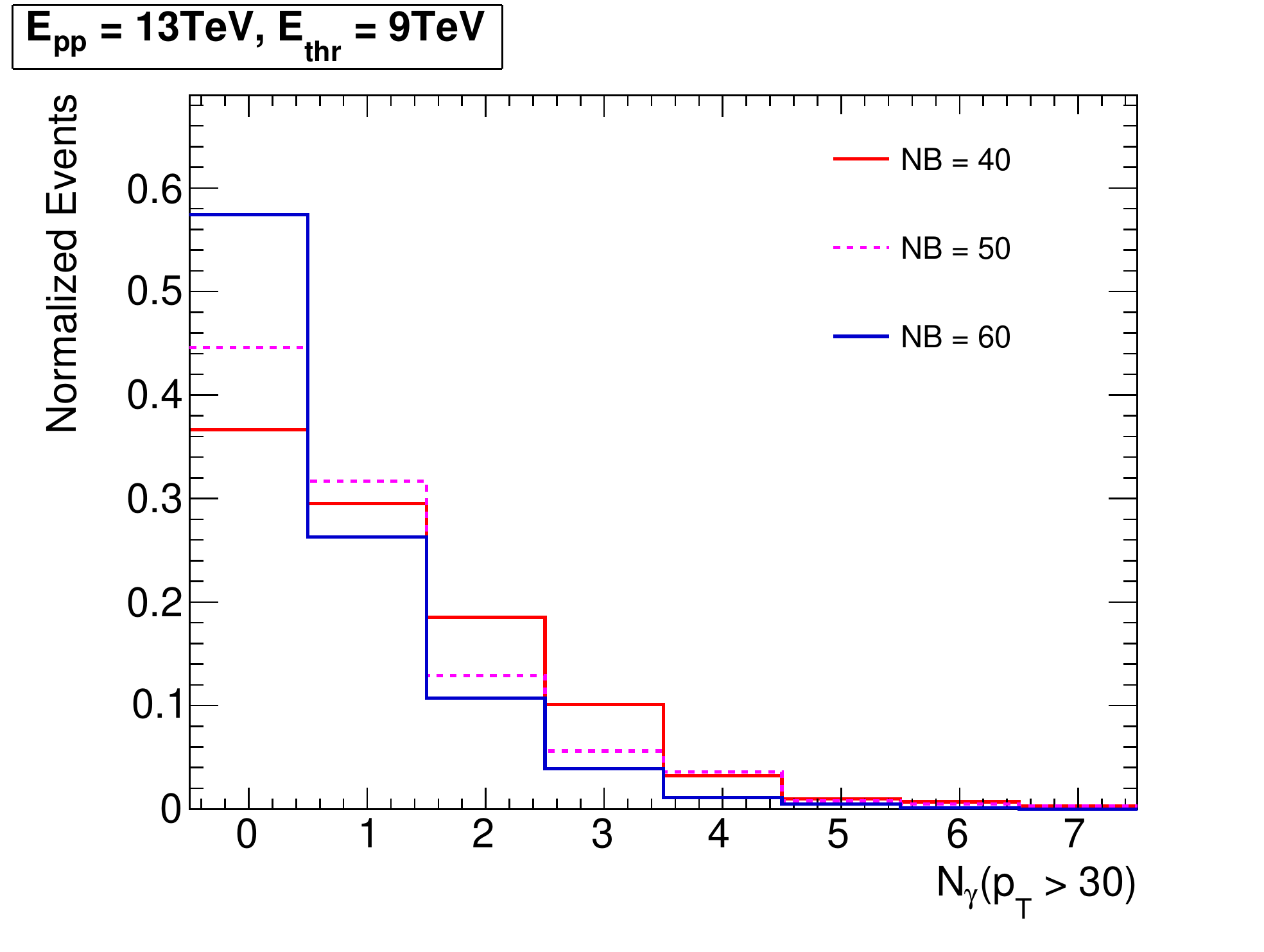}
\includegraphics[width=.45\textwidth,clip]{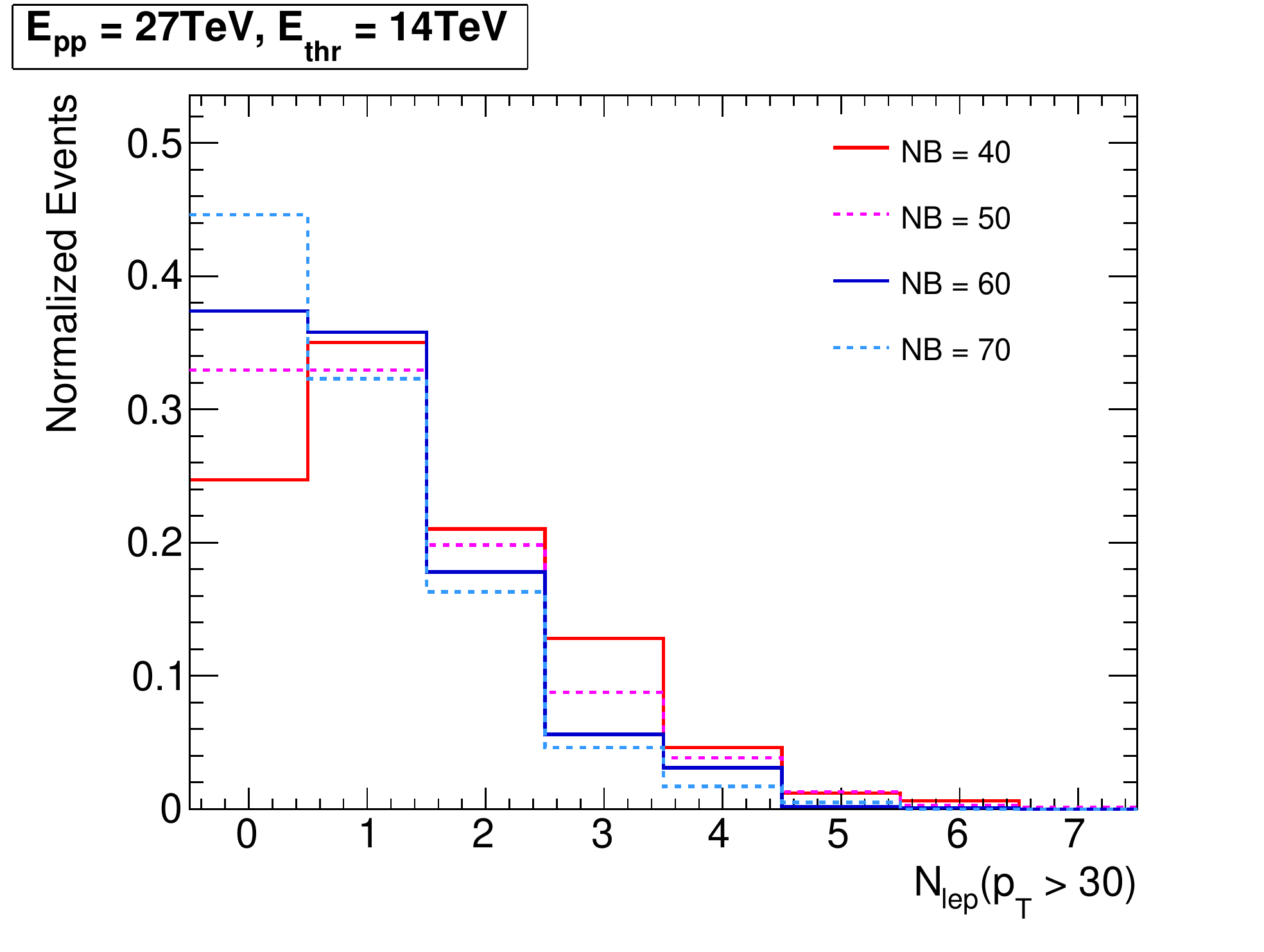}
\includegraphics[width=.45\textwidth,clip]{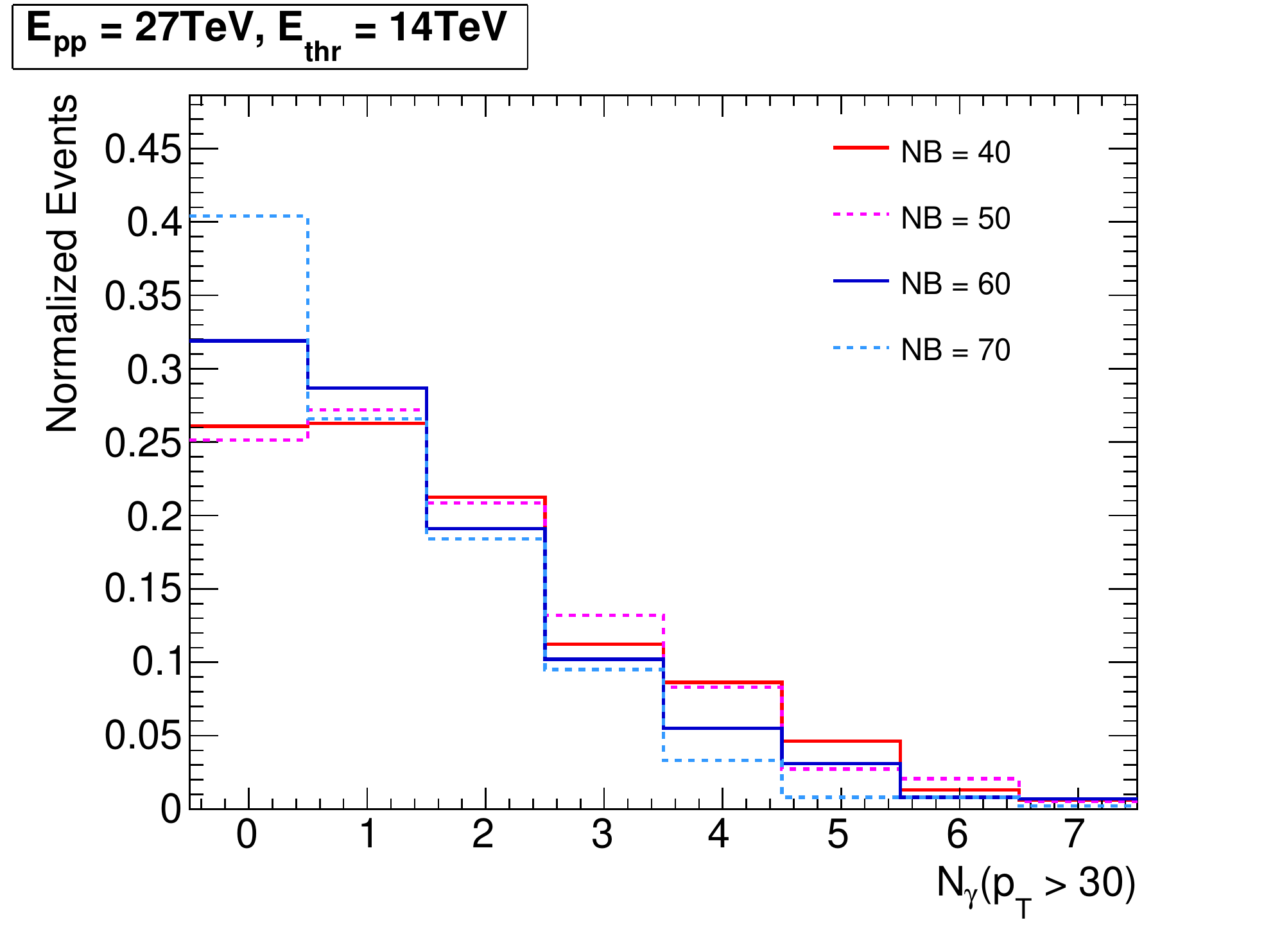}
\includegraphics[width=.45\textwidth,clip]{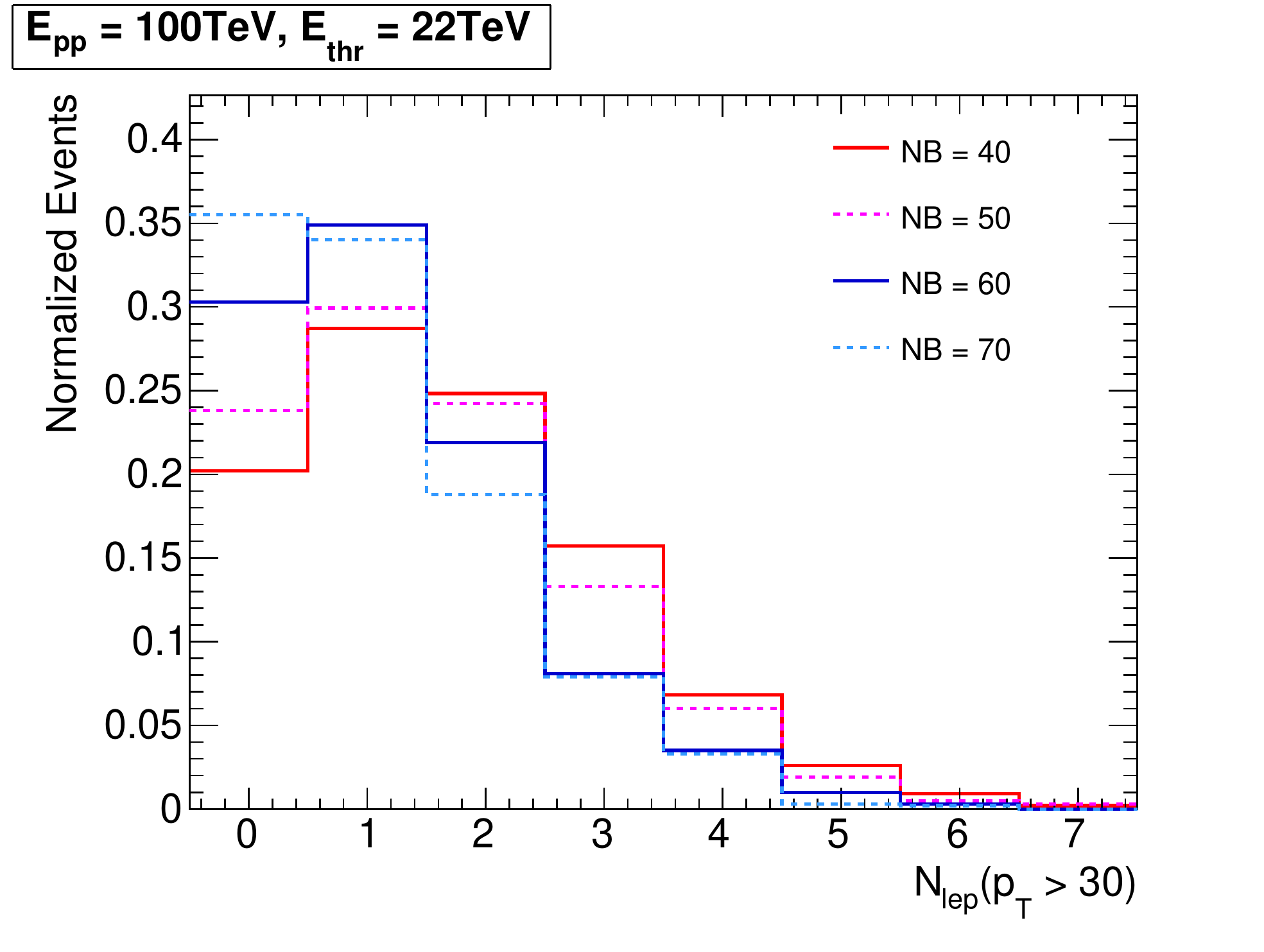}
\includegraphics[width=.45\textwidth,clip]{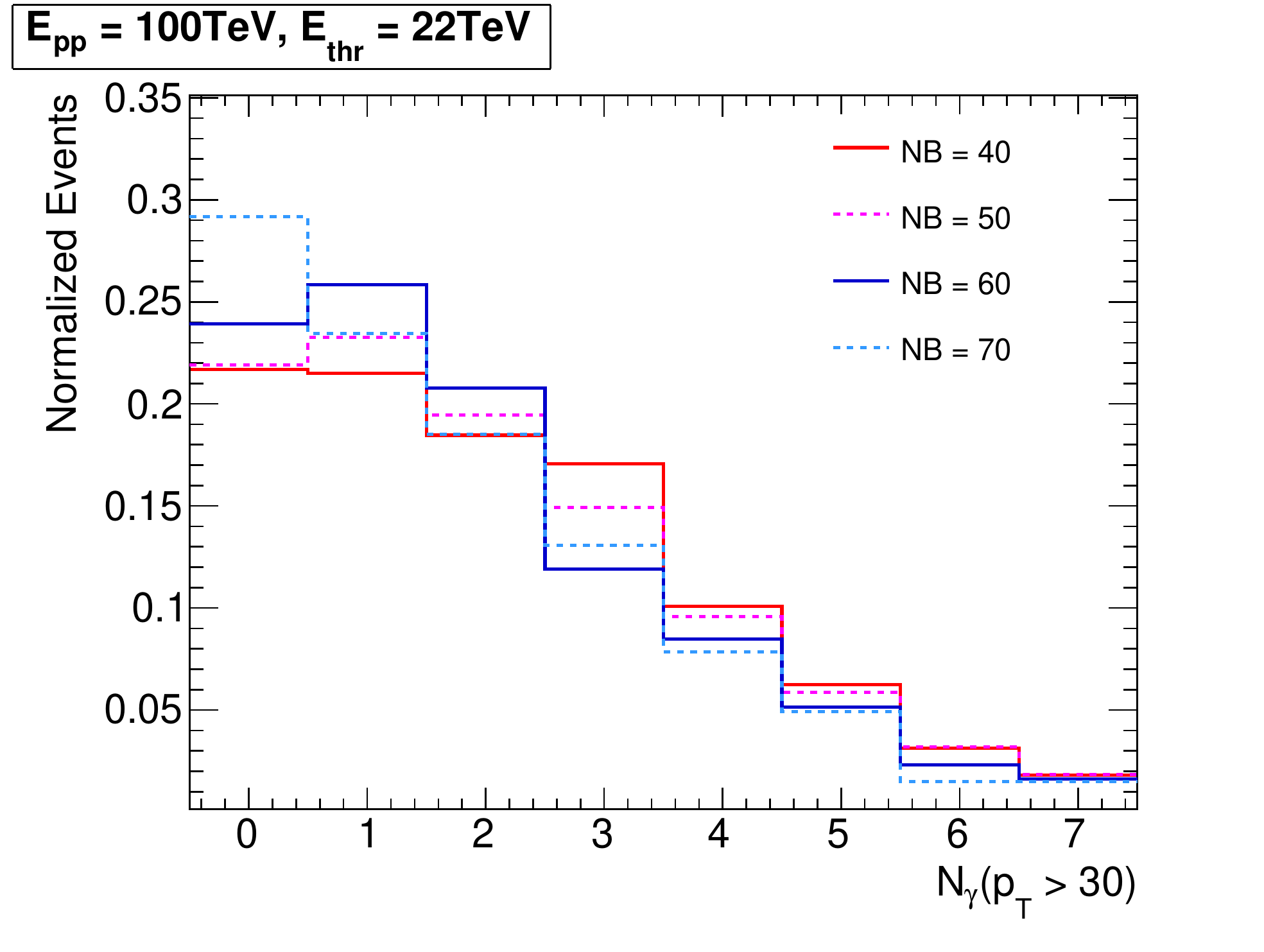}
\caption{\label{fig:nlep_nphot_nb} 
The same as Fig.~\ref{fig:nlep_nphot} but with the total numbers of bosons instead fixed to 
$n_B = 40$ (red-solid), 50 (pink-dashed) and 60 (blue-solid) for the 13 TeV LHC.
In addition to these $n_B = 70$ (cyan-dashed) is shown for $\sqrt{s} =$\;27 and 100\;TeV. 
The threshold energy is taken as $E_{\rm thr} = 9$, 14\;TeV and 22 TeV for
$\sqrt{s} = 13$, 27\;TeV and 100 TeV.
}
\end{figure}

In Fig.~\ref{fig:nlep_nphot_nb} we show the same $N_{\rm lep}$ and $N_\gamma$ distributions 
at fixed $n_B$.
The same line-styles and colour schemes are used as in Fig.~\ref{fig:n100_ptj1_nb} to discriminate the different $n_B$.
In these plots we can consistently see that the lepton and photon multiplicities are
larger in general for smaller $n_B$.
This is in agreement with our expectations of the EW gauge boson multiplicity and the isolation criteria.
Unlike the case in Fig.~\ref{fig:nlep_nphot}, the multiplicity becomes larger at larger collider energies.
This is because, for a fixed $n_B$, the effect of increasing $\sqrt{s}$ is
to make the leptons and photons more energetic on average, which helps in satisfying the $p_T$ cut.
We note that for the 100 TeV collider with a lower boson multiplicity ($n_B \sim 40$), the events tend 
to have many photons.  
There will be $N_\gamma \geq 2$ more than half of the time in this particular case ($\sqrt{s} = 100$ TeV, $n_B = 40$).
Exploiting this fact may help to reduce the SM background in future searches.

\subsection{Projected sensitivities}
\label{sec:projection}

In the present subsection we estimate the sensitivities of observing sphaleron-induced processes 
at proton-proton colliders.
We consider three cases; the 13 TeV high-luminosity LHC (HL-LHC13), the 27 TeV high-energy LHC (HE-LHC27)
and a future 100 TeV circular collider (FCC100).
To facilitate the comparison of the capability of each collider, we assume the envisioned integrated luminosity of the HL-LHC13, 3 ab$^{-1}$, in all three cases, although future collider experiments will likely collect more data. 

Our analysis is inspired by the CMS analysis of Ref.~\cite{Sirunyan:2018xwt}, on searches for 
sphaleron and mini black hole production in 13 TeV data with $L = 36$ fb$^{-1}$.
Following this analysis, we apply cuts on two variables; $N(p_T > 100)$ and $S_T^{100}$. 
The former variable is defined as before, as the number of reconstructed objects (jets, isolated leptons and photons)
with $p_T > 100$\;GeV and the latter variable is the scalar sum of the transverse missing energy ($\slashed{E}_T$)
and $p_T$ of all reconstructed objects with $p_T > 100$ GeV. 
Our signal selection requirement is given by
%
%
\begin{eqnarray}\label{eq:cuts}
N(p_T > 100) \geq 11, & S_T^{100} > 4\,{\rm TeV} & \cdots \mathrm{HL-LHC13} \\ \nonumber
N(p_T > 100) \geq 15, & S_T^{100} > 7\,{\rm TeV} & \cdots \mathrm{HE-LHC27, FCC100}
\end{eqnarray}

It has been reported in the CMS analysis \cite{Sirunyan:2018xwt} that the SM background is reduced 
to less than 1 event, with the above HL-LHC13 selection criteria, while the signal efficiency remains greater than 90\,\% 
\cite{Ringwald:2018gpv}.  
For the HE-LHC27 and FCC100, we have increased these requirements upwards.
We expect that the SM background for these cases is also reduced to less than 1 event.

The impact of the $N(p_T > 100)$ cut on the signal efficiency can be deduced from examining the left panel of
Fig.~\ref{fig:n100_ptj1} and \ref{fig:n100_ptj1_nb}.
One can see that signal events are localised above $N(p_T > 100) \gtrsim 11$ for $\sqrt{s} = 13$ TeV
and $\gtrsim 15$ for $\sqrt{s} = 27$ and 100 TeV.
Therefore, the signal selection efficiency is almost 100\% for this cut.

\begin{figure}[tbp]
\centering 
\includegraphics[width=.45\textwidth,clip]{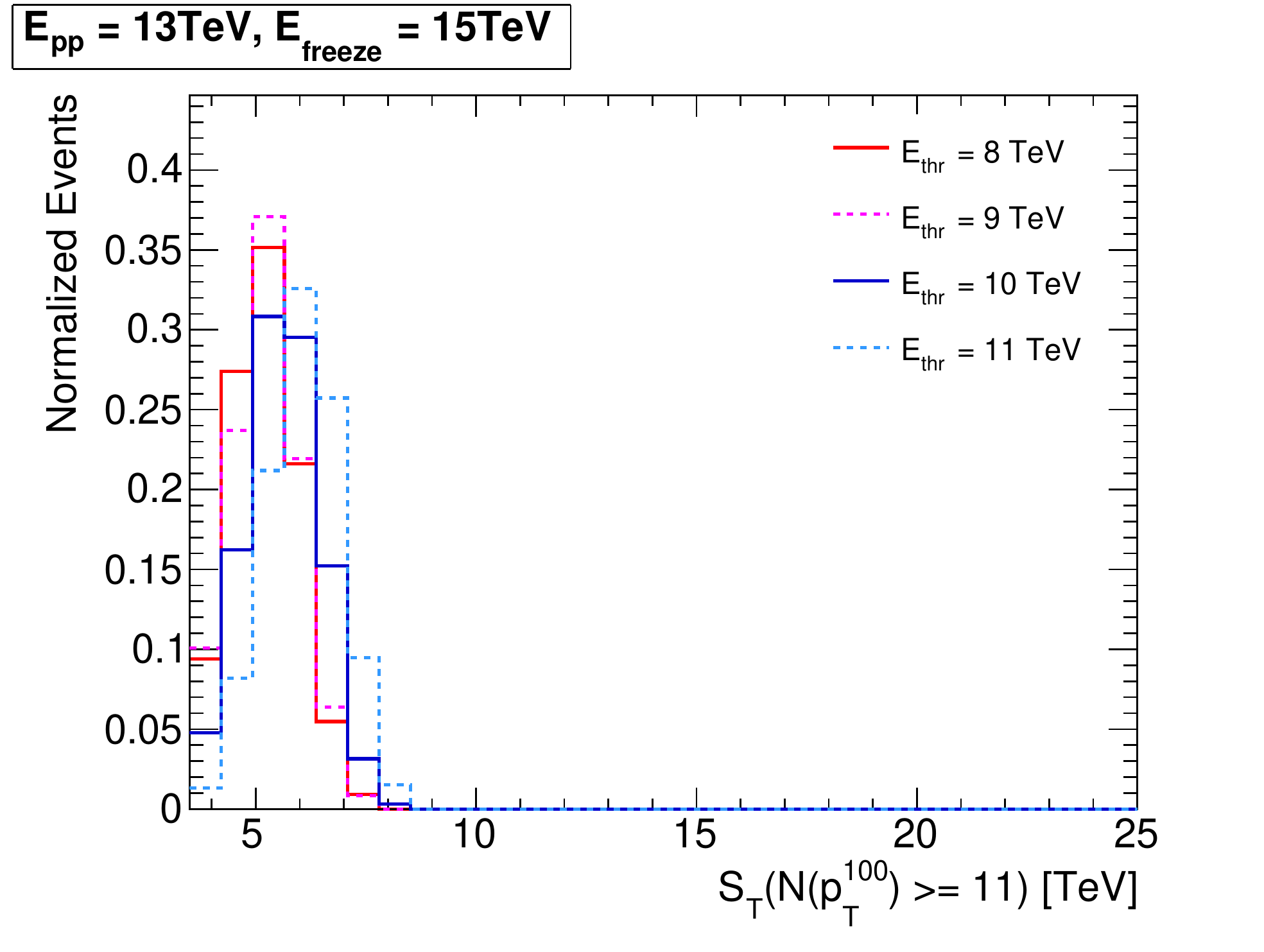}
\includegraphics[width=.45\textwidth,clip]{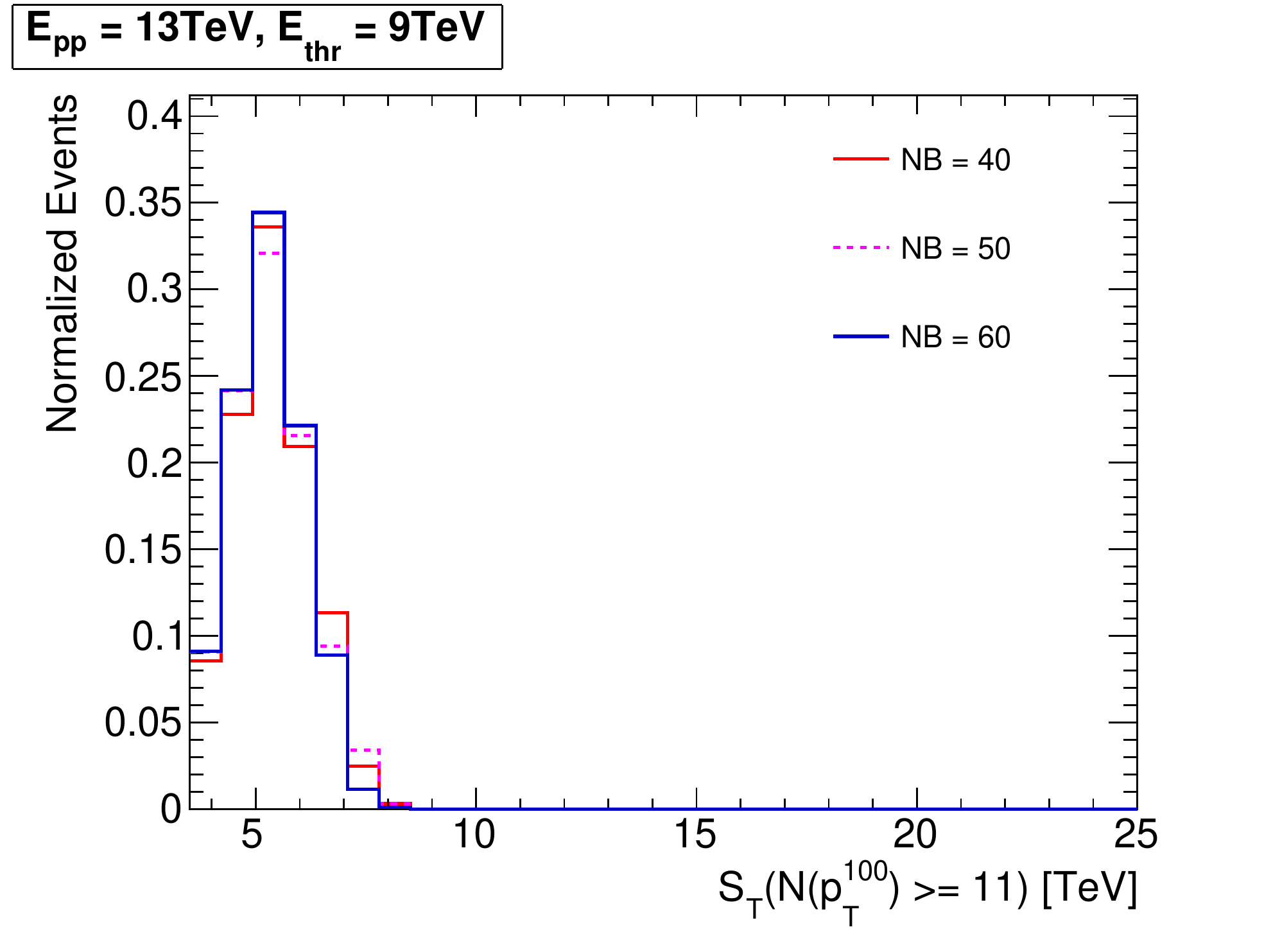}
\includegraphics[width=.45\textwidth,clip]{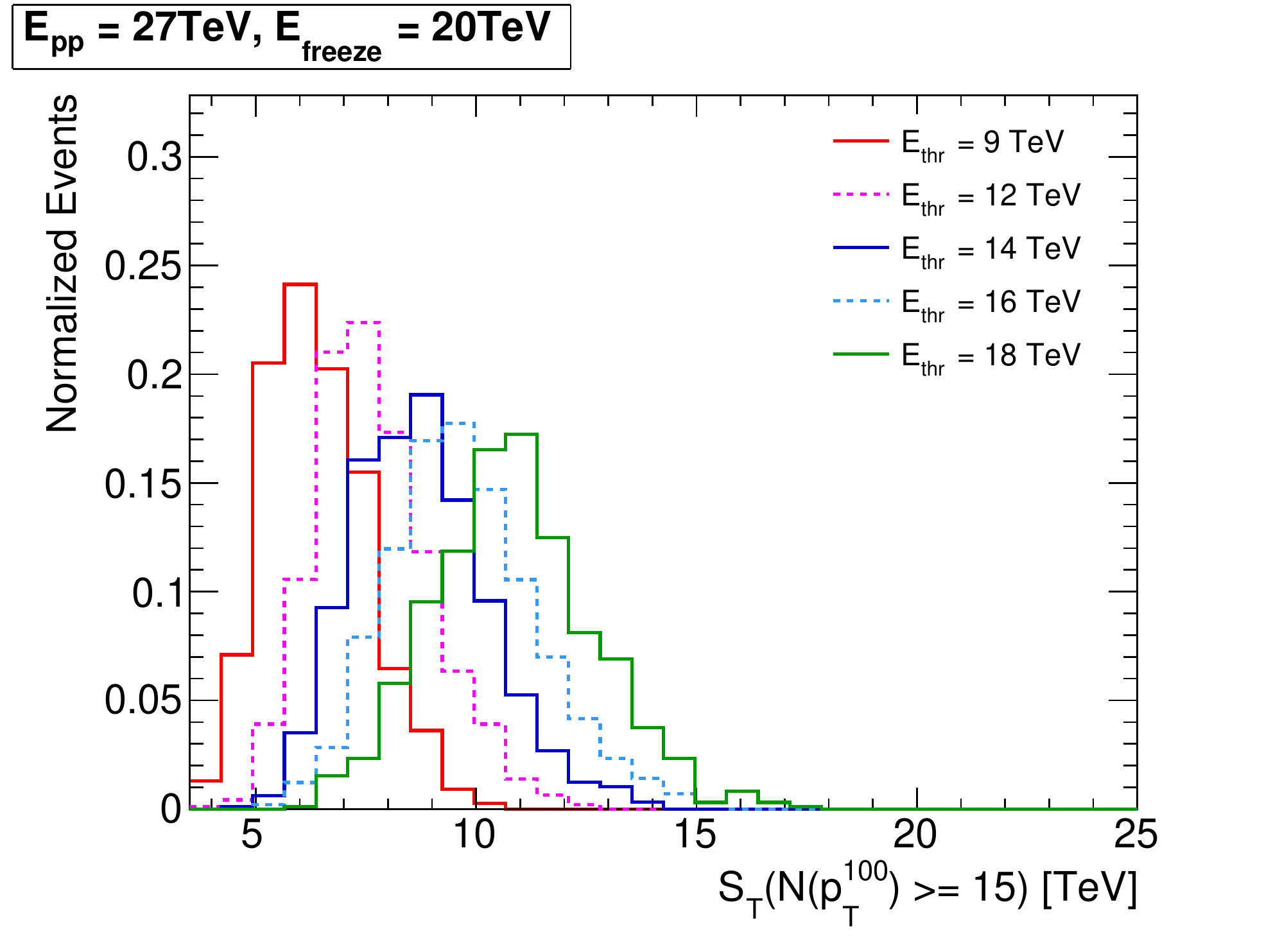}
\includegraphics[width=.45\textwidth,clip]{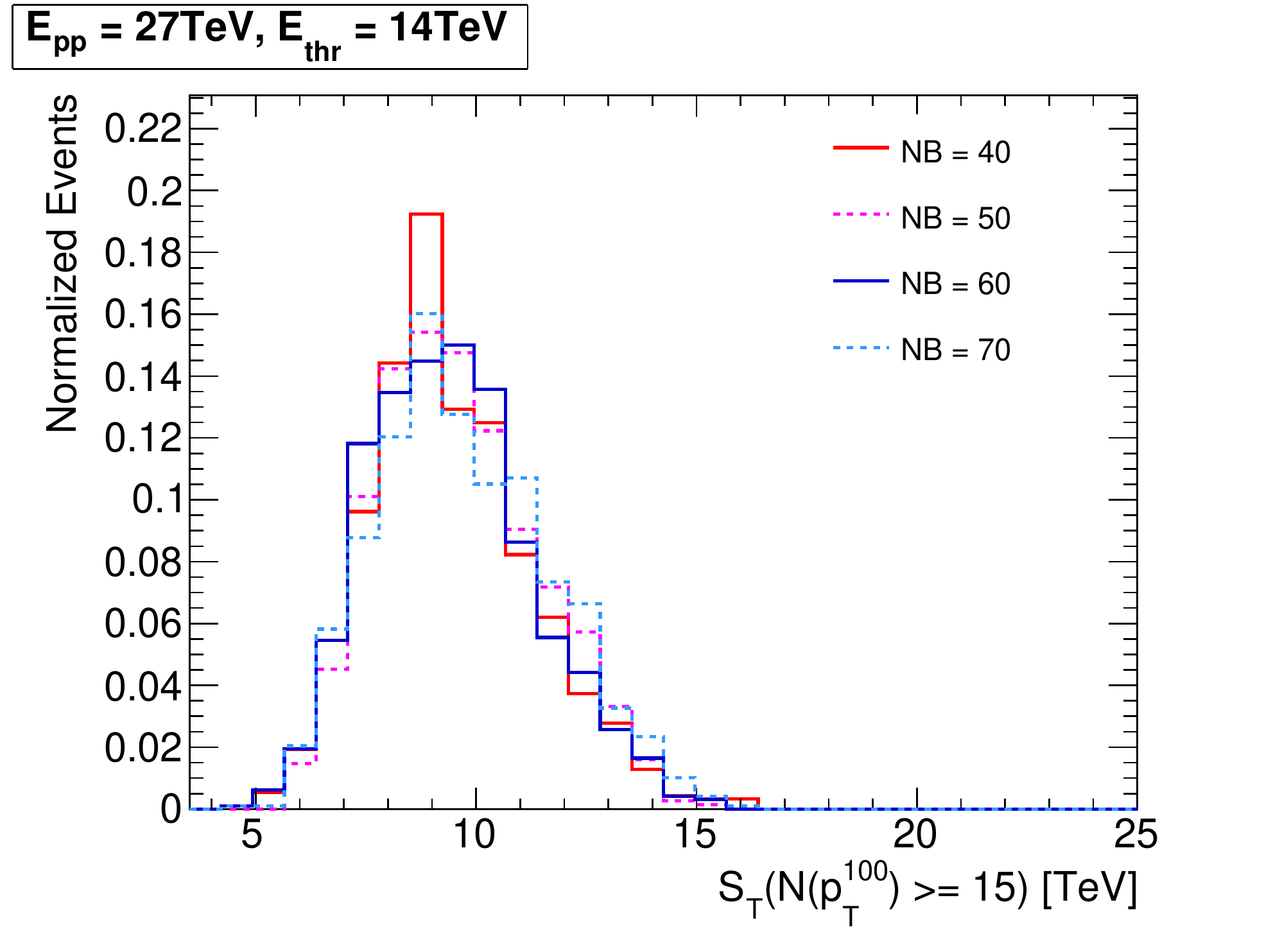}
\includegraphics[width=.45\textwidth,clip]{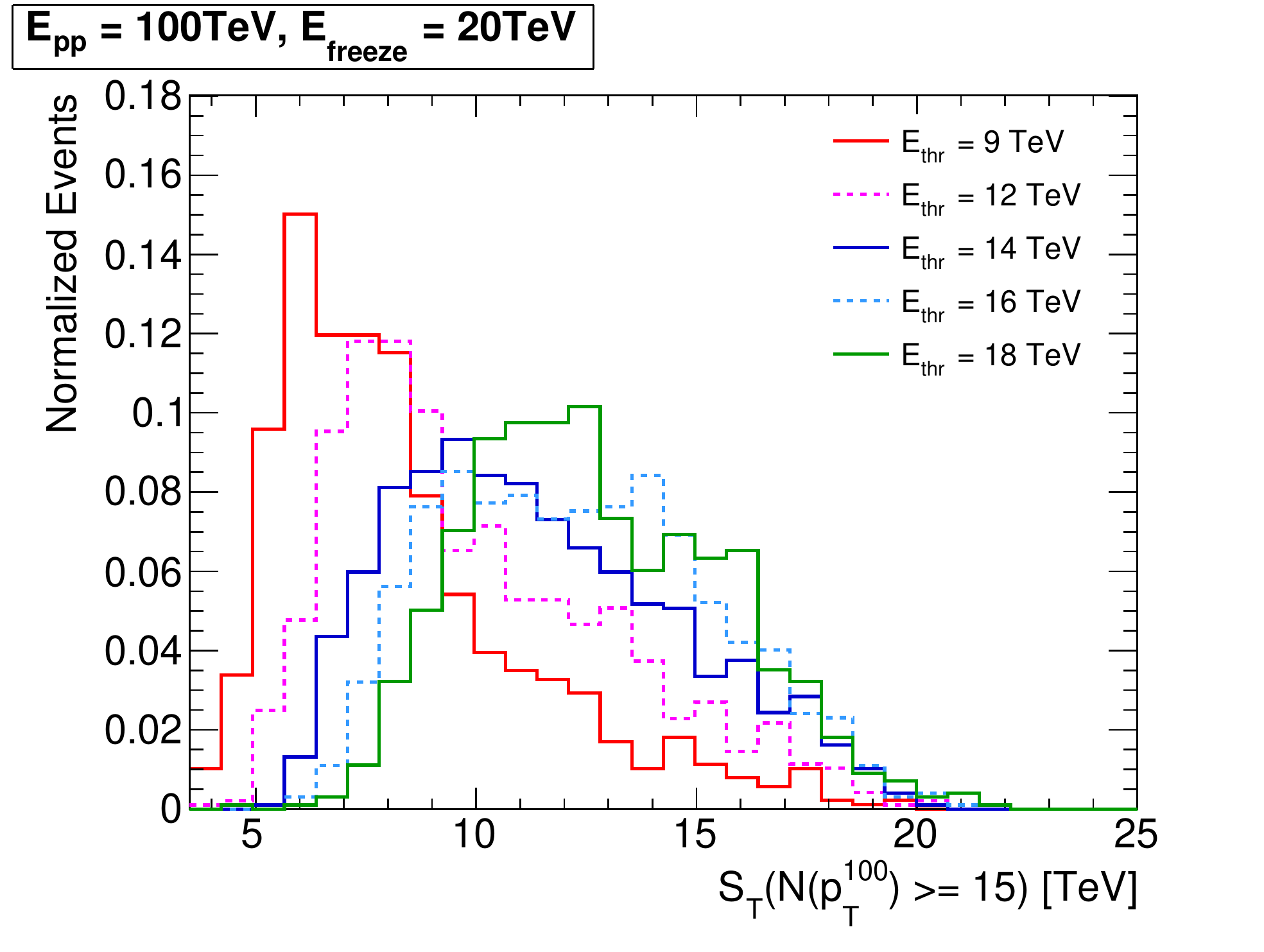}
\includegraphics[width=.45\textwidth,clip]{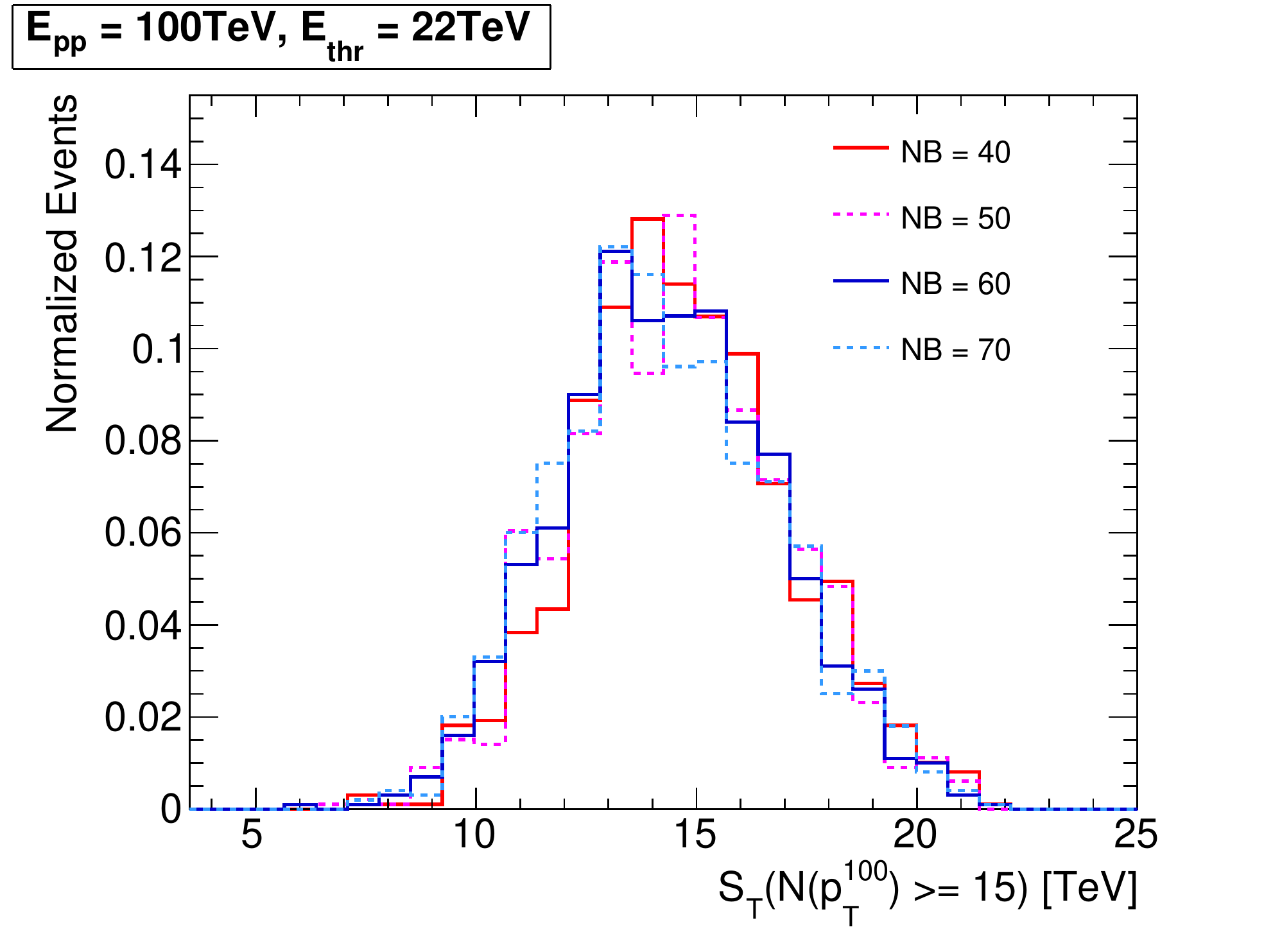}
\caption{\label{fig:ST} The $S_T$ distributions for $\sqrt{s} = 13$\;TeV (top), 27\;TeV (middle) and 100\; Tev(bottom).
On the left-hand-side plots, multiplicity distributions have been generated based on the LOME formula,
while the boson multiplicity was fixed on the right-hand-side panels.
The same line-styles and colour scheme are used to discriminate between the different $E_{\rm thr}$ and $n_B$ as in Fig.~\ref{fig:n100_ptj1}
and Fig.~\ref{fig:n100_ptj1_nb}. 
The $N_{100} \geq 11$ (15) cut is applied for $\sqrt{s} = 13$ (27 and 100) TeV.
}
\end{figure}

To see the efficiency for the $S_T^{100}$ selection, 
we show the distributions of $S_T^{100}$ in Fig.~\ref{fig:ST}.
In the left-hand side plots, multiplicity distributions are constructed based on the LOME prescription,
while the boson multiplicity is fixed in the right-hand-side plots.
Before plotting, events are required to satisfy $N(p_T > 100) \geq 11$ for $\sqrt{s} = 13$ TeV and $N(p_T > 100) \geq 15$ for $\sqrt{s} = 27$ and 100 TeV so that the area of the histogram with $S_T^{100} > 4$ (7) TeV corresponds the signal efficiency of our event selection.
In the top two plots, we see that the signal efficiency for the HL-LHC13 selection is $\gtrsim 90$\,\%.  
In the middle-left and bottom-right plots, one can see that the $S_T^{100}$ distribution is sensitive to
$E_{\rm thr}$.
For $\sqrt{s} = 27$ and 100 TeV, the efficiency can be larger than 80\,\% 
for $E_{\rm thr} \gtrsim 12$ TeV.
The signal efficiency is still as large as 50\% even for $E_{\rm thr} \sim 9$ TeV.

In the plots on the right-hand side, we see that the $S_T^{100}$ distribution is not sensitive to $n_B$.
This is expected because the variable is defined as the sum of all transverse momenta and, e.g.\ splitting one particle into two does not change the sum of the $p_T$, as long as all particles are counted correctly.

\begin{figure}[tbp]
\centering 
\includegraphics[width=.32\textwidth,clip]{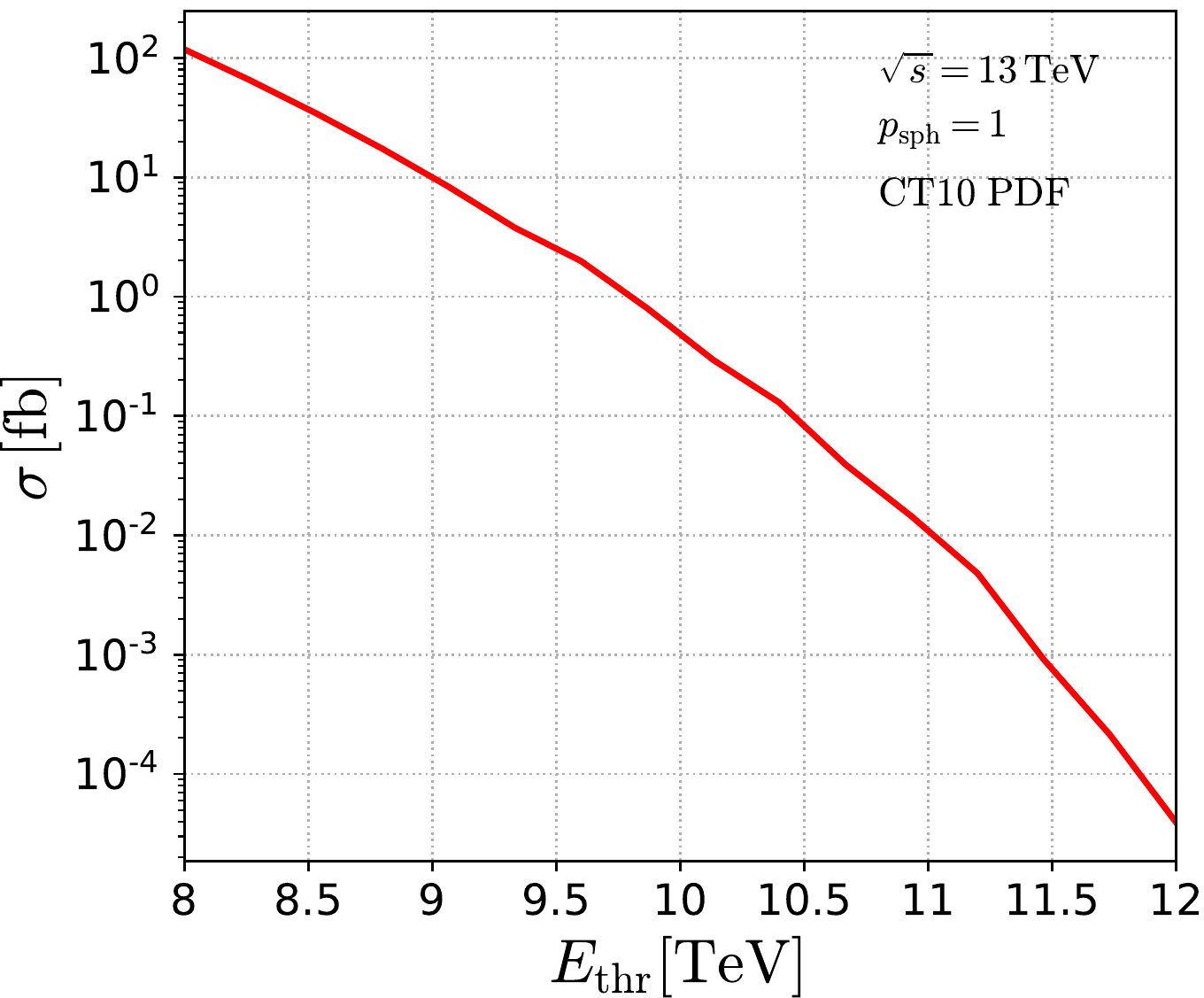}
\includegraphics[width=.32\textwidth,clip]{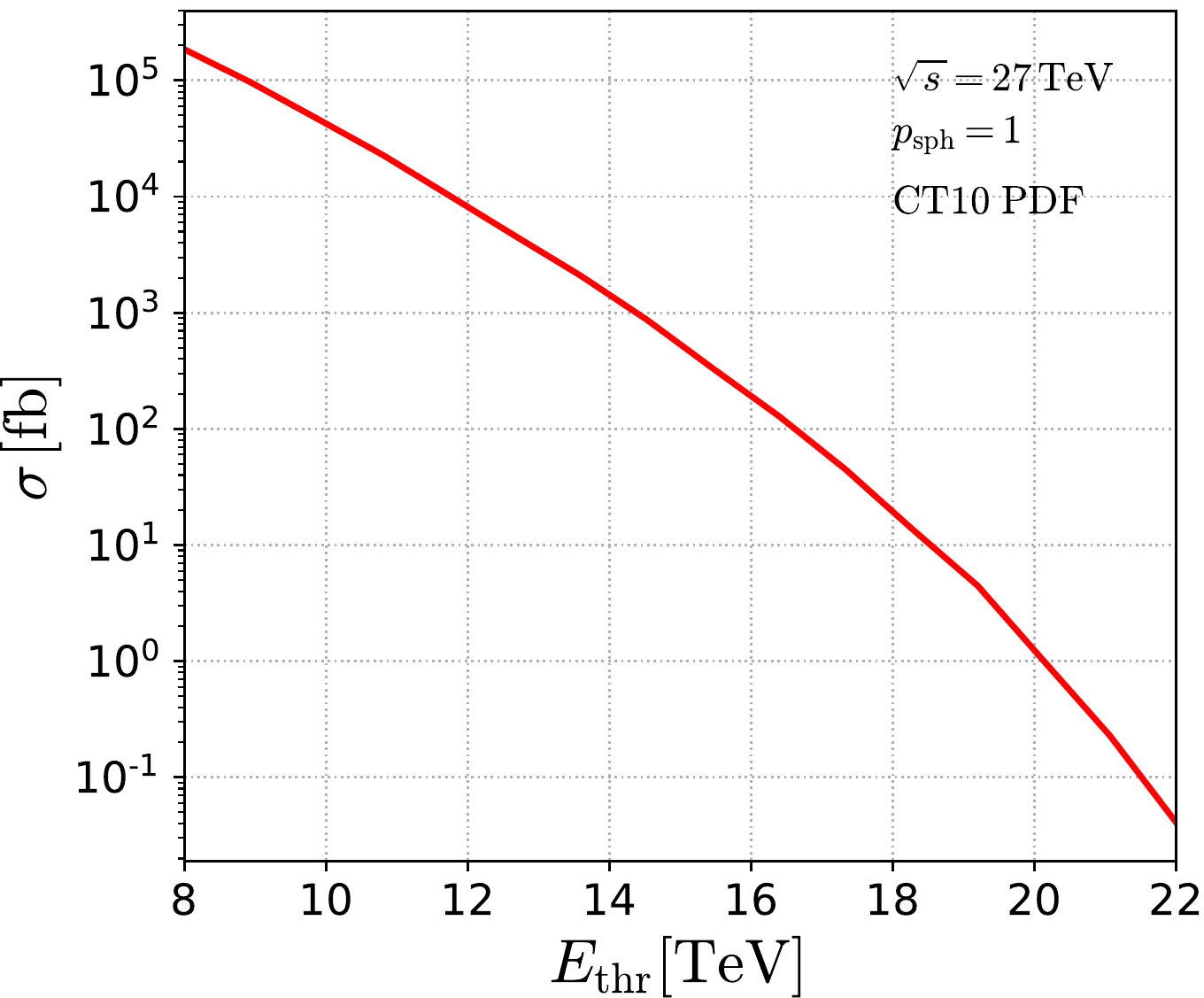}
\includegraphics[width=.32\textwidth,clip]{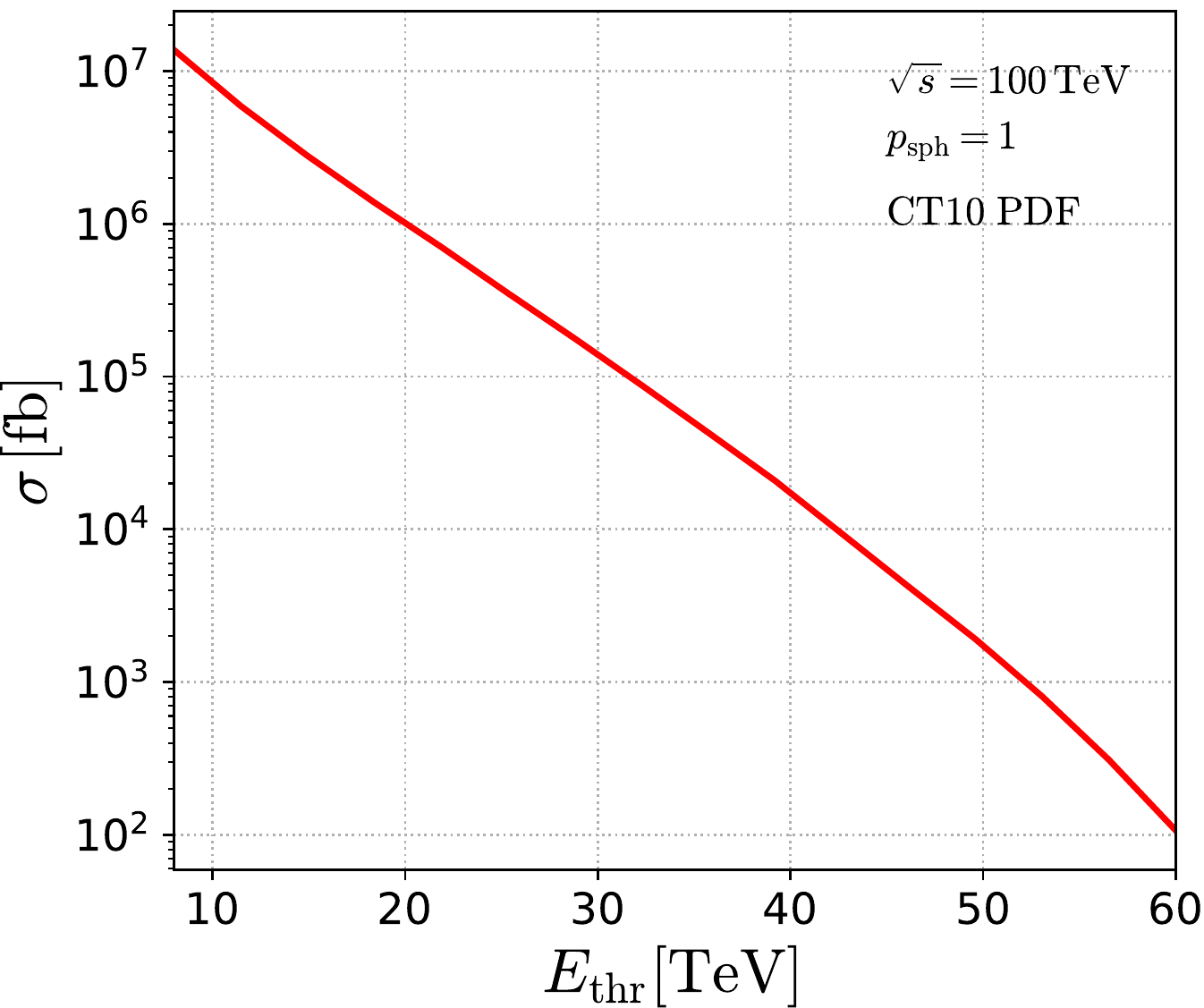}
\caption{\label{fig:xsec} The cross section for the sphaleron-induced processes as a function of $E_{\rm thr}$ 
with $p_{\rm sph}=1$. 
The proton-proton collision energy is taken to be 13\;TeV (left), 27\;TeV (middle) and 100\;TeV (right).
The leading order {\tt CT10} PDF set was used. }
\end{figure}

Our goal is to derive the projected limit on the $E_{\rm thr}$ vs $p_{\rm sph}$ plane.
For this purpose, we first calculate the cross section of sphaleron-induced processes in terms of $E_{\rm thr}$
setting $p_{\rm sph} = 1$.
We convolve the partonic cross section defined by Eqs.~\eqref{eq:sighat} and \eqref{eq:sig0} with the PDFs.
In particular, the leading order {\tt CT10} PDF set is used here~\cite{Guzzi:2011sv}.
The resulting hadronic cross sections for $p_{\rm sph} = 1$ are shown in Fig.~\ref{fig:xsec}. 
We see that the hadronic cross sections are quite large for smaller $E_{\rm sph}$.
At $E_{\rm sph} = 9$ TeV, we find $\sigma \simeq 10$, $10^5$ and $10^7$ fb for $\sqrt{s} = 13$, 27 and 100 TeV, respectively.
The cross section falls rapidly as $E_{\rm thr}$ increases, but this reduction 
is faster (slower) for smaller (larger) collider energies.
For example, for $\sqrt{s} = 13$ TeV, the cross section is reduced to $10^{-(4-5)}$ fb at $E_{\rm thr} = 12$ TeV.
On the other hand, for $\sqrt{s} = 100$ TeV, the cross section remains as large as $10^5$ fb even for $E_{\rm thr} = 30$ TeV.

With these considerations at hand, we can derive the projected sensitivities.
We estimate the signal yield by
\beqn
N_s = \sigma \cdot \epsilon \cdot L_{\rm int},
\eeqn  
where $\sigma$ is hadronic cross section, $\epsilon$ is the signal efficiency 
and $L_{\rm int}$ is the integrated luminosity.

Since precise estimation of the multi-particle SM background is theoretically challenging and depends substantially on 
detector performance, we follow here a simplified approach.
We assume that the SM background is well-suppressed below ${\cal O}(1)$ due to the selection cuts, 
and we highlight the region between $N_s = 3$ and 10 in the $E_{\rm thr}$ vs $p_{\rm sph}$ plane.  
We expect that the boundary of this region will roughly correspond to the exclusion and discovery
of the sphaleron-induced processes.

\begin{figure}[tbp]
\centering 
\includegraphics[width=.45\textwidth,clip]{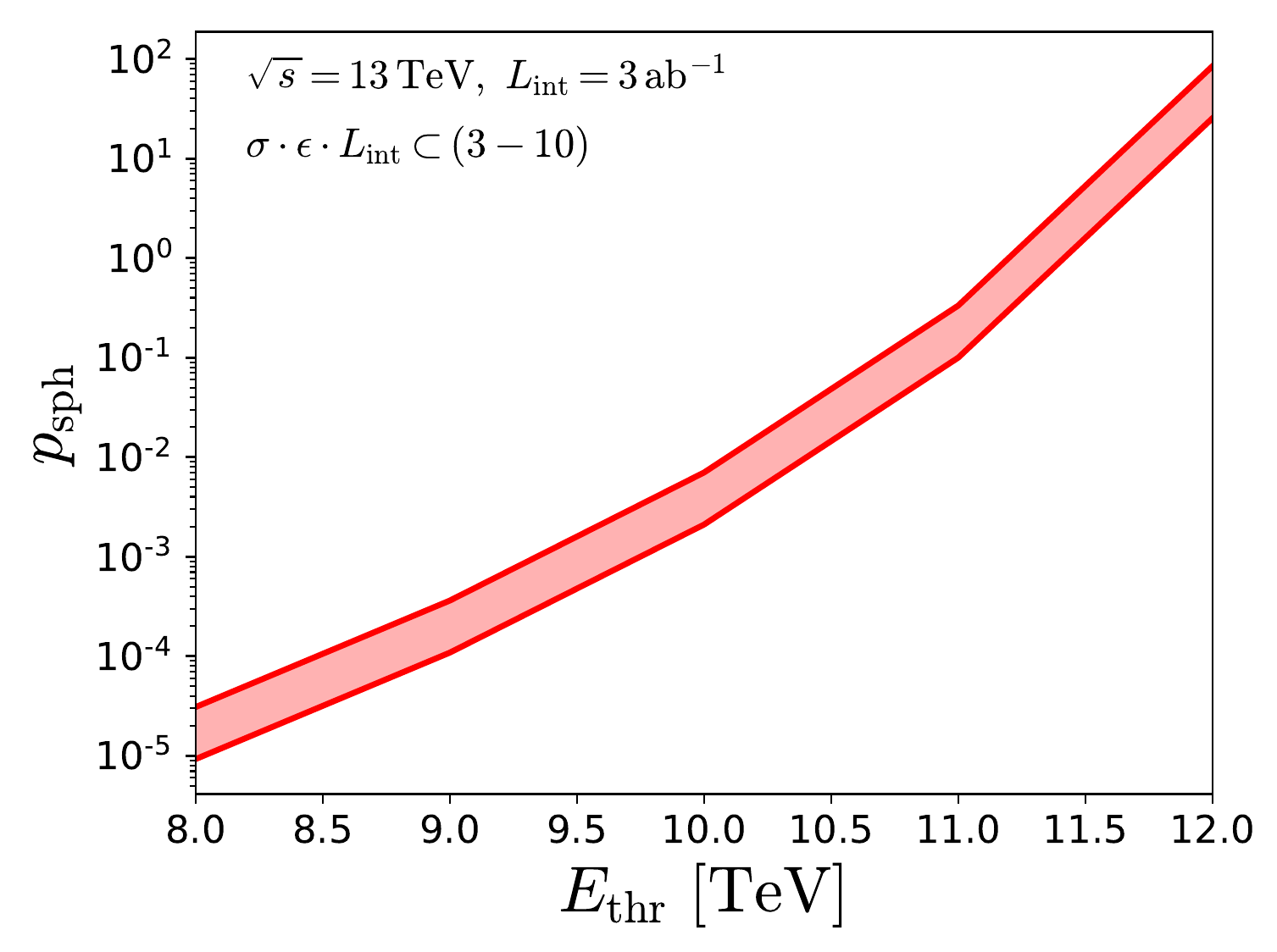}
\includegraphics[width=.45\textwidth,clip]{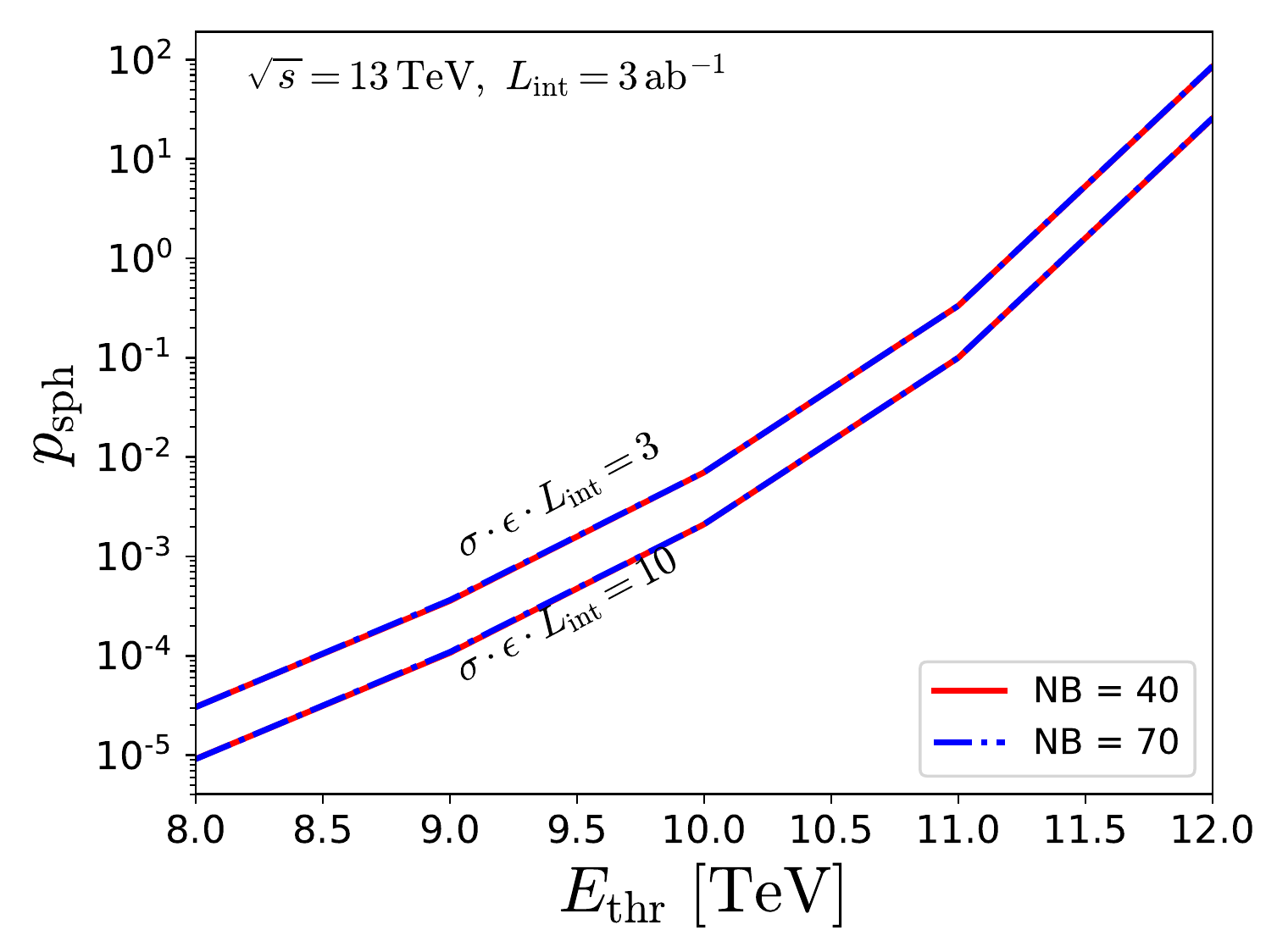}
\includegraphics[width=.45\textwidth,clip]{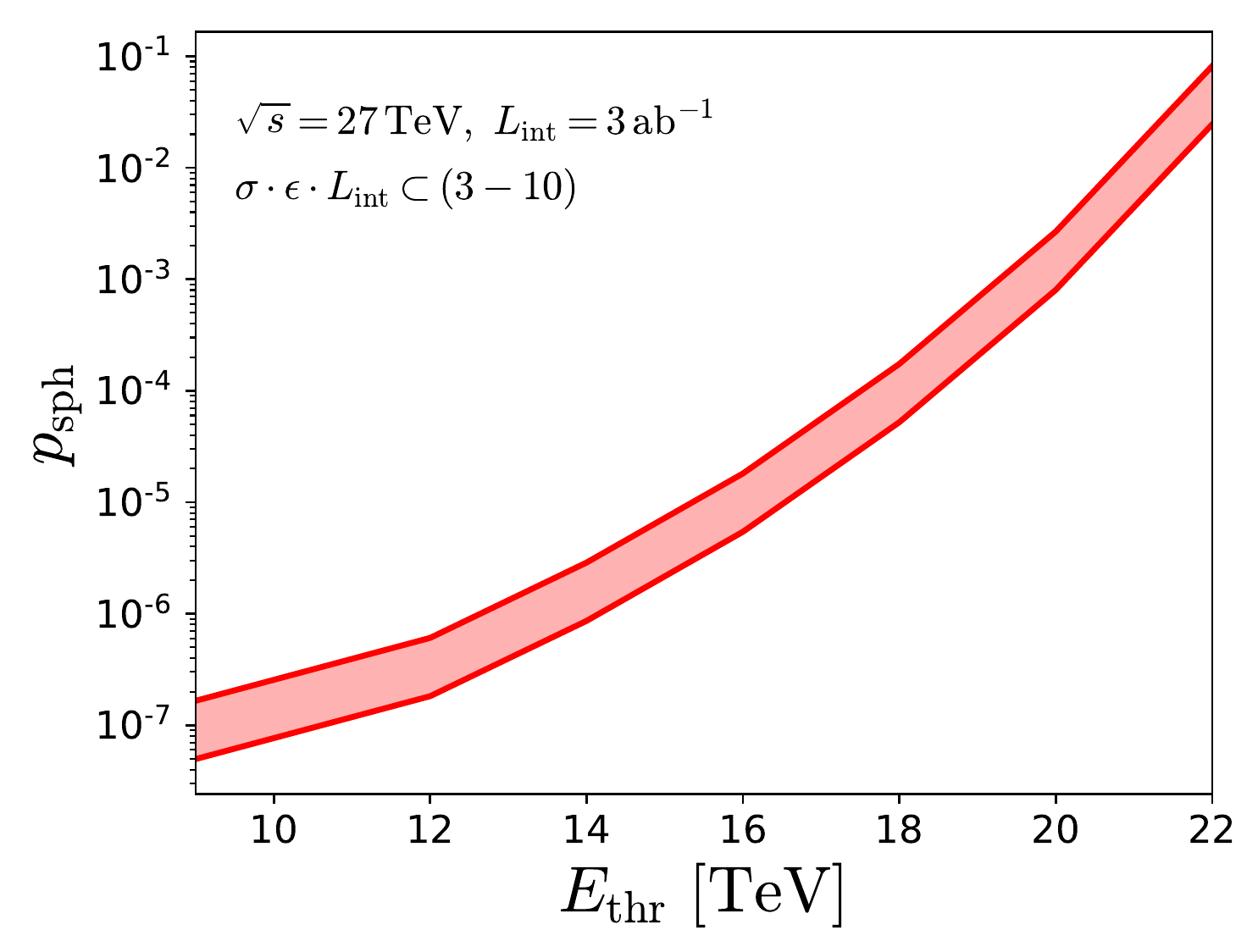}
\includegraphics[width=.45\textwidth,clip]{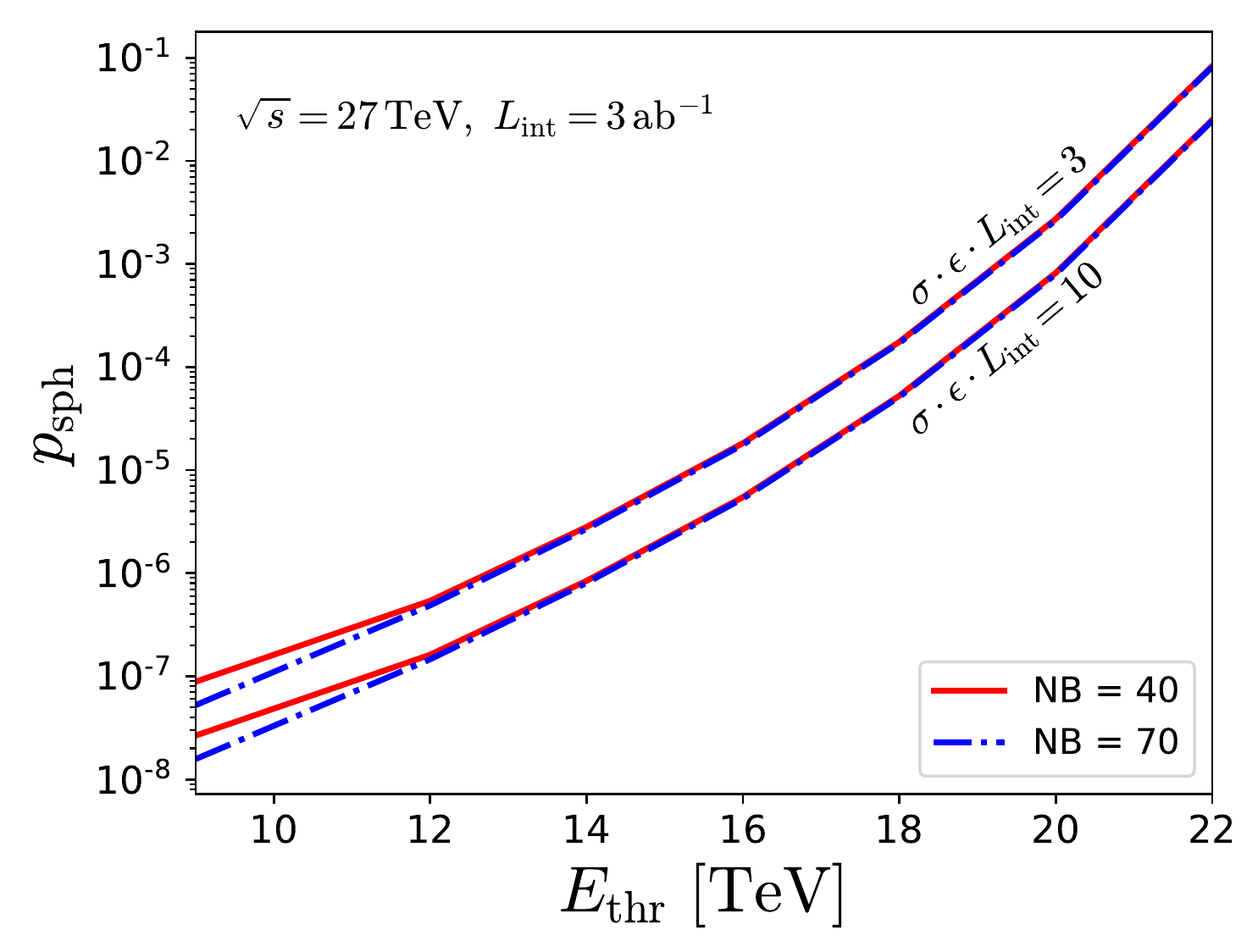}  
\includegraphics[width=.45\textwidth,clip]{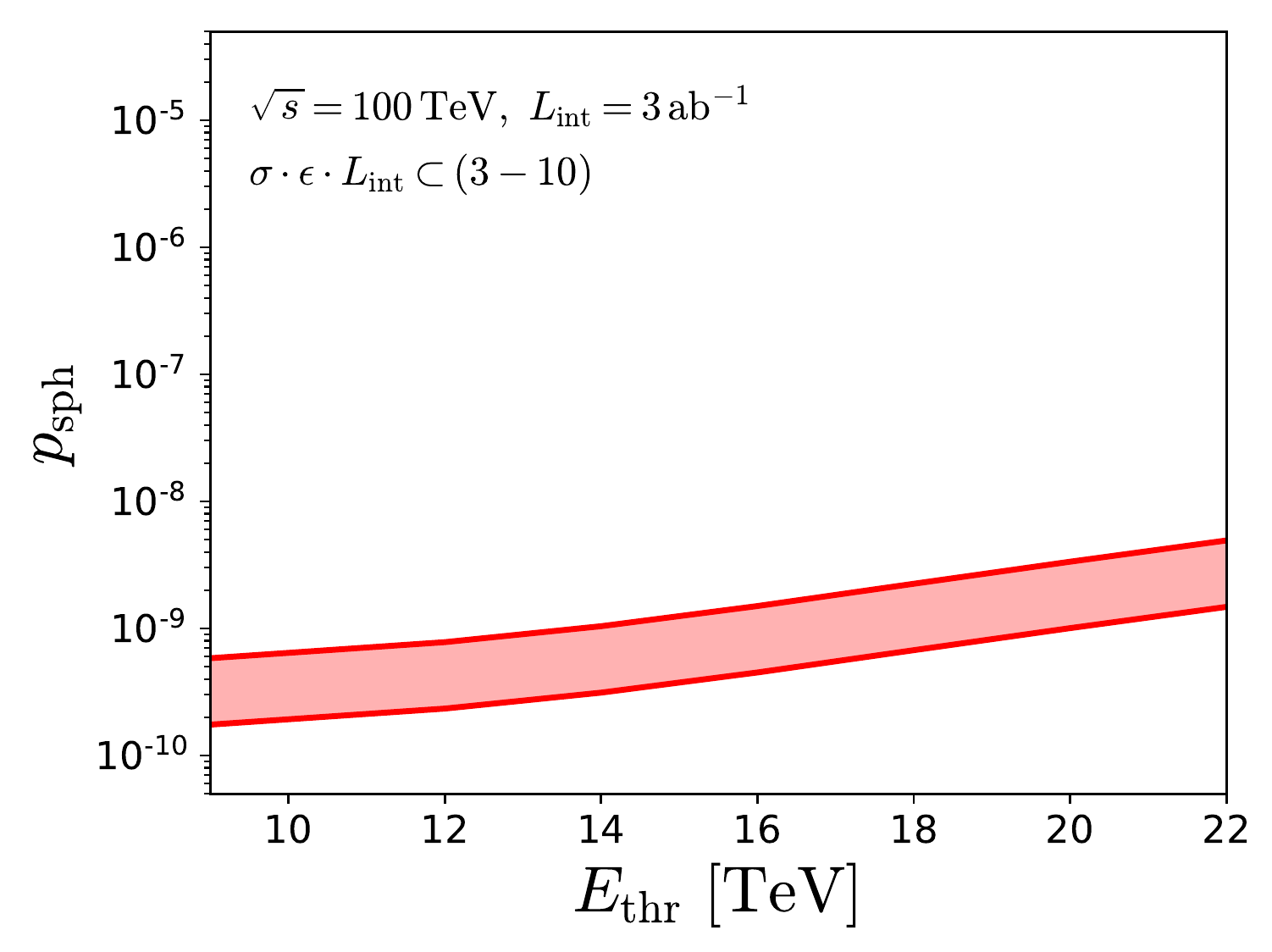}
\includegraphics[width=.45\textwidth,clip]{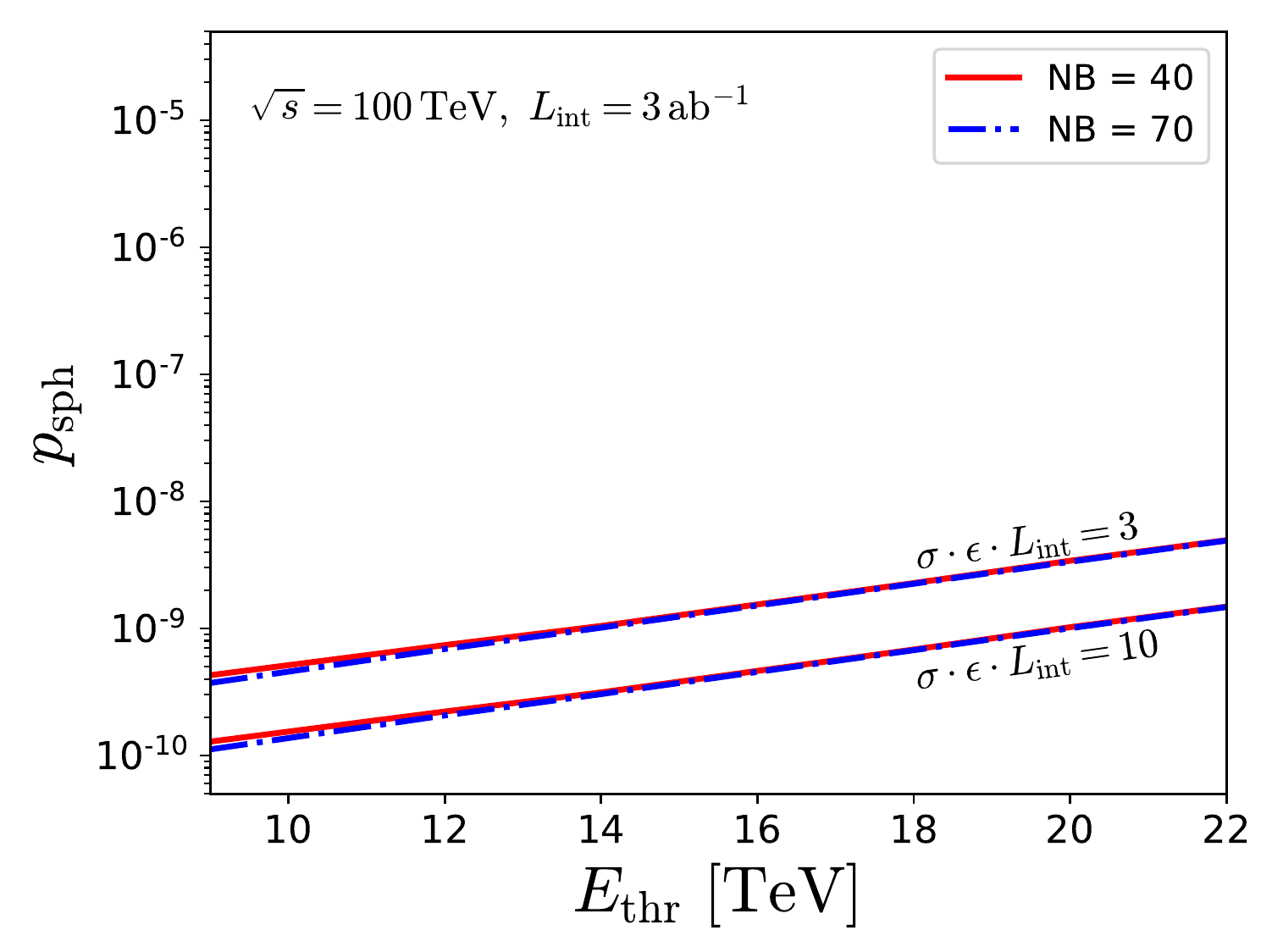}
\caption{\label{fig:limit} The projected sensitivities in the $E_{\rm thr}$ vs $p_{\rm sph}$ plane
at the HL-LHC13 (top), HE-LHC27 (middle) and FCC100 (bottom).
In the right panel, we generate boson multiplicity according to the LOME formula,
while we fix the boson multiplicity to $n_B = 40$ and 70.
On the left-hand side, $E_{\rm freeze}$ is taken to be 15 (20) TeV for $\sqrt{s} = 13$ (27, 100) TeV.
}
\end{figure}

Fig.~\ref{fig:limit} shows the sensitivity of the HL-LHC13 (top), HE-LHC27 (middle) and FCC100 (bottom)
to the sphaleron-induced processes in the $E_{\rm thr}$ vs $p_{\rm sph}$ plane.  
On the left-hand side, the boson multiplicity has been generated according to the LOME formula,
while on the right-hand side it has been fixed to either $n_B= 40$ or $70$.
The light red bands in the left-hand-side plots denote the region where the collider is sensitive to the sphaleron processes, $N_s \in (3-10)$.
On the right-hand side we show only the boundary of this region, $N_s = 3, 10$,
and distinguish different $n_B$ with different lines. 

First of all, it is clear that the impact of boson multiplicity on the sensitivity is very mild.
This is evident by noticing the similarities between left-hand- and right-hand-side plots, as well as between different lines in the right-hand-side plots.
There exists only a mild dependence on $n_B$ in the region around $E_{\rm thr} \sim 9$ TeV for the HE-LHC27 (middle-right plot).
This is because the efficiency of the $N(p_T > 100) \geq 15$ cut differs slightly between
$n_B = 40$ and 70, as can be seen in the middle-left plot in Fig.~\ref{fig:n100_ptj1}.

In the top two plots, one can see that HL-LHC13 can probe values of $p_{\rm sph}$ up to $\sim 10^{-4}$
for $E_{\rm thr} \sim 9$ TeV. It is interesting to note that recent studies estimated the threshold energy to lie around this scale~\cite{Tye:2015tva,Tye:2017hfv}.
The high-luminosity LHC will therefore provide a meaningful constraint on this type of estimation.
In the middle two plots, we see that the reach of HE-LHC27 (middle plots) performs much better than HL-LHC13. 
In particular, one can exclude $p_{\rm sph} \gtrsim 10^{-3}$ at $E_{\rm thr} \lesssim 20$ TeV.
In fact, some studies have suggested that $E_{\rm thr}$ may be around 20 TeV \cite{Ringwald:2002sw,Ringwald:2003ns},
and therefore the HE-LHC27 will be capable of providing important information on the rate of sphaleron processes up to this limit. 
Finally we turn our attention to the bottom plots, showing the results for the FCC100.
The reach of this collider is much higher, probing e.g.\ values of the parameter $p_{\rm sph} \gtrsim 10^{-9}$ even up to threshold scales of $E_{\rm thr} \sim 22$ TeV.
Therefore, a 100 TeV collider would provide a strong constraint to the rate of EW sphaleron processes in the SM.

\section{Conclusions}
\label{sec:conclusions}

We have developed a modern event generator that models sphaleron-induced processes within the \texttt{Herwig} general-purpose Monte Carlo framework.\footnote{This event generator acts as a plug-in to \texttt{Herwig} and can be retrieved from the repository~\cite{sphaleronrepo}.}  This event generator captures the gross theoretical features of these baryon- and lepton-number-violating processes, inspired by theoretical considerations of the fermionic (flavour) and bosonic content of the final states. We have employed this development to perform phenomenological studies at hadron colliders such as the LHC (13\;TeV), as well as a high-energy upgrade (27\;TeV) and at the Future Circular Collider (100\;TeV). We have examined various relevant reconstructed distributions, describing in detail their behaviour with respect to variations of the model parameters. Our analysis demonstrates that for a wide range of the parameters of the model considered, all three colliders will provide meaningful constraints on the rates of baryon- and lepton-number violating processes at various energy scales with 3~ab$^{-1}$ of integrated luminosity. In particular, through the $p_{\rm sph}$ parameter, appearing in the parametrisation of the cross sections of Eqs.~\ref{eq:sighat} and~\ref{eq:sig0}, the HL-LHC at 13\;TeV can constrain $p_{\rm sph}$ up to $\sim 10^{-4}$ for a threshold scale of $9$ TeV, the HE-LHC at 27\;TeV could provide $p_{\rm sph} \gtrsim 10^{-3}$ at $E_{\rm thr} \lesssim 20$ TeV and the FCC-hh at 100\;TeV could probe values of this parameter $p_{\rm sph} \gtrsim 10^{-9}$ up to threshold scales of $E_{\rm thr} \sim 22$ TeV. Given these limits, our study motivates the inclusion of the study of sphaleron-induced processes in current or future collider programmes; it is quite likely that such an endeavour will illuminate the observed matter-anti-matter asymmetry of the Universe.

\acknowledgments

We would like to thank Bryan Webber for stimulating discussions in the duration of this project. AP is also grateful for interesting discussions with Bert Schellekens, Marieke Postma, Chris Korthals-Altes and Jan Smit and acknowledges support by the ERC grant ERC-STG-2015-677323.  This work is supported by the Netherlands National Organisation for Scientific Research (NWO) that is funded by the Dutch Ministry of Education, Culture and Science (OCW). In particular, AP is supported by the NWO D-ITP consortium. 
The work of KS is partially supported by 
the National Science Centre, Poland, under research grants 2017/26/E/ST2/00135 and
the Beethoven grants DEC-2016/23/G/ST2/04301. SP is grateful for the kind hospitality of Mainz Institute for
Theoretical Physics (MITP) of the DFG Cluster of Excellence PRISMA$^+$ (project ID 39083149), where this work has been 
finalised. This work has been supported in part by the European
Union’s Horizon 2020 research and innovation programme as part of the
Marie Skłodowska-Curie Innovative Training Network MCnetITN3 (grant
agreement no. 722104). SP acknowledges partial support by the COST
actions CA16201 ``PARTICLEFACE'' and CA16108 ``VBSCAN''.


\bibliographystyle{JHEP}
\bibliography{sphaleron.bib}

\end{document}